\documentclass[useAMS,usenatbib,A4paper]{mn2e}

\usepackage{times}
\usepackage{amsmath}
\usepackage{amssymb}
\usepackage{graphicx}  

\long\def\Ignore#1{\relax}

\newcommand{\Rd}        {R_{\rm d}}
\newcommand{\zd}        {z_{\rm d}}

\newcommand{\degrees} {^\circ}
\newcommand{\kms}{\mbox{$\>{\rm km\, s^{-1}}$}}
\newcommand{\Msun}{\mbox{$\rm M_{\odot}$}}
\newcommand{\sig}{\mbox{$\sigma_{e}$}}
\newcommand{\Mbh}{\mbox{$\rm M_{\bullet}$}}

\newcommand{\Mnc}{\mbox{$\rm M_{\rm NC}$}}
\newcommand{\Mcmo}{\mbox{$\rm M_{\rm CMO}$}}
\newcommand{\Msig}{\mbox{$\rm M_{\bullet}-\sig$}}
\newcommand{\Mcsig}{\mbox{$\rm M_{CMO}-\sig$}}
\newcommand{\re}{\mbox{$R_{\rm eff}$}}
\newcommand{\vrms}{\mbox{$\rm V_{rms}$}}

\newcommand{\eg}{{\it e.g.}}

\newcommand{\ie}{{\it i.e.}}

\title[Constraining the role of star cluster mergers in nuclear cluster formation]
{Constraining the role of star cluster mergers in nuclear cluster
formation: Simulations confront integral-field data}

\author[M. Hartmann et al.]{
Markus Hartmann$^{1}$\thanks{E-mail: {\tt mhartmann@uclan.ac.uk}}, 
Victor P. Debattista$^{1}$\thanks{E-mail: {\tt vpdebattista@uclan.ac.uk}, RCUK Fellow},
Anil Seth$^{2}$\thanks{E-mail: {\tt aseth@cfa.harvard.edu}, CfA Fellow},
Michele Cappellari$^{3}$\thanks{E-mail: {\tt cappellari@astro.ox.ac.uk}},
\newauthor
Thomas R. Quinn $^{4}$\thanks{E-mail: \tt{trq@astro.washington.edu}}
\\
$^{1}$Jeremiah Horrocks Institute, University of Central Lancashire, Preston, PR1 2HE, UK \\
$^{2}$Harvard-Smithsonian Center for Astrophysics, 60 Garden Street Cambridge, MA 02138, USA \\
$^{3}$Sub-department of Astrophysics, University of Oxford, Denys Wilkinson Building, Keble Road, Oxford OX1 3RH. UK \\
$^{4}$Astronomy Department, University of Washington, Box 351580, Seattle, WA 98195, USA}

\begin{document}

\date{Accepted xxx Received xxx ; in original form \today}

\pagerange{\pageref{firstpage}--\pageref{lastpage}} \pubyear{2011}

\maketitle

\label{firstpage}

\begin{abstract}
We present observations and dynamical models of the stellar nuclear
clusters (NCs) at the centres of NGC~4244 and M33. We then compare
these to an extensive set of simulations testing the importance of
purely stellar dynamical mergers on the formation and growth of NCs.
Mergers of star clusters are able to produce a wide variety of
observed properties, including densities, structural scaling
relations, shapes (including the presence of young discs) and even
rapid rotation. Nonetheless, difficulties remain, most notably that
the second order kinematic moment $\vrms = \sqrt{\rm{V}^2 + \sigma^2}$ of the models
is too centrally peaked to match observations. This can be remedied by 
the merger of star clusters onto a pre-existing nuclear disc, but the 
line-of-sight velocity $\rm V$ 
is still more slowly rising than in NGC~4244.  Our results therefore 
suggest that purely stellar dynamical mergers cannot form NCs, and that gas
dissipation is a necessary ingredient for at least $\sim 50\%$ of a NC's mass.
The negative vertical anisotropy found in NGC~4244 however requires at least $10\%$ 
of the mass to be accreted as stars, since gas dissipation and in situ star formation 
leads to positive vertical anisotropy.
\end{abstract}



\section{Introduction} 
\label{sec:intro}

Studies of the centres of galaxies across the Hubble sequence have
shown that they frequently host central massive objects (CMOs) such as
supermassive black holes (SMBHs) and massive stellar nuclear clusters
(NCs).  NCs are very common: the ACS Virgo Cluster Survey found a NC
in $66 - 88\%$ of low and intermediate luminosity early-type galaxies
\citep{Grant2005,Cote2006}.  A similar fraction, $\sim 75\%$, of
late-type spiral galaxies contain a NC \citep{Boeker2002}.  However
NCs do not appear to populate the most massive early-type galaxies
brighter than $M_B\approx-20.5$ \citep{Grant2005,Cote2006}.  Both
\citet{Rossa2006} and \citet{Cote2006} found that the luminosity of
the NC correlates with that of the host galaxy in elliptical and 
early-type spiral galaxies.  Remarkably, both NCs
and SMBHs follow the same correlation between their mass, \Mcmo, and
that of their host galaxy, with NCs on the lower end and SMBHs on the
upper end of the relation \citep{Ferrarese2006a, Wehner2006}.
Likewise, both NC and SMBH masses follow an \Mcsig\ relation with the
velocity dispersion of their host spheroid \citep{Ferrarese2006a,
Wehner2006}, with the relation for NCs parallel to, but
$\sim 10\times $ more massive at a given \sig, that for SMBHs.  These
observational facts suggest that SMBHs and NCs share a similar growth
history \citep{McLaughlin2006b}.  This motivates us to understand SMBH
growth by using the formation of NCs as their visible proxies.  An 
important advantage of NCs is that there are additional observables which can 
be measured for them, in addition to their mass. In particular one can 
characterise their kinematics and stellar population. This provides 
further constraints to their formation, which we try to exploit in 
this paper.

Two main scenarios have been proposed to explain the formation of NCs.
One scenario relies on NCs forming in situ out of gas falling into the
centre.  A number of mechanisms have been proposed for driving gas to
galactic centres, including the magneto-rotational instability
\citep{Milosavljevic2004}, gas cloud mergers \citep{Bekki2007} or the
action of instabilities \citep{Shlosman1989, Maciejewski2002,
Schinnerer2003, Schinnerer2008}.  Alternatively, NCs may form from
the merging of star clusters (SCs) after sinking to the centre under
the action of dynamical friction \citep{Tremaine1975,Miocchi2006}.
Indeed \citet{Lotz2001} found a depletion of bright globular clusters
within the inner region of dEs and similar colours for the globular
cluster and NC population which, however, are bluer than the
underlying stellar population of the host.  Analysis of NC colours in 
dE galaxies by \citet{Lotz2004} and \citet{Cote2006} suggest that many, 
but not all dE nuclei could be explained by a cluster merger scenario.  
Self-consistent simulations have found that mergers can occur and the 
resulting NC masses and sizes are consistent with those observed 
\citep{Bekki2004, Dolcetta2008, Dolcetta2008b}.

NC formation in late-type galaxies is an ongoing process, with
star formation histories that are extended, possibly constant 
\citep{Rossa2006, Walcher2006,Seth2010a}, and stars younger 
than $100$~Myrs present.  Mergers and accretion of {\it old} SCs 
is thus not a viable formation scenario, regardless of whether 
enough such SCs are available.

However, mergers and accretion of young SCs formed near the centres of 
galaxies, such as those observed in the Milky Way \citep{Figer1999a,
Figer2002}, NGC~2139 \citep{Andersen2008}, and NGC~253
\citep{Kornei2009}, could still provide a viable formation mechanism.  
\citet{Agarwal2011} used analytic modelling to show that 
infalling SCs from an empirical SC population produce NCs of the right 
mass in isolated spheroidal and late-type galaxies.  Such SCs must 
form quite close to the galaxy centres ($\lesssim$1 kpc), otherwise 
the timescales for infall due to
dynamical friction are prohibitively long \citep{Milosavljevic2004}.  In 
a sample of observed bulgeless spiral galaxies \citet{Neumayer2011} find 
that the 
dynamical friction timescales are $<2$~Gyrs for SCs with masses
$>2\times10^5$~\Msun\ within $500$~pc.  The supply of young SCs 
to the inner regions may be
enhanced because tidal forces are compressive within 10\% of the
effective radius when the density profile has a S\'{e}rsic index $n<2$
\citep{Emsellem2008}.  However, a galaxy with a constant density dark 
matter core would inhibit infall altogether \citep{Read2006, Goerdt2008b,Goerdt2010}.

The morphology and kinematics of NCs in disc galaxies provide
additional constraints on their formation.  From observations of
edge-on galaxies, \citet{Seth2006} find that their NCs are typically
elongated along the plane of the galaxy disc.  These NCs are often
compound structures, with a younger, bluer thin disc (NCD) embedded
within an older spheroidal component (NCS).  In the case of the NC in
NGC~4244, spectra reveal young ($<100$~Myr) stars in the NCD amounting
to $\sim5\%$ of the total NC mass.  Furthermore, integral-field
spectroscopy of this NC shows rotation in the same sense as the galaxy
\citep{Seth2008b}.  As we show below, it is not yet possible to
determine whether this rotation is restricted to the NCD.

Accretion without gas dissipation need not result in slowly rotating
systems: \citet{Read2008} showed that satellite galaxies that accrete
onto disc galaxies can get dragged down on to the plane of the disc and
settle directly into a rapidly rotating thick disc with $(V/\sigma)_*$
as large as 4.  Likewise, \citet{Eliche-Moral2006} find that low
density satellites can accrete onto bulges and end up rapidly
rotating.  Can NCDs form through an analogous process?  Since SCs
falling into the nucleus are likely to have formed in the mid-plane of
the galaxy, they will generally share in its rotation.

In this Paper we explore in detail whether the SC merger/accretion
scenario is able to produce the density and kinematic properties
observed in NCs. In Section \ref{sec:observations} we present the
observations of the NCs in NGC~4244 and M33.  In Section \ref{sec:jam}
we model integral field data of NGC~4244 and M33 using the
Jeans Anisotropic MGE method of \citet{Cappellari2008}.  In Section
\ref{sec:methods} we describe our simulations and methods. We present
the results of NCs formed by mergers/accretions of SCs in Section
\ref{sec:mergers}. In Section \ref{sec:accretions} we introduce an
initial NCD and follow its evolution as it accretes SCs.  In Section
\ref{sec:jams} we apply the Jeans Anisotropic MGE modelling also to
the simulations.  We discuss the implications of our simulations and
show that a purely collisionless merger origin of NCs is not
consistent with the observations in Section \ref{sec:conclusions}.


\section{Observational constraints}
\label{sec:observations}

We compare our simulations with observations of the NCs in two nearby
galaxies, M33 and NGC~4244.  These two galaxies provide us with
well-resolved nuclei for which integral field kinematic data are 
available.  While M33 is inclined by $i = 49\degrees$ \citep{Corbelli1997}, 
NGC~4244 is an edge-on galaxy.  These two objects thus give us two different
perspectives for studying the morphology and kinematics of NCs in disc
galaxies.  In this paper we assume a distance of $4.4$~Mpc for NGC~4244 
\citep{Seth2005} and $0.8$~Mpc for M33 \citep{Lauer1998}.

\subsection{Spectroscopy of NGC~4244 and M33}

\citet{Seth2006} presented {\it Hubble Space Telescope} F814W
photometry of the NC in NGC~4244, which is resolved into a nuclear
cluster spheroid (NCS) and a bluer nuclear cluster disc (NCD).  The
half mass radius, \re, of the NC obtained by fitting a King profile 
to the ACS/F814W images \citep{Seth2006} is about $5$~pc.

\begin{figure}
\includegraphics[angle=270,width=1.0\columnwidth]{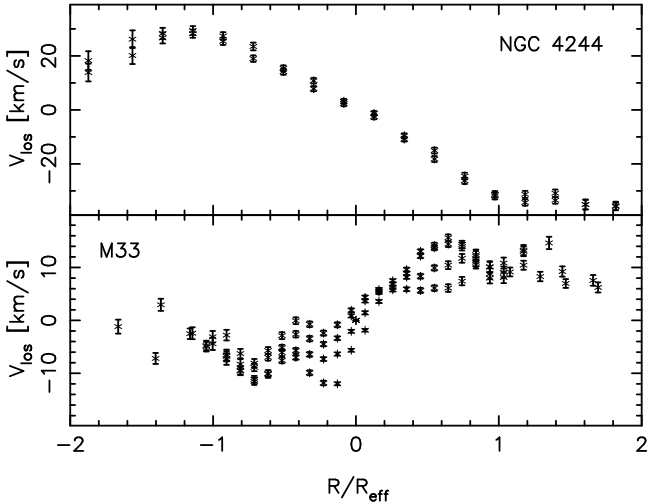}
\caption{Profiles from the NIFS data \citep{Seth2008b} of line-of-sight 
velocities along the major axis in NGC~4244 (top) and M33.  The kinematics 
were extracted from the integral-field data within a slit of $|z|<0.4$~pc.  
This corresponds to $0.08$ \re\ for NGC~4244 and $0.2$ \re\ for M33.}
\label{fig:VCurves}
\end{figure}

Here we use $K$-band spectra of NGC~4244 \citep{Seth2008b} and 
M33 (Seth et al, in prep) obtained at Gemini
North with the Near-Infrared Integral Field spectrograph (NIFS), an
image slicing field unit spectrograph.  The PSF core in both observations 
is $\sim0.1$\arcsec FWHM, with final data cubes sampled with 0$\farcs$05 
pixels.  In NGC~4244, this corresponds to $1.06$~pc pixel$^{-1}$, while in 
M33 the pixel scale is $0.19$~pc pixel$^{-1}$.  The imperfect adaptive optics 
corrections result in nearly half of the light being scattered PSF halo 
($\sim0\farcs7$ FWHM).  This leads to problems separating the rotation in 
the NCS and NCD in NGC~4244 (see Section \ref{sec:models}).  In both clusters, the 
rotation amplitude increases with radius out to the effective radius, and 
decreases at larger radii.  However, whether or not a NCD is present in the
NC of M33 cannot be determined because of the galaxy's inclination. In
Figure~\ref{fig:VCurves} we show the line-of-sight velocity profiles
of M33 and NGC~4244. 

\subsection{Isophotal shape of the NC in NGC 4244}
\label{sec:n4244}

\begin{figure}
\centering
\includegraphics[width=0.35\hsize,angle=270]{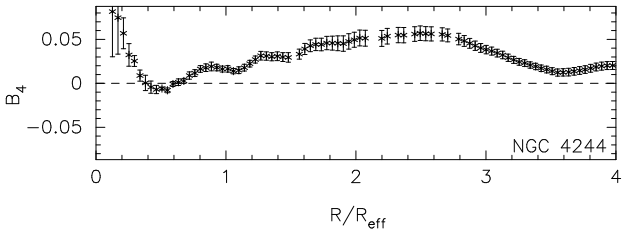}
\caption{$B_4$ for the NC in NGC~4244 derived from the integrated
$K$-band flux data \citep{Seth2008b}.  }
\label{fig:harmonics}
\end{figure}

The shape of an isophote can be quantified by the Fourier coefficients
of the expansion
\begin{equation}
I\left(\phi\right)=I_0+\sum_{n=1}^\infty A_n\sin(n\phi)+B_n\cos(n\phi)
\end{equation}
where $I_0$ is the mean intensity along the ellipse, $\phi$ is the
azimuthal angle and $A_n$, $B_n\left(n=1,2,...\right)$ are harmonic
amplitudes.  The ellipse which best fits the isophote has the
coefficients $A_n$, $B_n\left(n=1,2\right)$ equal to zero.  The
deviations from the best-fit isophote are then given by the higher
order coefficients $A_n$, $B_n\left(n=3,4,...\right)$.  The leading
residual term generally will be the $n=4$ term, which determines
whether the isophote is discy $\left(B_4>0\right)$ or boxy
$\left(B_4<0\right)$.  We use the task {\sc ellipse} in IRAF to 
measure $B_4$.  The parameter $B_4$ has been shown to correlate
with kinematic properties of the host galaxy in observations
\citep{Bender1988b, Kormendy1989, Kormendy1996} and in simulations of
galaxy mergers \citep{Naab1999, Naab2003, Naab2006}.  This result was strongly 
confirmed with the large ATLAS$^{\rm 3D}$ sample of early-type galaxies 
\citep{Cappellari2011}  which demonstrated that a significant disciness is an 
unambiguous indication of a fast rotator galaxy \citep{Emsellem2011}.  Figure
\ref{fig:harmonics} shows $B_4$ for the NC in NGC~4244, which is clearly
discy.

\subsection{Axisymmetry of the M33 nucleus}
\label{sec:m33}

The face-on axial symmetry is an important constraint 
on the formation of any stellar system.  In general the merger of stellar systems 
leads to triaxial systems, \eg\ elliptical 
galaxies are found to be triaxial \citep{Wagner1988, Naab1999, Naab2003, Naab2006}. Very 
little data are available for the face-on axial symmetry of NCs.  For 
the NC in M33 we find evidence for axisymmtry.  The inner
$0.5\arcsec$ ($= 1.8$ pc) of the NC in M33 has an apparent $b/a =
0.84$, with an average PA~$= 19.3 \degrees$ \citep{Lauer1998}, which
is close to the PA of the inner disc PA~$=23\degrees \pm 1\degrees$
within $R < 4.0$ kpc \citep{Zaritsky1989}.  Thus we can assume that 
the NC is in the same plane as the inner disk and therefore it is possible to 
determine the NC's apparent axis ratio $b/a$.  The galaxy is inclined 
by $i = 49\degrees$ \citep{Corbelli1997}, so that an oblate spheroid of
apparent axis ratio $b/a$ at an inclination $i$ would have an
intrinsic vertical-to-horizontal axis-ratio $q_0$ given by
\begin{equation}
q_0 = \sqrt{\frac{(b/a)^2 - \cos^2i}{\sin^2i}}
\label{eq:q_0}
\end{equation}
\citep{Hubble1926}.  The $q_0 \simeq 0.7$ that Eqn. \ref{eq:q_0}
implies for M33's NC is fully consistent with the range of values of
$q_0$ found in edge-on galaxies by \citet{Seth2006} (NGC~4244 having
$q_0\sim 0.5$).  Moreover we measured the kinematical PA$_{\rm kin}=36\pm18$ 
(1$\sigma$ error) from the NIFS integral-field kinematics, with the 
routine \textsc{fit\_kinematic\_pa} described in Appendix C of 
\citet{Krajnovic2006}.  Thus the kinematical misalignment is consistent 
with zero, consistent with the NC of M33 being axisymmetric.  


\section{Modelling the Observations}
\label{sec:jam}

We want to compare the kinematics of the observed NCs to those of the
simulations.  Given that integral-field kinematics are available for
the observed NCs, we use two methods which make full use of the
two-dimensional data.  One method is based on the Jeans equations and
the other is based on the
$\left(\rm V/\sigma,\varepsilon\right)$ diagram.  In the
following we show that consistent results are obtained with both
approaches.

\subsection{The Jeans Anisotropic MGE dynamical models}

The integral-field stellar kinematics for the NCs in NGC~404 
\citep{Seth2010a}, NGC~4244 \citep{Seth2006} and M33 (Seth et al, in prep), 
the few NCs for which this type of data are available, suggest
that NCs are likely not far from oblate axisymmetric systems (as we
also argued in Section \ref{sec:m33}).  A classic reference model to
quantify the rotation of axisymmetric galaxies is the {\em isotropic
rotator} \citep{Binney1978}, to which real galaxies have often been
compared \citep[e.g.][]{Satoh1980, Binney1990, vanDerMarel1991,
vanderMarel2007, vanderWel2008}.

Recent dynamical modelling studies have found that real axisymmetric
galaxies are generally best matched by models which have a nearly
oblate velocity ellipsoid, flattened along the direction of the
symmetry axis $\sigma_z \la \sigma_R \approx\sigma_\phi$
\citep{Cappellari2007, Cappellari2008, Thomas2009}.  A useful
generalisation of the isotropic rotator to quantify the rotation of
these anisotropic systems is then a {\em rotator with oblate velocity
ellipsoid} ($\sigma_z\le\sigma_R=\sigma_\phi$), which provides a good
approximation for the observed integral-field kinematics of real
galaxies \citep{Cappellari2008,Scott2009}.

To perform the measurement of the degree of rotation we used the Jeans
Anisotropic MGE (JAM) software\footnote[1]{Available from
http://purl.org/cappellari/idl/\label{jam}} which
implements an efficient solution of the Jeans equations with oblate
velocity ellipsoid \citep{Cappellari2008}.  Under that assumption the
model gives a unique prediction for the observed first two velocity
moments $\rm V$ and $\rm V_{rms}=\sqrt{V^2+\sigma^2}$, where $\rm V$
is the observed mean stellar velocity and $\sigma$ is the mean stellar 
velocity dispersion.

The luminous matter likely dominates in the high-density nuclei of the
studied galaxies. The same is true by construction for the
simulations.  For this reason the dynamical models assume that light
traces mass.  To parametrise the surface brightness distribution of
either the galaxies or the $N$-body simulations we adopted a
Multi-Gaussian Expansion \citep[MGE;][]{Emsellem1994}, which we fit to
the images with the method and software\footnotemark[1] of 
\citet{Cappellari2002}.
For a given inclination the models then have one free nonlinear
parameter, the anisotropy $\beta_z=1-\sigma_z^2/\sigma_R^2$, and two
linear scaling factors: (i) the dynamical $M/L$ and (ii) the amount of
rotation $\kappa\equiv L_{\rm obs}/L_{\rm obl}$, which is the ratio
between the observed projected angular momentum $L_{\rm obs}$ and the
one for a model with oblate velocity ellipsoid $L_{\rm obl}$
\citep[see][for details]{Cappellari2008}.  To find the best fitting
model parameters we constructed a grid in the non-linear parameter
$\beta_z$ and for each value we linearly scaled the $M/L$ to match the
$\rm V_{rms}$ data in a $\chi^2$ sense.  At the best-fitting
$(\beta_z,M/L)$ we then computed the model velocity V, further
assuming $\sigma_R=\sigma_\phi$, and measure the amount of rotation
$\kappa$ from the observed velocity.

\subsection{JAM models of observed NCs}
\label{sec:models}
\begin{figure}
\centering
  \includegraphics[width=1.0\columnwidth]{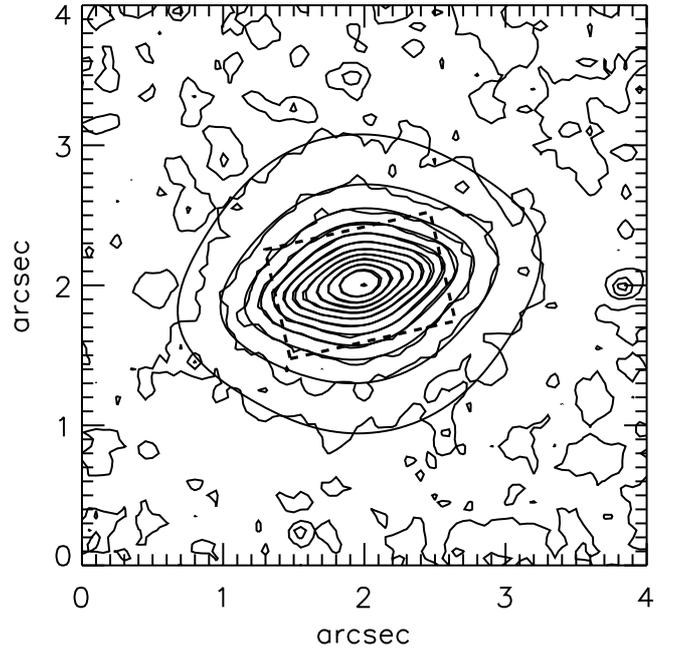}
  \caption{The contours of the surface brightness of the ACS/F814W
  image of the NC in NGC~4244, in steps of 0.5 mag, are overlaid on
  the PSF-convolved MGE model.  Both the NCD and the NCS are well
  described by the model. The scale is $\sim 21.1$~pc/arcsec with a
  half-mass radius $\simeq~0.27''$ \citep{Seth2008b}.  The 
  dashed box shows the region within which kinematic data are observed.}
  \label{fig:mge}
\end{figure}

\begin{figure*}
 \centering
  \includegraphics[width=2.0\columnwidth]{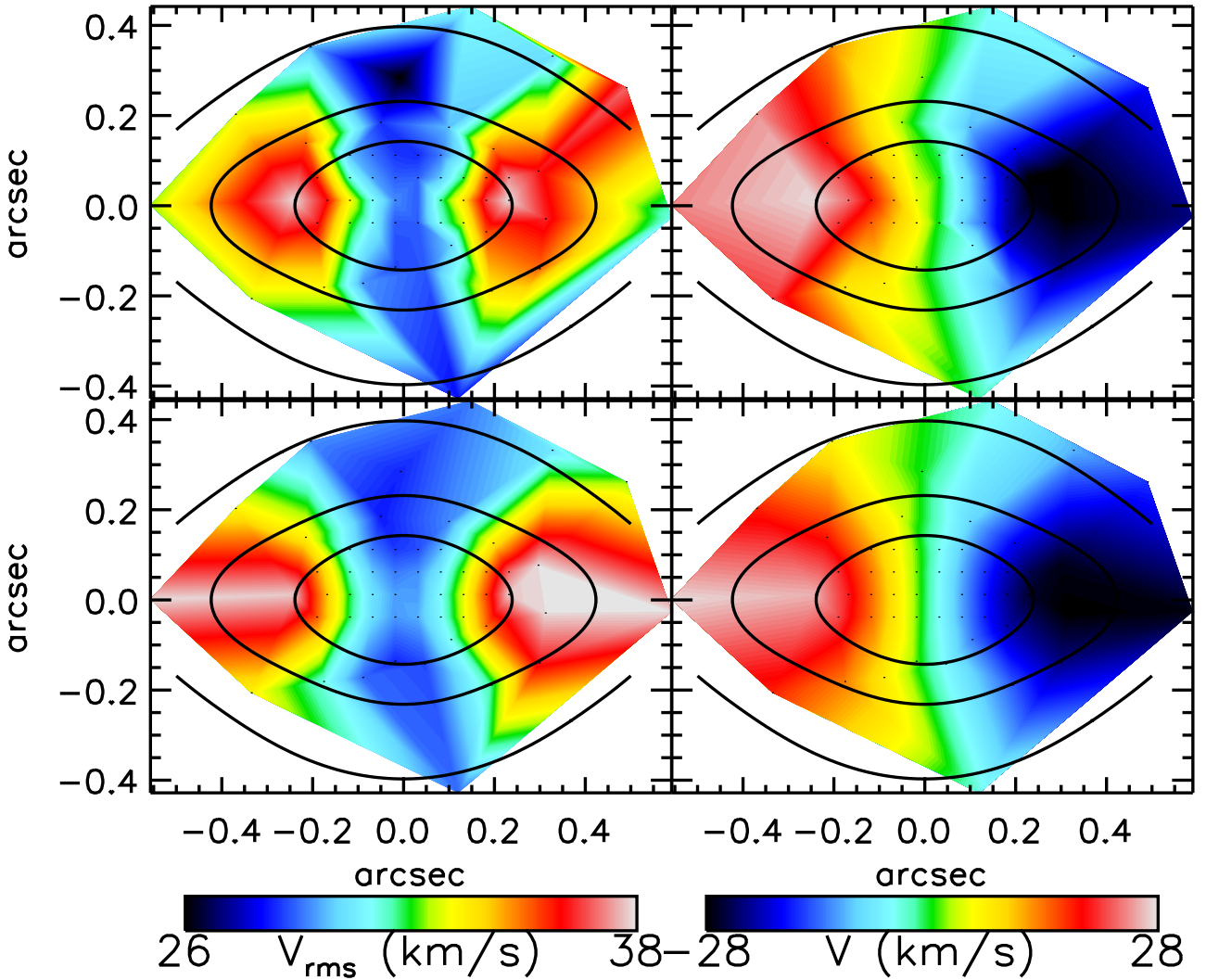}
  \caption{JAM model for the stellar kinematics of NGC~4244.  The top
  two panels show the bi-symmetrized NIFS $\rm V_{rms}$ (left) and $\rm V$
  (right).  The two bottom panels show
  the corresponding kinematics of the best fitting JAM model.  The
  contours of the surface brightness are overlaid in steps of 0.5 mag.
  The dots indicate the position of the centroids of the Voronoi bins
  for which the kinematics were extracted.  The best fitting model has
  $\beta_z=-0.2\pm0.1$ and $\kappa=0.99\pm0.05$.  The NC of NGC~4244
  has an almost perfect oblate velocity ellipsoid.}
  \label{fig:jam}
\end{figure*}

We applied the procedure to model the Gemini NIFS stellar kinematics
of the NC in NGC~4244 \citep{Seth2008b}.  The MGE fit to the nuclear
part of the ACS/F814W image is shown in Figure~\ref{fig:mge}.  We used
the ACS/F814W image for the MGE fit, because of the better known point
spread function compared to the $K$-band NIFS data.  The NC was
assumed to be a dynamically distinct component, in equilibrium in the
combined potential due to the galaxy and the NC itself.  We used all
MGE Gaussians, of both the NC and the main galaxy disc, in the
calculation of the gravitational potential, but only the nuclear
Gaussians were used to parametrise the surface brightness of the NC.
Thus our JAM models of NCs are not self-consistent.  Although this 
assumption is physically motivated and can sometimes produce significant 
differences in the kinematics of the model, in this case the results are 
indistinguishable from a self-consistent model.  The
best-fitting JAM model for NGC~4244, which has an edge-on inclination
($i=90^\circ$), is shown in Figure~\ref{fig:jam}.  It has a
best-fitting anisotropy parameter $\beta_z=-0.2\pm0.1$ and a rotation
parameter $\kappa=0.99\pm0.05$.  This implies that the NC rotates
almost perfectly as the reference rotator with oblate velocity
ellipsoid (for which $\kappa=1$).  The best-fitting JAM model has no
central intermediate mass black hole (IMBH).  However from our NIFS
data we can place an upper limit of 
$\Mbh\la1\times10^5\ \Msun$ 
on the mass of a possible IMBH inside the NC.  This is
$\la1\%$ of the mass of the cluster, ${\rm M_{NC}}=(1.1\pm0.2)\times10^7$~\Msun.  
For larger \Mbh\ the model shows a clear central peak in $\rm V_{rms}$ which 
is strongly excluded by the data.

The NC of NGC~4244 is composed of a NCD and a NCS \citep{Seth2006},
which have distinct stellar populations \citep{Seth2008b}.  While
\citet{Seth2008b} argued that both the disc and spheroid are rotating
due to the change of line strengths and optical colour, the models
suggest that the rotation signal may come only from the disc.  For
this we fitted two MGE models separately for the spheroidal and disc
component defined by \citet{Seth2006}.  We computed the best-fitting
JAM model for the $\rm V_{rms}$ like before, but then we fit $\rm V$
by setting the rotation to zero ($\kappa=0$) for the MGE Gaussians
describing the NCS.  This model with an unrotating NCS reproduces the
NIFS data as well as the model with constant anisotropy and rotation
for both the NCS and the NCD.  This is because the flat disc component
of the NC dominates the light and the rotation of the model in the
region where NIFS data are available.  More spatially
resolved/extended kinematics would be required to measure the rotation
of the spheroidal component of the NC.

We constructed a similar MGE model for the NC of M33 using the
WFPC2/F814W photometry.  For the JAM model we adopted the inclination
of the main galaxy disc ($i\approx49^\circ$; \citealt{Corbelli1997}).
In this case the observed distribution of stellar $\sigma$ from the
NIFS data is quite irregular and presents significant asymmetries,
which cannot be reproduced by an axisymmetric model.  The irregularity
in the velocity field is possibly due to granularity in the velocity
field, with individual bright AGB stars having significant influence
on the velocity and dispersion measurements.  The velocity field of 
M33 and other nearby NCs will
be discussed in more detail in an upcoming paper (Seth et al. {\em in
prep}).  As the anisotropy cannot be reliably inferred from the data,
we assumed isotropy and derived the dynamical $M/L$ from a fit to the
observed \vrms.  The rotation parameter derived from the observed V is
then $\kappa=1.02\pm0.10$, which, as for NGC~4244, is consistent with
the rotation of the reference rotator with oblate velocity ellipsoid.
The total mass of the NC is ${\rm M_{NC}}=(1.4\pm0.2)\times10^6\ \Msun$.

\subsection{Rotation from the $\left(\rm V/\sigma,\varepsilon\right)$ diagram}

\begin{figure}
\includegraphics[width=\columnwidth]{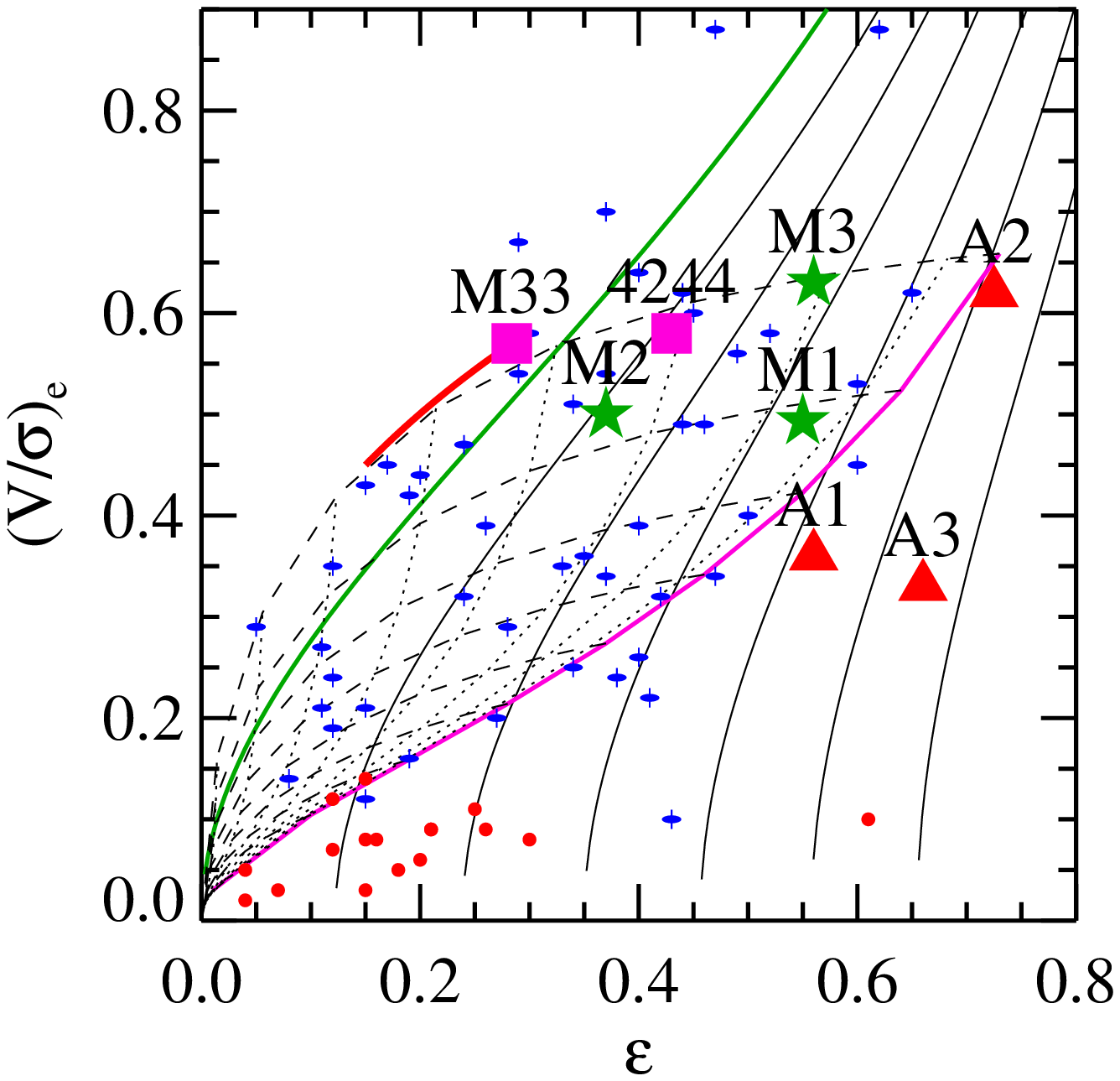}`
\caption{Real and simulated NCs on the
$\left(\rm V/\sigma,\varepsilon\right)$ diagram of
\citet{Binney2005}. The thick green line indicates the location of
edge-on isotropic models, while the other solid lines are anisotropic
models with global anisotropy 
$\delta\equiv1-2\sigma_z^2 / \left(\sigma_\phi^2 + \sigma_R^2\right)$, spaced 
at 0.1
intervals. The magenta line is the lower envelope for 
fast-rotating galaxies defined in \citet{Cappellari2007}. The dashed 
lines indicate how the
magenta line transforms at lower inclinations in 10$^\circ$ steps,
indicated by the dotted lines.  The green stars and red triangles 
indicate the location of the simulated NCs, while the
magenta squares are the observed NCs. The observed location for M33 has
been projected (red line) assuming an inclination of $i=49^\circ$. For
comparison we also plot the fast-rotator (blue
dots) and slow-rotator (red circles) early-type galaxies from
\citet{Cappellari2007}. Both the simulated and real NCs are fast
rotating and relatively close to the isotropic line.}
\label{fig:vs}
\end{figure}

An alternative classic way of quantifying rotation is given by the
$(\rm V/\sigma,\varepsilon)$ diagram \citep{Illingworth1977,
Binney1978}. Traditionally the observed $\rm V/\sigma$ quantity was
computed from the central velocity dispersion and the maximum
rotational velocity. \citet{Binney2005} updated and improved the
formalism to compute the quantity in a more robust way when
integral-field data are available. Here the availability of
integral-field kinematics for both the observations and the
simulations allow us to apply this improved method. We use the updated
formulae and define
\begin{equation}
\left(\frac{{\rm V}}{\sigma}\right)_e^2\equiv\frac{\langle {\rm V^2} \rangle}{\langle\sigma^2 \rangle} = \frac{\sum_{n=1}^{N} F_n\, {\rm V}_{n}^2}{\sum_{n=1}^{N} F_n\, \sigma_{n}^2}
\label{eq:vsigma}
\end{equation}
which we applied within 2 \re\ as a luminosity-weighted 
quantity, which we
estimate from the integral-field kinematics. Here $F_n$ is the flux
contained inside the $n$-th Voronoi bin and ${\rm V}_n$ and 
$\sigma_n$ the corresponding measured mean line-of-sight velocity and velocity 
dispersion. The ellipticity is defined following 
\citet{Cappellari2007} by a similar expression as
\begin{equation}
(1-\varepsilon)^2=q^2=\frac{\langle y^2 \rangle}{\langle x^2 \rangle} = \frac{\sum_{n=1}^{N} F_n\, y_n^2}{\sum_{n=1}^{N} F_n\, x_n^2},
\label{eq:eps}
\end{equation}
where the $(x,y)$ coordinates are centred on the galaxy nucleus and
the $x$ axis is aligned with the NC photometric major axis. We
estimate $\varepsilon$ from the individual pixels, inside a given
galaxy isophote, within the same region used for the computation of
$\left(\rm V/\sigma\right)_e$.  The main disadvantage of the
$\left(\rm V/\sigma,\varepsilon\right)$ diagram, with respect to the
JAM models, is that it does not rigorously take into account of
multiple photometric systems, like a disc and a spheroid. Moreover,
while the diagram can quantify the anisotropy, it does not provide any
information on whether it is mostly radial or tangential. Still the
diagram provides an important independent test of more detailed models
and provides an easy way of comparing simulations and observations.

We used Eqns. \ref{eq:vsigma} and \ref{eq:eps} to place the NC of M33
and NGC~4244 on the $\left(\rm V/\sigma,\varepsilon\right)$ diagram,
using our NIFS data. Given that the diagram is defined for edge-on
orientations, while M33 has an inclination of $i\approx49^\circ$, we
projected the $\left(\rm V/\sigma,\varepsilon\right)$ values for the
NC to an edge-on view following
\citet{BinneyTremaine}. The NC in NGC~4244 is seen at a nearly
edge-on orientation and is weakly anisotropic. The location of M33 is
slightly more uncertain, given the non edge-on view, but the NC is
consistent with isotropy.


\section{Numerical Methods}
\label{sec:methods}

The half mass relaxation time for the NC in NGC~4244, with
$\re\simeq5$~pc and total mass $1.1\times10^7\ \Msun$ is
$\sim10$~Gyrs.  NCs in the Virgo Cluster Survey have relaxation times
ranging from $1-10$~Gyrs \citep{Merritt2008}.  Therefore it is
reasonable to approximate NCs as collisionless systems on timescales
of $\la 1$~Gyr, allowing us to use standard collisionless codes to
simulate their evolution.

In this section we describe the $N$-body initial conditions for the
simulations.  Because we are interested in evolution at the inner
$\sim 100$~pc region of such galaxies, we neglect the dark matter
halo.

\subsection{Galactic disc model}

\begin{figure}
\centerline{
\includegraphics[width=0.5\hsize,angle=270]{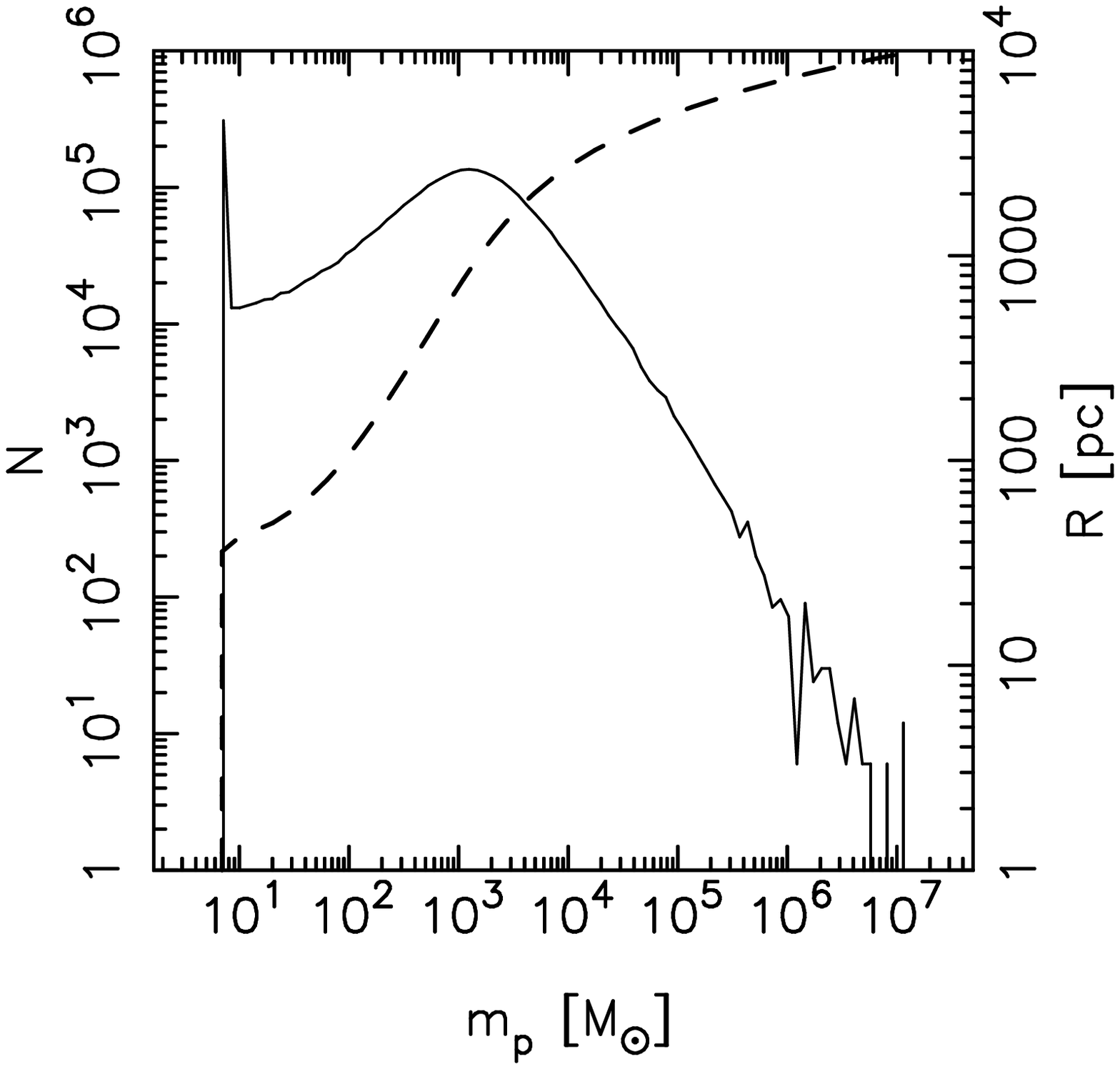}
\includegraphics[width=0.5\hsize,angle=270]{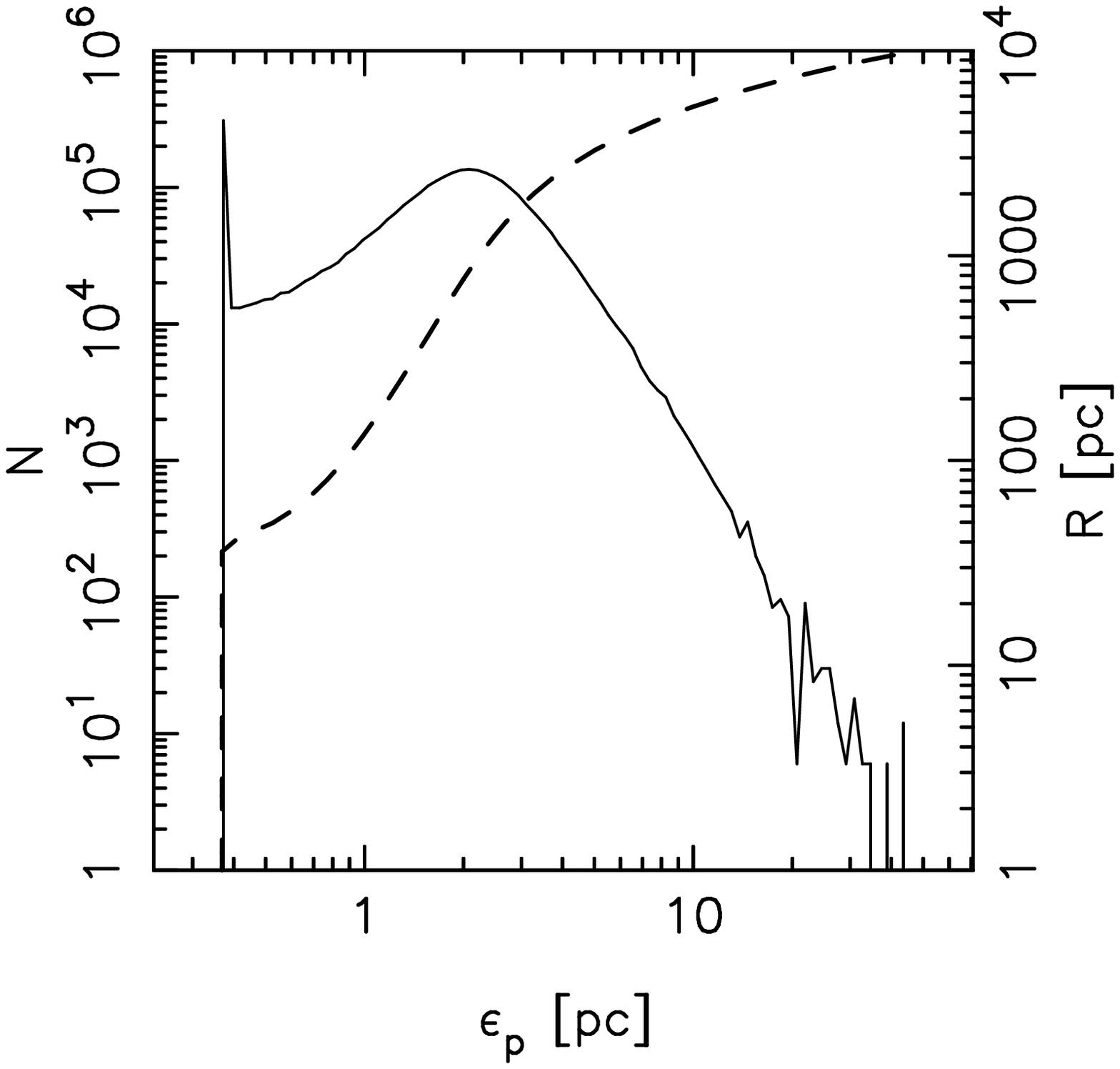}
}
\caption{The unrelaxed initial conditions of the main galactic disc.
Left: The solid line shows the number of particles (left axis) and the
dashed line the radius $R$ (right axis) versus the particle
mass. Right: The solid line represents the number of particles with a
certain softening $\varepsilon$ and the dashed line the softening of the
particles versus the radius.}
\label{fig:initialconditions}
\end{figure}

\begin{figure}
\centerline{
\includegraphics[width=\hsize,angle=270]{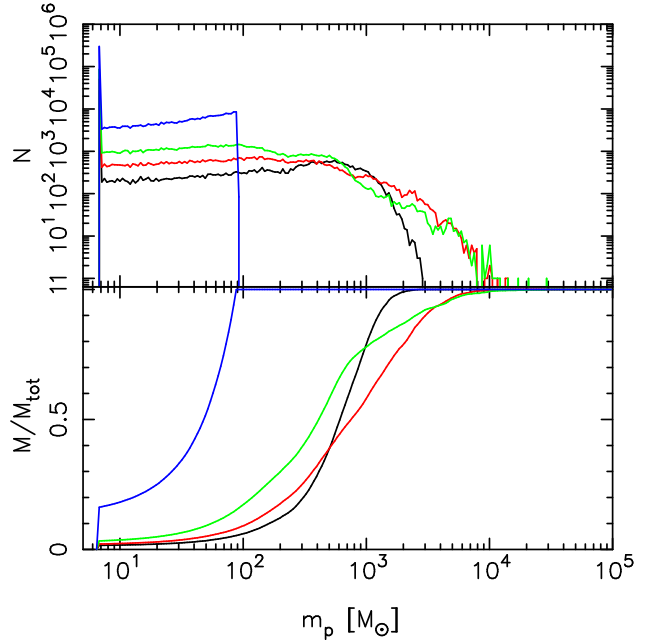}
}
\caption{
The masses of particles in the inner 100~pc of the main disc in run
M2.  Blue: initial setup; black: initial conditions after relaxing the
disc for 10~Myr; red: after 17.5~Myr; green: after 35~Myr.  The top
panel shows the number of particles of a given mass while the bottom
panel shows the cumulative mass distribution.
}
\label{fig:massdistribution}
\end{figure}

NGC~4244 and M33 are late-type, bulgeless galaxies.  We consider only
the main galactic disc and adopt an exponential profile:
\begin{equation}
\rho(R,z) = \frac{M_d}{2\pi\Rd^2}
e^{-R/\Rd}\frac{1}{\sqrt{2\pi}\zd}e^{-{z^2/2 \zd^2}}
\label{eqn:expdisc}
\end{equation}
where $M_d$ is the disc mass, $\Rd$ is the scale-length of the disc
and $\zd$ the scale-height.  We use $M_d = 10^{10}~ \Msun$, $R_d =
2.0$~kpc, $z_d = 200$~pc and truncate the disc at 5~$R_d$.  The disc
is represented by $4\times10^6$ particles.  If we had used equal mass
particles for the galactic disc, this would correspond to particle
masses of 2500~\Msun.  Such high masses would inhibit dynamical
friction on objects of mass $\sim 10^4$~\Msun, and lead to excessive
heating of any in-falling clusters.  In order to reduce such effects,
we use multi-mass disc particles, with masses ranging from 7~\Msun\
within the inner 20~pc increasing to $1.2\times10^7$~\Msun\ in the
disc's outskirts.  Figure~\ref{fig:initialconditions} shows the
distributions of masses and softenings of disc particles; the latter
is related to particle mass via $\epsilon_p \propto m_p^{1/3}$.  We
use the epicyclic approximation to set Toomre-$Q = 1.2$.  This setup
is imperfect and needs to be relaxed; moreover simulations using such
initial conditions can only be run for a few crossing times of the
main disc, since the radial migration of stars induced by disc
instabilities, including bars and spirals, would introduce massive
particles to the central regions \citep{Sellwood2002,Victor2006,
Roskar2008b}.  In order to check that the system does not rapidly
homogenise, we relaxed the system for 10~Myrs, after which main disc
particles in the central $100$~pc have masses ranging from 7~\Msun\ to
$2.8\times10^3$~\Msun.  In Figure~\ref{fig:massdistribution} we plot
the mass distribution for the disc particles within the inner 100~pc
at three different times for the longest merger simulation (M2,
described below).  The cumulative distribution (bottom panel) shows
that the mass at the centre is dominated by particles with masses
smaller than 2500~\Msun, with more than half the mass in this region
coming from particles with masses less than 1000~\Msun.

\subsection{Bulge model}

\begin{figure}
\centerline{
\includegraphics[width=0.9\hsize,angle=270]{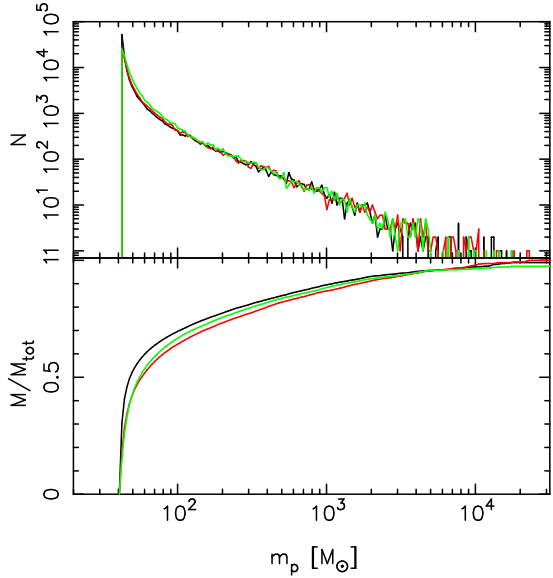}
}
\caption{ The masses of particles in the inner 30~pc of the bulge in
simulation A1.  Black: initial setup; red: initial conditions after 
the 65 Myr needed for the infall of the NCS seed; green: 
after $0.275$~Gyrs. The top panel shows the number of 
particles of a given mass while the bottom panel shows the 
cumulative mass distribution.}
\label{fig:histmassMA}
\end{figure}

On timescales comparable to the crossing time of the galaxy, dynamical
instabilities, such as bars and spirals, move particles to the central
regions.  Multi-mass particle simulations within discs therefore are
no longer possible in these cases.  We therefore set up a bulge, again
neglecting the dark matter halo, and evolve the systems in this
environment. The bulge model has a \citet{Hernquist1990} profile:
\begin{equation}
\rho(r) = \frac{aM_b}{2\pi r\left(r+a\right)^3},
\label{eqn:bulge}
\end{equation}
where $M_b$ is the bulge mass and $a$ is the scale radius.  We use
$M_b=5\times10^9$ \Msun\ and $a=1.7$~kpc.  The bulge is truncated by
eliminating all particles with enough energy to reach $r > 15a$,
therefore the density drops gently to zero at this radius
\citep{Sellwood2009}.  The bulge is populated by $3.5\times10^6$
particles with masses ranging from $40\ \Msun$ to $3.9\times10^5\
\Msun$; particle masses are selected by the weighting function $w(L)
\propto 3 + 5000 L^2$, with $L$ the specific angular momentum, 
ensuring a high resolution within the inner 160~pc \citep{Sellwood2008}.  
The softening is related to the particle
mass via $\epsilon_p \propto m_p^{1/3}$.  Unlike the disc, the bulge
has no strong instabilities, therefore the distribution of particles
remains unchanged even on timescales of a few Gyrs.  We need these
timescales to model multiple accretions of SCs.  Figure
\ref{fig:histmassMA} shows the particles distribution for simulation A1, 
described below, within 32~pc for three
different times.  This simulation shows the largest changes in the mass
distribution.  Although accretion delivers higher mass particles 
into the central region, the distribution of particles does not change
substantially.

\subsection{Star cluster models}

\begin{figure}
\centering{
\includegraphics[angle=270,width=0.9\hsize]{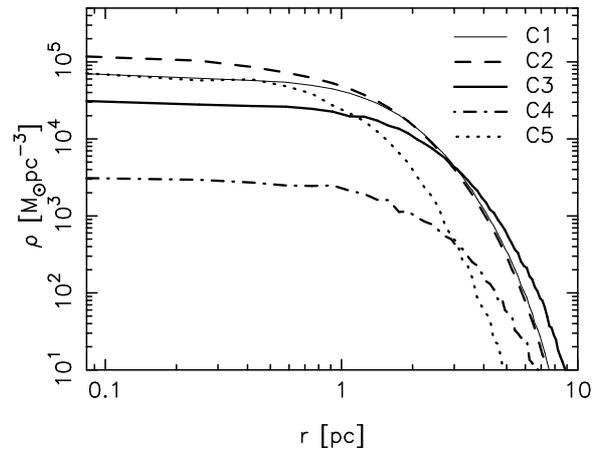}
}
\caption{Volume density of the SC models C1 - C5.}
\label{fig:VDGC}
\end{figure}

We set up model SCs, ranging in mass from $2\times10^5$~\Msun\ to
$2\times10^6$~\Msun, using an isotropic distribution function (DF):
$f(x,v) = \mathcal{F}(E)$.  The specific form we choose is a lowered
polytrope DF
\begin{equation}
f(x,v) \propto [-2E(x,v)]^{n-3/2} - [-2E_{max}]^{n-3/2}
\label{eqn:polytropedf}
\end{equation}
with polytrope index $n=2$ in all cases.  We produce equilibrium
models through the iterative procedure described in
\citet{Victor2000}.  We set up five such models, C1-C5.  All SC models
have particles of equal mass ($1.1\ \Msun$ for C1-C2, $5.0\ \Msun$ for
C3-C4 and $15\ \Msun$ for C5) and equal softening ($\varepsilon =
0.104$ pc for C1 and C2 and $\varepsilon = 0.13$~pc for C3-C5).  IMBHs
may be present in some SCs (\citealt{Gebhardt2000b, Gebhardt2005,
Gerssen2002, Gerssen2003, Noyola2008, Noyola2010}, but see also
\citealt{vanderMarel2010}).  In model C2 we include an IMBH at the
centre of the cluster C1 by adiabatically growing the mass of a single
particle with softening $\varepsilon_p = 0.042$~pc over $100$~Myrs.
Table \ref{tab:scs} lists the properties of the SC models.  The
concentration $c$ is defined as $c\equiv \log(\re/R_c)$ where \re\ 
is the half mass radius (effective radius) and $R_c$ is the core radius, 
where the surface density drops to half of the central.  
Figure~\ref{fig:VDGC} plots the volume density
profiles of the SCs.  The central density $\rho_0$ ranges from
$3\times10^3$ to $1\times10^5\ \Msun$ pc$^{-3}$; the masses and \re\
(see Tab. \ref{tab:scs}) are comparable to young massive star clusters
in the Milky Way \citep{Figer1999a,Figer2002}, in the LMC
\citep{Mackey2003}, in the Fornax Cluster \citep{McLaughlin2005}, in
irregular galaxies \citep{Larsen2004} and in interacting galaxies
\citep{Bastian2006}.

\begin{table}
\centering
\begin{minipage}{0.5\textwidth}
\caption{The SCs used in the simulations.  $\rm M_*$ is the stellar
mass of the SC, \re\ is the effective (half-mass) radius, $c$ is the
concentration (see text for definition) and \Mbh\ is the mass of the
black hole if one is present.}
\centering
\begin{tabular}{@{}cccccc@{}}
\hline
\hline
Model	 & $\rm M_*$ & \re & c & \Mbh    \\
	& [\Msun] & [pc] &   & [\Msun]   \\
\hline
C1    & $2.2\times10^6$ & 1.72 & 0.07 & -  \\
C2    & $2.2\times10^6$ & 1.60 & 0.07 & $4.4 \times 10^4$ \\
C3   & $2.0\times10^6$ & 2.18 & 0.12 & -  \\
C4   & $2.0\times10^5$ & 1.11 & 0.12 & - \\
C5   & $6.0\times10^5$ & 1.11 & 0.16 & - \\
\hline
\label{tab:scs}
\end{tabular}
\end{minipage}
\end{table}

\subsection{Bare NCD model}

If direct formation of a NCD via gas inflows precedes the full
formation of a NC, how does the accretion of SCs alter the properties
of a NCD?  In order to explore this, we generated a bare NCD model by
adiabatically growing, over a period of $0.5$~Gyr, a disc at the
centre of the bulge model.  The NCD is exponential of the type in
Eqn.~\ref{eqn:expdisc}, with a scale-length $R_d = 9.5$~pc and
scale-height $z_d = 0.1 R_d$.  The disc, truncated at a radius of $5\ R_d$, 
consists of $2\times10^5$ particles each with softening
$\varepsilon = 0.13$~pc .  The final total mass of the NCD is $1
\times10^6$~\Msun.  We set the kinematics of the grown disc to give
constant $z_d$ and Toomre-$Q = 1.2$, as described in
\citet{Victor2000}.  For this we calculated the potential using a
hybrid polar-grid code \citep{Sellwood2003}.

\subsection{Numerical parameters}
\label{ssec:simulations}

All the simulations in this paper were evolved with \textsc{pkdgrav}
\citep{Stadel2001}, an efficient, parallel tree-code.  We used an 
opening angle $\theta=0.7$ in all simulations.  {\sc pkdgrav}
is a multi-stepping code, with timesteps refined such that $\delta t =
\Delta t/2^n < \eta (\varepsilon/a)^{1/2}$, where $\varepsilon$ is the
softening and $a$ is the acceleration at a particle's current
position.  We set $\eta = 0.1$ in all cases.  Simulations A1-A3 used
base timestep $\Delta t = 10^5$ years, whereas all other simulations
used half this value $\Delta t = 0.5 \times 10^5$ years.


\section{Results of the Merger Simulations}
\label{sec:mergers}

We ran three simulations in which eight massive star clusters were
allowed to merge at the centre of the disc to form a NC.  We use SC
models C1 and C2 in these simulations because these SCs are
sufficiently massive and concentrated to not evaporate too rapidly
\citep{McLaughlin2008} and to have orbit decay times due to dynamical
friction less than 3~Gyr from a radius of 1~kpc
\citep{Milosavljevic2004}.  The SCs were placed at radii ranging from
$14$ to $92$~pc with velocities between $8$ and $13$~\kms, which are
$\sim 60\%$ of the local circular velocity.  In run M1 the SCs are all
initially in the mid-plane; in run M2 instead the SCs are vertically
offset from the mid-plane by up to $67$~pc, with tangential velocities
similar to M1 and vertical velocities up to $1$~\kms.  Finally, run M3
is identical to run M1, but uses SC C2 instead of C1.  Details of
these simulations are listed in Table \ref{tab:mergsims}.

SCs merge after $13-36$~Myr, with the longest time needed in M2.  One
SC failed to merge in run M1 and was excluded from all analysis.

\begin{table}
\centering
\begin{minipage}{0.5\textwidth}
\caption{The merger simulations.  N(SC) gives the number of star
clusters used and column SC lists which star cluster from those in
Table \ref{tab:scs} are used in the simulation. Column host shows 
which host galaxy model is used as initial conditions.}
\centering
\begin{tabular}{@{}rcccl@{}}
\hline
\hline
Run & N(SC) & SC & host & Comments \\
\hline
M1 & 8 &  C1 & disc &  SCs at mid-plane of main disc \\
M2 & 8 &  C1 & disc &  6 SCs offset from main disc mid-plane \\
M3 & 8 &  C2 & disc &  SCs at mid-plane of main disc \\
A1 & 10 & C5 & bulge &  multiple accretion of SCs onto NCS \\
A2 & 20 & C4 & bulge &  accretion of SCs onto a NCD \\
A3 & 20 & C4 & bulge &  like A2 with $50\%$ retrograde orbits\\
\hline
\label{tab:mergsims}
\end{tabular}
\end{minipage}
\end{table}

\subsection{Structural properties}

\begin{figure} 
\centering
\includegraphics[width=1.0\hsize]{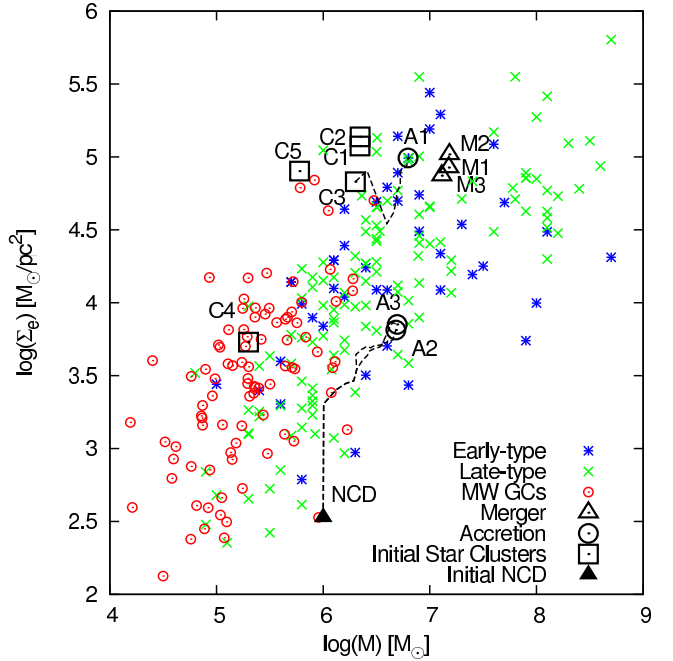}
\caption{Comparison of the simulated and observed NCs in the mean
surface density within \re\ versus total NC mass plane. We compare
with the observed NCs of early-type and late-type spiral galaxies as
well as Milky Way globular clusters. The initial SCs are shown by the
open black squares while the remnant NCs are shown by black triangles
for multiple mergers and black circles for multiple accretions.  
The tracks of the evolving NC in the multiple accretion 
simulations are indicated by dashed lines.}
\label{fig:ScalRel} 
\end{figure}

We measured mass surface density profiles of the remnant NCs
viewed face-on and obtained \re\ by fitting S\'{e}rsic or King
profiles.  The King profile clearly fits the 
profiles better and we present results only of this fit throughout. 
The King profile is given by \citep{King1962}
\begin{equation}
\Sigma \left(R\right)=k\left[X^{-1/2}-C^{-1/2}\right]^2
\label{eqn:king}
\end{equation}
with normalisation constant $k=\Sigma_0\left[1-C^{-1/2}\right]^{-2}$,
$X\left(R,R_c\right)=1+\left(R/R_c\right)^2$ and
$C\left(R_t,R_c\right)=1+\left(R_t/R_c\right)^2$, where $R_c$ is the
core radius and $R_t$ is the tidal radius at which the projected
density drops to zero.  Integration yields the cumulative form of the
King profile for $R\leq R_t$:
\begin{equation}
M\left(R\right)=\pi R_c^2 k\left[\ln{X}-4\frac{X^{1/2}-1}{C^{1/2}}+\frac{X-1}{C}\right]
\label{eqn:kingms}
\end{equation}
which is the mass in projection within a cylinder of radius R.  \re\ can be 
approximated by 
\begin{equation}
\re=R_c \left[ e^{\left(M/2\pi R_c^2k\right)-1}\right]^{1/2}.
\label{eqn:re}
\end{equation}
The merger remnants have \re\ in the range $4.3 - 4.7$~pc with mass
fractions from $87 - 97\%$ of the total {\it merged} SC mass.  These
masses are consistent with those of observed NCs \citep{Cote2006,
Geha2002, Boeker2004, Walcher2006, Seth2006}.  Figure
\ref{fig:ScalRel} plots the mean surface density within \re\ versus
the total mass of the merger remnants and compares them to NCs in
early-type galaxies \citep{Cote2006}, late-type spiral galaxies
\citep{Carollo1997, Carollo1998, Carollo2002, Boeker2002, Seth2006}
and Milky Way GCs \citep{Walcher2005}.  As shown by
\citet{Dolcetta2008b}, we find that SC mergers produce remnants which
have structural properties in good agreement with observed NCs.

\begin{figure}
\centering
\includegraphics[width=\hsize]{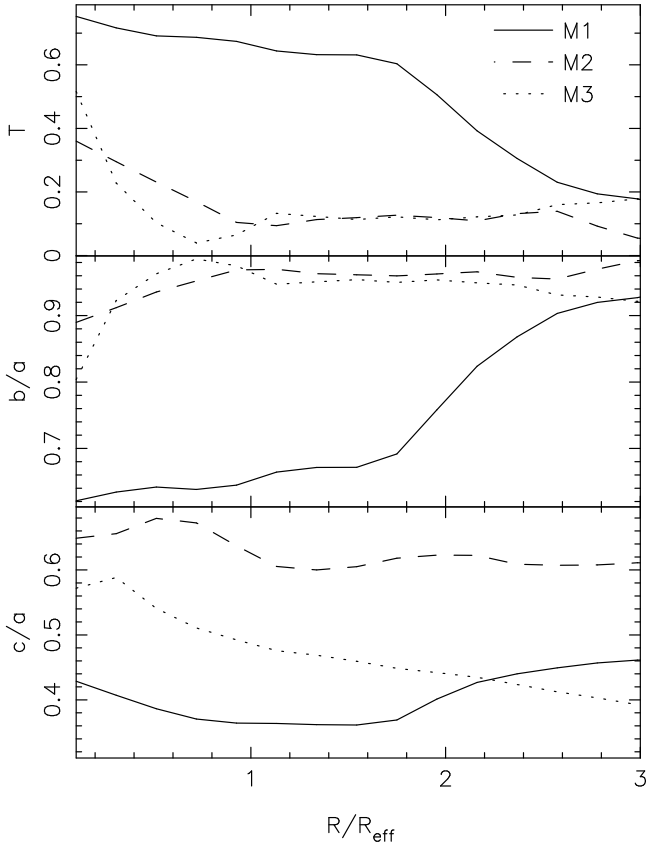}
\caption{The 3-D shape of the NC in runs M1 (solid lines), M2 (dashed
lines) and M3 (dotted lines).  The top panel shows the triaxiality $T$,
the second row $b/a$ and the third $c/a$.}
\label{fig:shape}
\end{figure}

As was shown already by \citet{Bekki2004}, the merger remnants can be
triaxial.  We measured the ellipticities viewed face-on and
edge-on of the simulated NCs using Eqn. \ref{eq:eps}, obtaining 
the isophote at \re\ using the task {\sc ellipse} in IRAF.  The NC 
in M1 is significantly non-axisymmetric, with
face-on mean ellipticity $\varepsilon_{FO} \simeq 0.37$.
We also measured the 3-D shape using the moment-of-inertia tensor as
described in \citet{Victor2008}.  Figure~\ref{fig:shape} plots the
density axes ratios and triaxiality, $T = (a^2 -b^2)/(a^2 - c^2)$
\citep{Franx1991}, of the remnant NCs.  In run M1 the NC is triaxial
within 2~\re.  Models M2 and M3 explore two ways of producing more
axisymmetric NCs.  In M2 we start the SCs off the mid-plane.  This
makes the NC oblate, with $\varepsilon_{FO} \simeq 0.05$.
In run M3 instead we add $2\%$ IMBHs to the SCs which again results in
an oblate NC, also with $\varepsilon_{FO} \simeq 0.03$.

We find remnant edge-on ellipticities at 3 \re\
$\varepsilon_{EO}$ in the range of $0.36 - 0.56$.  These 
values are consistent with the range of observed ellipticities 
$0.39 - 0.89$ \citep{Seth2006}.

\begin{figure}
\centering
\includegraphics[width=0.7\hsize,angle=270]{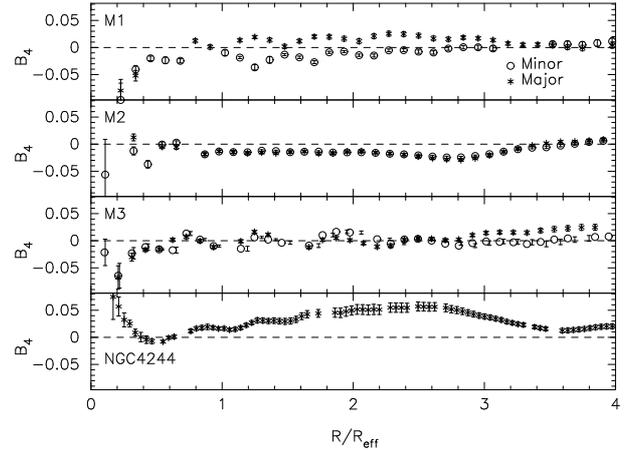}
\caption{The $B_4$ parameter for the merger simulations.  From top to
bottom these are M1, M2, M3 and the NC of NGC~4244 for comparison.
The remnant NCs have $B_4\la 0$, whereas the observed NC in NGC~4244
has clearly discy isophotes, i.e. positive $B_4$.}
\label{fig:shaperesid}
\end{figure}

Figure~\ref{fig:shaperesid} shows $B_4$ for the edge-on view.  In the
triaxial NC of M1, $B_4$ varies with viewing angle but is always
negative, \ie\ the NC is boxy.  The NCs in M2 and M3 are also
boxy. The merger of SCs cannot produce isophotes as discy as
observed in the NC of NGC~4244.

\subsection{Remnant kinematics}

\citet{Bekki2004} found that his merger remnants were rotating, while 
\citet{Dolcetta2008} found that merger remnants are kinematically 
distinct from the main disc/bulge.  

\begin{figure*}
\centering
\begin{tabular}{ccc}
\includegraphics[width=0.31\textwidth]{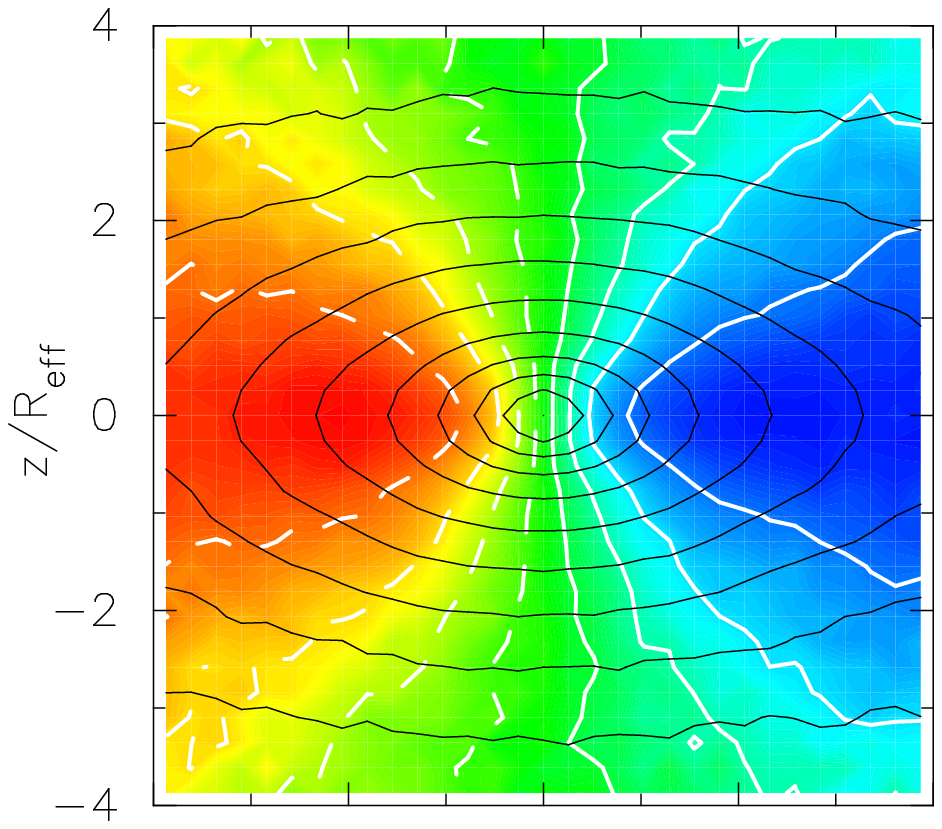} &
\includegraphics[width=0.26\textwidth]{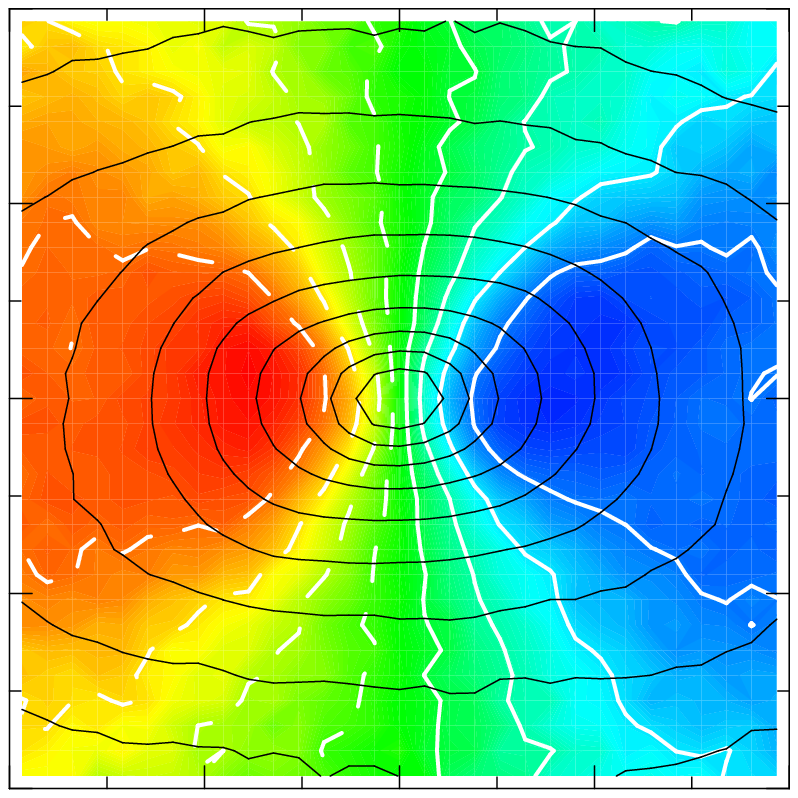} &
\includegraphics[width=0.34\textwidth]{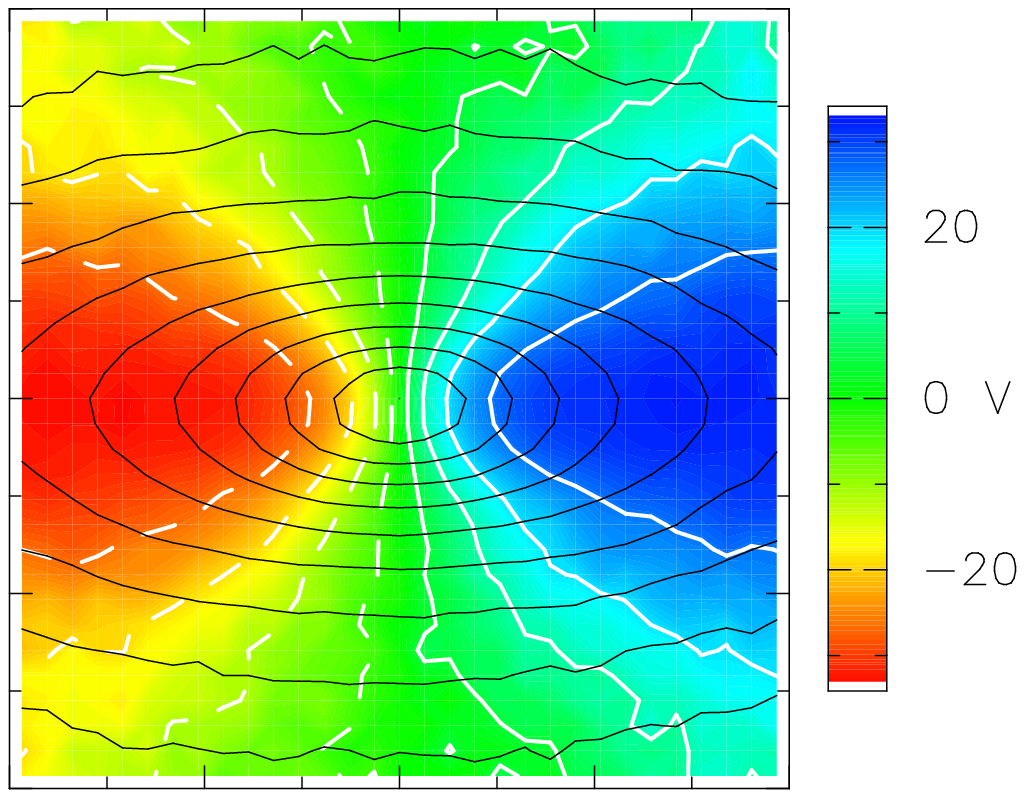} \\
\includegraphics[width=0.31\textwidth]{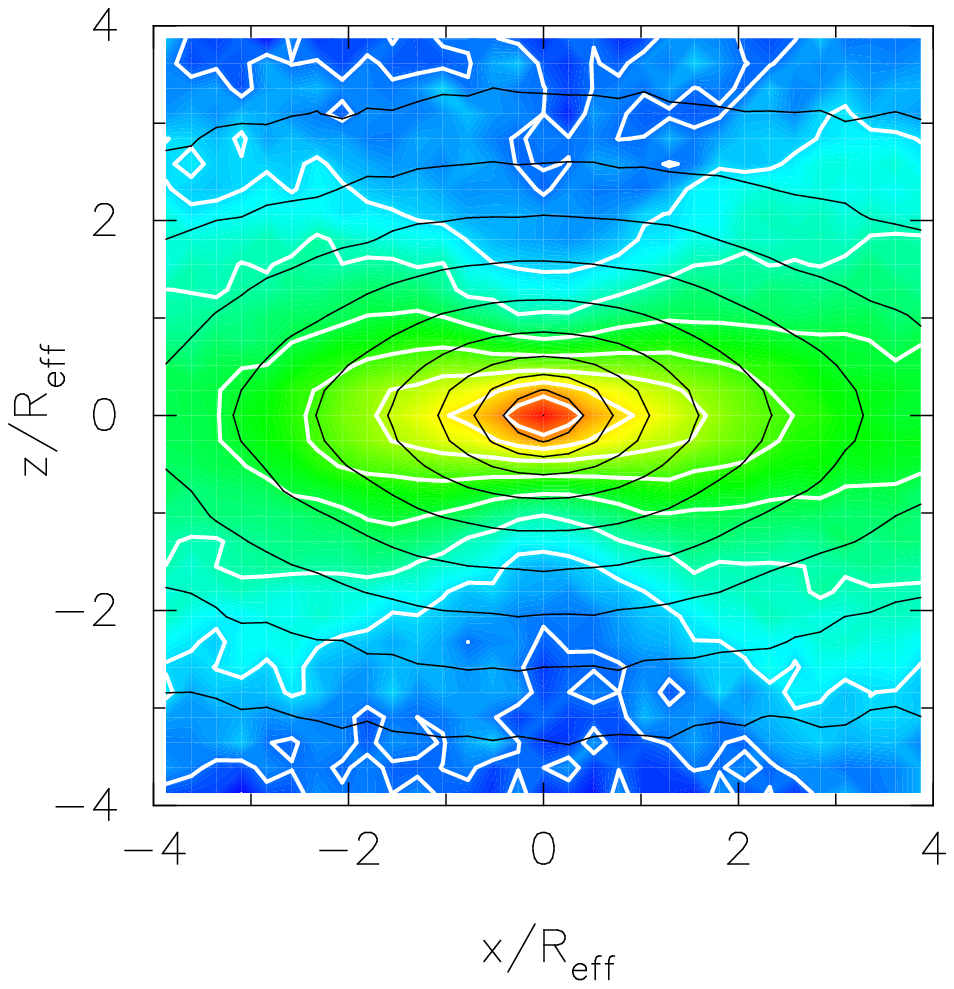} &
\includegraphics[width=0.27\textwidth]{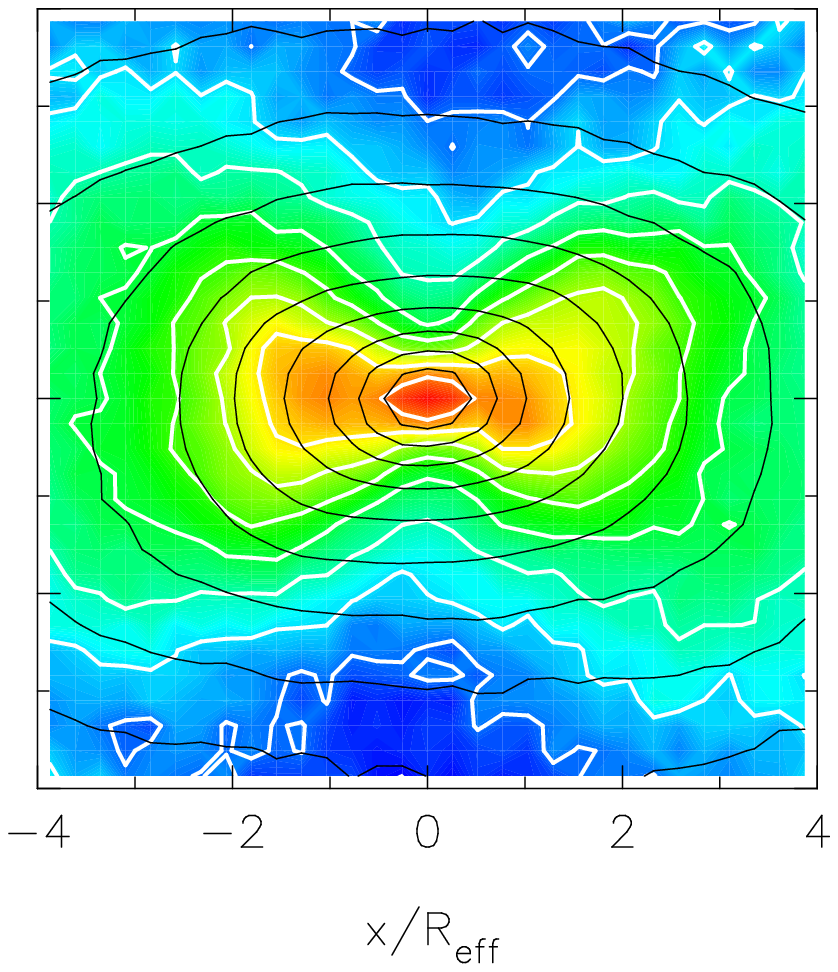} &
\includegraphics[width=0.365\textwidth]{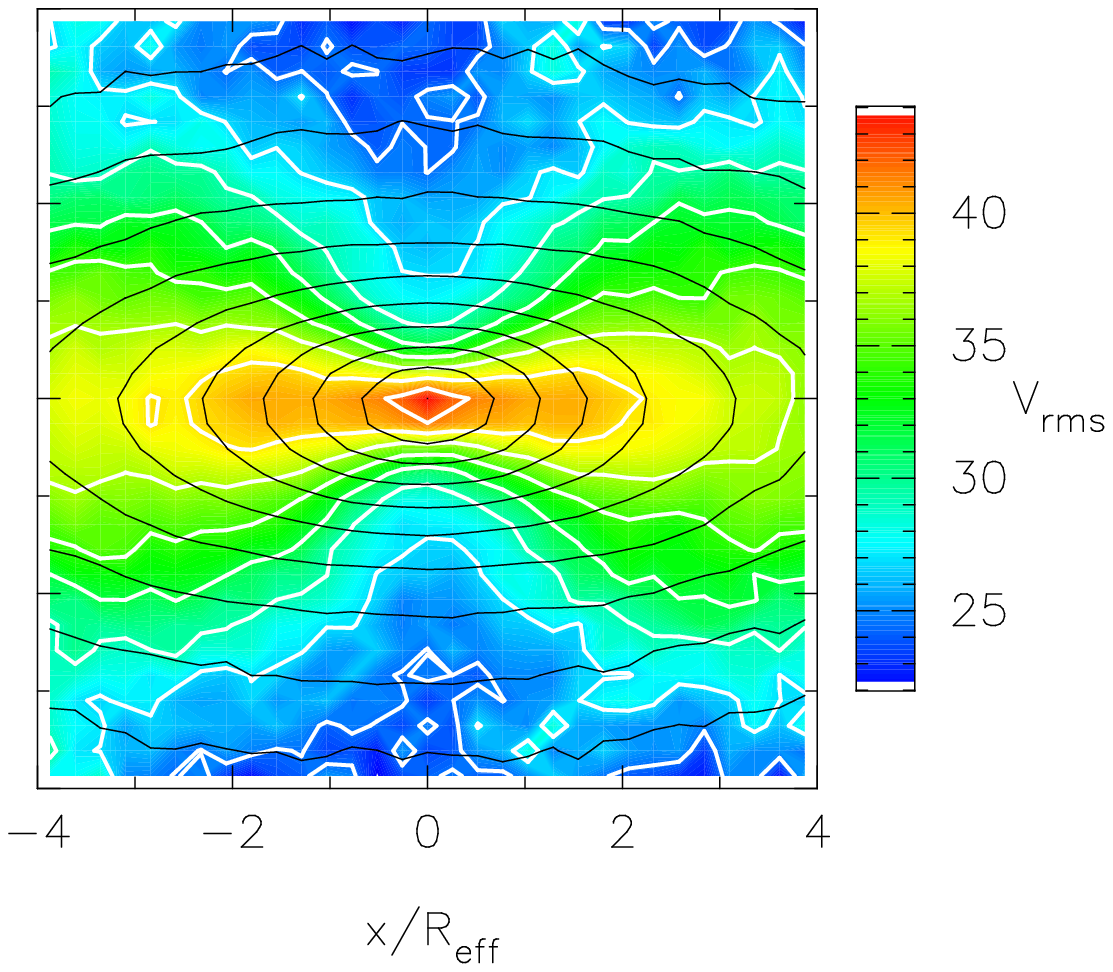} \\
\includegraphics[width=0.32\textwidth,angle=270]{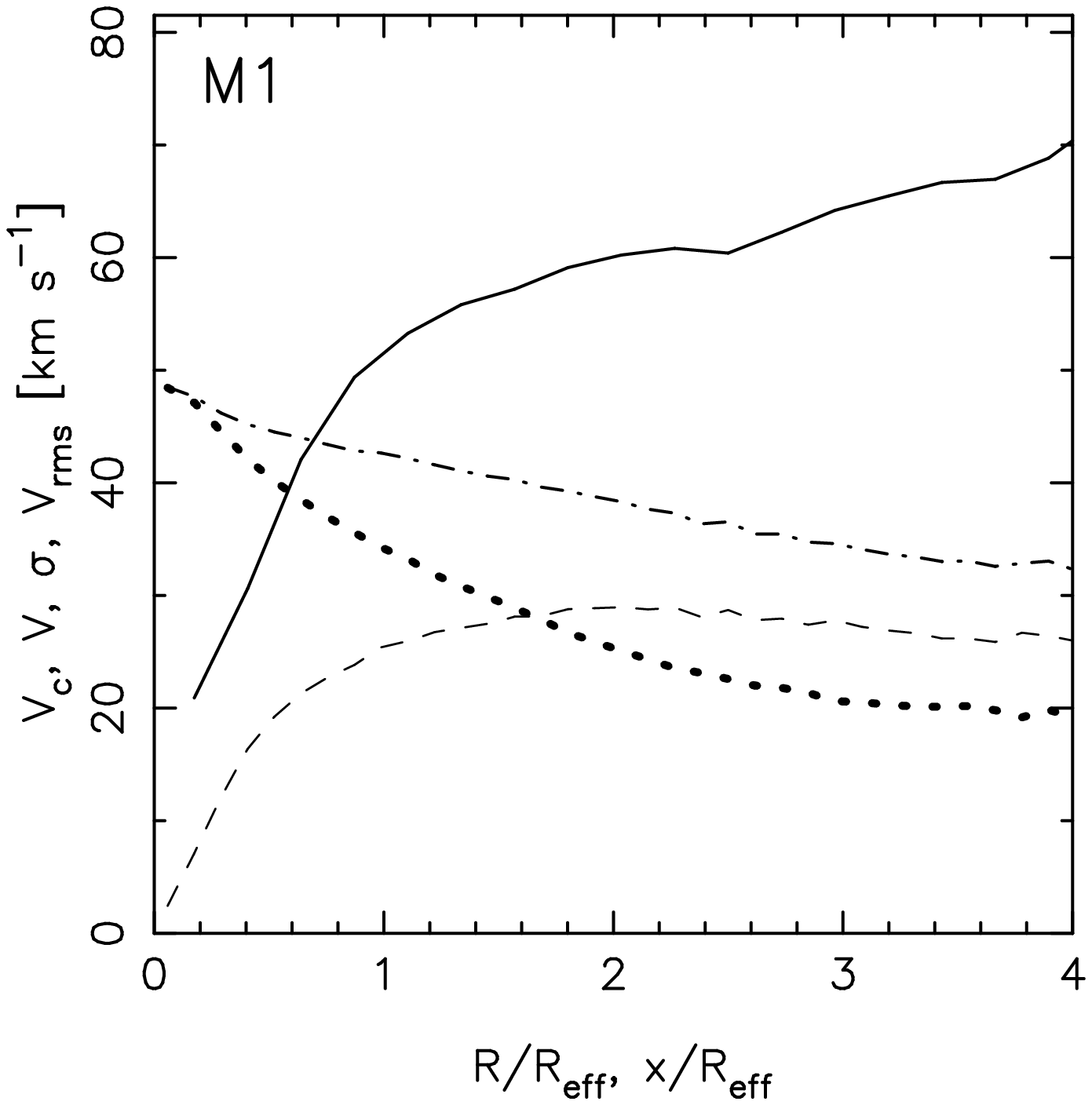} &
\includegraphics[width=0.32\textwidth,angle=270]{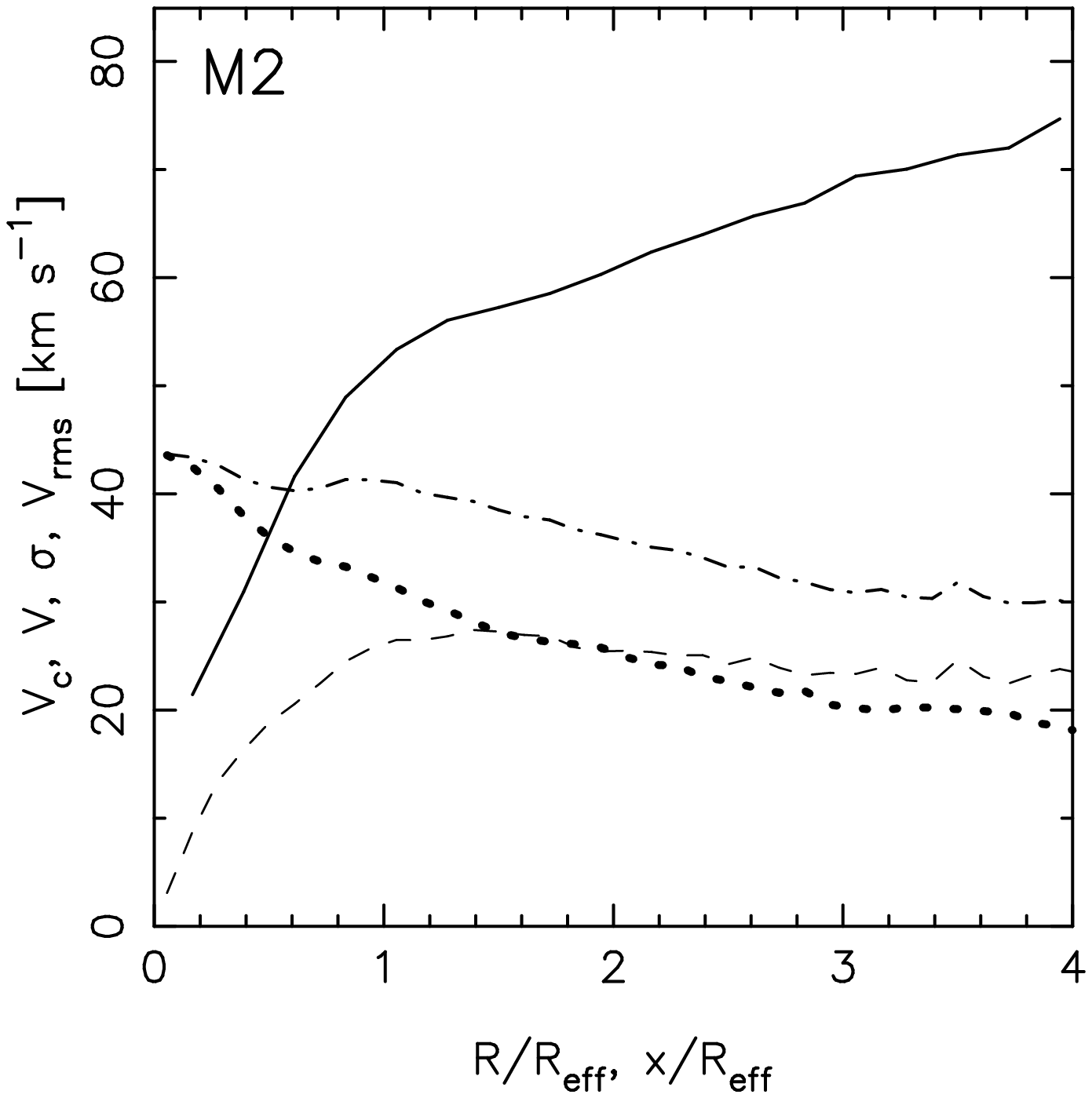} &
\includegraphics[width=0.32\textwidth,angle=270]{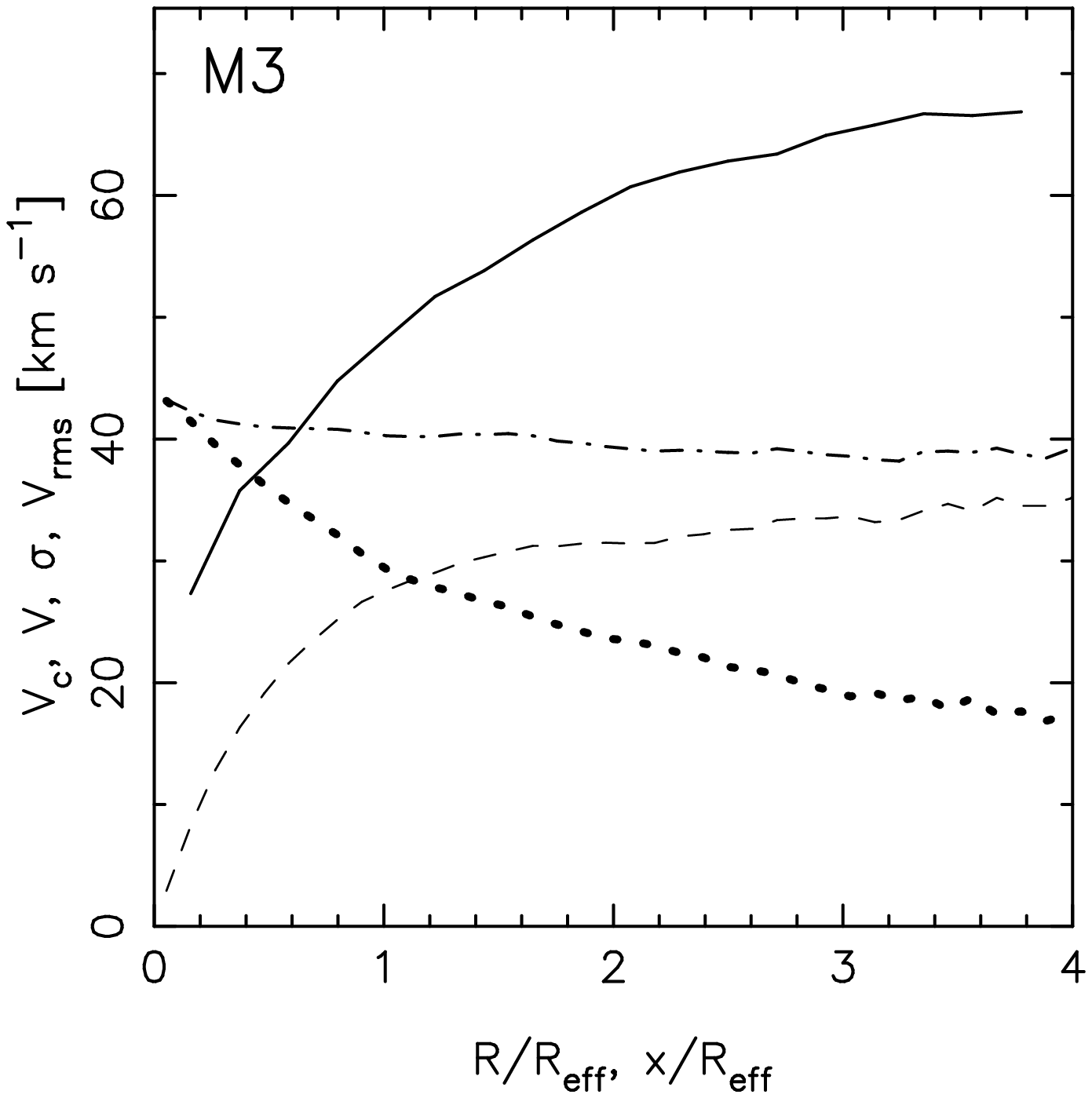} \\
\end{tabular}
\caption{Velocity (top row) and \vrms\ (middle row) fields within $4$ \re\ 
for M1 (left), M2 (middle), and M3 (right). The velocity fields show a large 
scale rotation.  In all cases, \vrms\ is centrally peaked.  In the top two rows black 
contours show log-spaced density while the white contours show the kinematic 
contours corresponding to each panel. The bottom row 
plots the rotation curve $\rm V_c\left(R\right)$ (solid lines), the line-of-sight 
velocities $\rm V\left(x\right)$ (dashed lines), the line-of-sight velocity dispersions 
$\sigma\left(x\right)$ (dotted lines) and the root-mean-square velocities 
$\rm V_{rms}\left(x\right)$ (dashed-dotted lines) along the major axis.}
\label{fig:vfieldmerger}
\end{figure*}

Figure~\ref{fig:vfieldmerger} shows the edge-on line-of-sight
kinematics of the remnant NCs. They are all clearly strongly
rotating. However the second moment of the velocity,
\vrms, shows that the merger remnants are so dominated by dispersion
at the centre that \vrms\ is centrally peaked, contrary to what is
seen in the NC of NGC~4244 (Figure~\ref{fig:jam}).  In the
bottom row we show the rotation curve ${\rm V_c}\left(R\right)$, the
line-of-sight velocity ${\rm V}\left(x\right)$, the line-of-sight velocity
dispersion $\sigma\left(x\right)$ and the root-mean-square 
velocity ${\rm V_{rms}\left(x\right)}$ of the merger remnants.  
Velocities in M1-M3 peak at larger radii than those observed in 
NGC~4244 and M33 (see Figure~\ref{fig:VCurves}). 

\begin{figure}
\centering
\includegraphics[angle=270,width=\hsize]{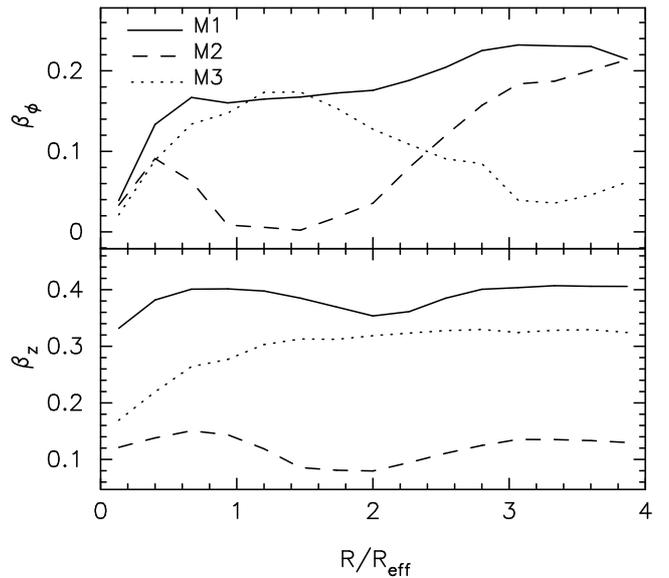}
\caption{Final anisotropy $\beta_\phi$ (top) and $\beta_z$ (bottom) in
runs M1, M2 and M3.}
\label{fig:cylanmerger}
\end{figure}

In Figure~\ref{fig:cylanmerger} the anisotropies $\beta_\phi =
1-\sigma_\phi^2/\sigma_R^2$ (top panel) and $\beta_z =
1-\sigma_z^2/\sigma_R^2$ (bottom panel) show that the remnant NCs are
all radially biased  within $4\ \re$.  M2 is less radially biased
than M3, which may seem surprising at first, but the radius of the
sphere of influence of the IMBH in M3 is less than 1~pc, explaining
the absence of a tangential bias at \re.  The vertical pressure
support is generally smallest.  The initial vertical energy of the SCs
in run M2 imparts larger vertical random motions, leading to the
smallest $\beta_z$ and smallest edge-on ellipticity
$\varepsilon_{EO}$.  Radial anisotropy has been noted in the past as a
signature of the merging process \citep[e.g.][]{Burkert2005,
Bournaud2007, Victor2008, Thomas2009}.  The presence of plausible
IMBHs does not alter this result.

\subsection{Accretion onto Super Star Clusters}

The super star cluster (SSC) found by \citet{Kornei2009} in the
nuclear region of NGC~253 seems destined to fall into the centre of
the galaxy and form the basis of a NC.  How would the accretion of
further SCs alter the structure and kinematics of such a seed NC?
Observed Milky Way globular clusters are found to be nearly isotropic
\citep{GebhardtPhD1994, Gebhardt1995b}.  How much mass needs to be
accreted to appreciably alter the isotropic distribution?  In run A1 we
study the accretion of SCs onto a NCS by introducing a spherical
isotropic SC inside the bulge model.  We form the NCS by letting a
massive star cluster fall to the centre.  We used SC C3 and started it
at $127$~pc on a circular orbit allowing it to settle at the centre over
65~Myrs, before we start the accretion of 10 SCs.  We use model C5 for
the accreted SCs, starting them on circular orbits at a distance of
$32$~pc from the centre.  In total the mass accreted corresponds to
$\sim3\times$ the NCS's initial mass.  Each accretion is allowed to
finish before a new SC is inserted.  A single accretion on average
requires $\sim 20$~Myrs.  Table \ref{tab:mergsims} gives further
details of this simulation.

\begin{figure}
\centering
\includegraphics[angle=270,width=\hsize]{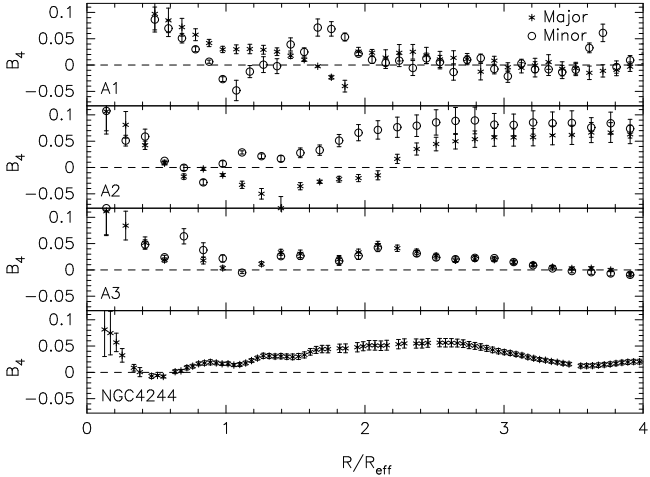}
\caption{$B_4$ for edge-on projections of runs A1, A2 and A3 compared
with the NC in NGC~4244 (bottom).  For the simulations we measured the
isophotes along the edge-on semi-principal axes.}
\label{fig:harma}
\end{figure}

After each accreted SC we measure \re\ by fitting a King profile (Eqn.
\ref{eqn:king}) to the mass surface density profile of the NC.  The
final remnant has $\re \sim 3.2$~pc and structural properties
consistent with observed NCs, as shown in Figure~\ref{fig:ScalRel},
which tracks the evolving NC.  The NC becomes triaxial after it has
doubled in mass.  The final  $\varepsilon_{FO} \simeq 0.17$ and 
$\varepsilon_{EO} \simeq 0.51$; the latter is in the observed
range \citep{Seth2006}.  The final triaxiality $T\simeq0.4$.  
Figure~\ref{fig:harma} shows that the remnant NC in A1 is discy 
($B_4 > 0$).

Multiple accretion of young SCs allows us to also explore the effect
of a different $M/L$ ratio for a {\it young} accreted SC.  For the 
NC in NGC~4244 the structural properties are obtained from $I$-band 
observations \citep{Seth2006} and the kinematics from $K$-band data 
\citep{Seth2008b}.  Therefore we obtain both the $M/L$ in 
the $I$-band and $K$-band for the NCD assuming a single stellar population
with an age about 70~Myrs and a metallicity of $\mathrm{[Fe/H]}=-0.4$
and for the NCS assuming two stellar populations, the first with an
age of 1~Gyr with the same metallicity and the second of 10~Gyrs and a
metallicity of $[\mathrm{Fe/H}]=-1.4$ in NGC~4244.  Using the stellar
evolution code of \citet{Maraston1998,Maraston2005} this gives
$M/L\approx0.2$ for the NCD and $M/L\approx1.6$ for the NCS in the 
$I$-band and $M/L\approx0.1$ for the NCD and $M/L\approx0.8$ for 
the NCS in the $K$-band.  Throughout the rest of this paper, we adopt 
the $M/L$ values in the $I$-band for the analysis of structural 
properties and $M/L$ values in the $K$-band for the kinematics, 
assuming that stars from the last accreted SC are young while the 
rest of the stars are old, to obtain a luminosity-weighting.

Adopting these $M/L$ values, the final edge-on ellipticity becomes
$\varepsilon_{EO}\simeq 0.56$.

We produce a luminosity-weighted density map of the final NC and  
fit two components, an elliptical King and an exponential 
disc profile, as in \citet{Seth2006}, to measure the structural 
properties of the NCS and NCD component.  The NCS has $\re\sim3.2$~pc 
and a flattening $\varepsilon\sim0.59$ while the NCD has $z_0\sim15.8$~pc 
and scale length $R_d\sim1.6$~pc.  The NCD constitutes $3\%$ of the 
NC luminosity which is a factor of $6\times$ smaller than in NGC~4244.  
The NCD is $\sim10\times$  thicker than in NGC~4244. 

\begin{figure}
\centering
\includegraphics[angle=270,width=\hsize]{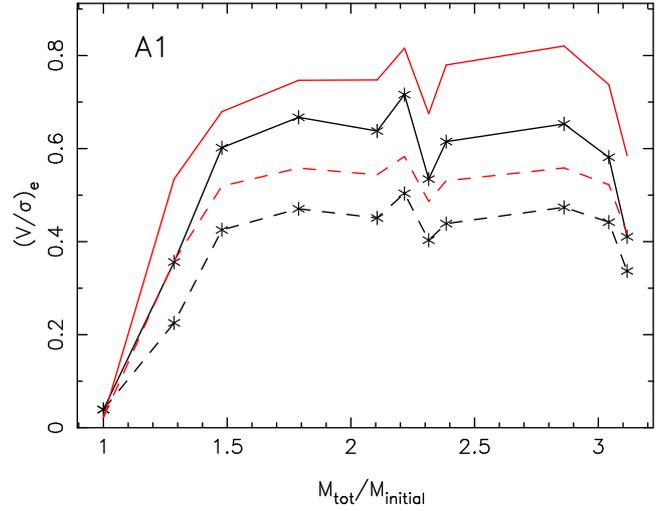} 
\caption{The evolution of $\left(V/\sigma\right)_e$ in run A1 within
\re\ (black lines) and 3~\re\ (red lines).  The dashed lines represent 
mass-weighted and the solid lines luminosity-weighted measurements 
as described in the text.}
\label{fig:CVSSH2}
\end{figure}

\begin{figure*}
\centering
\begin{tabular}{lll}
\includegraphics[width=0.325\hsize]{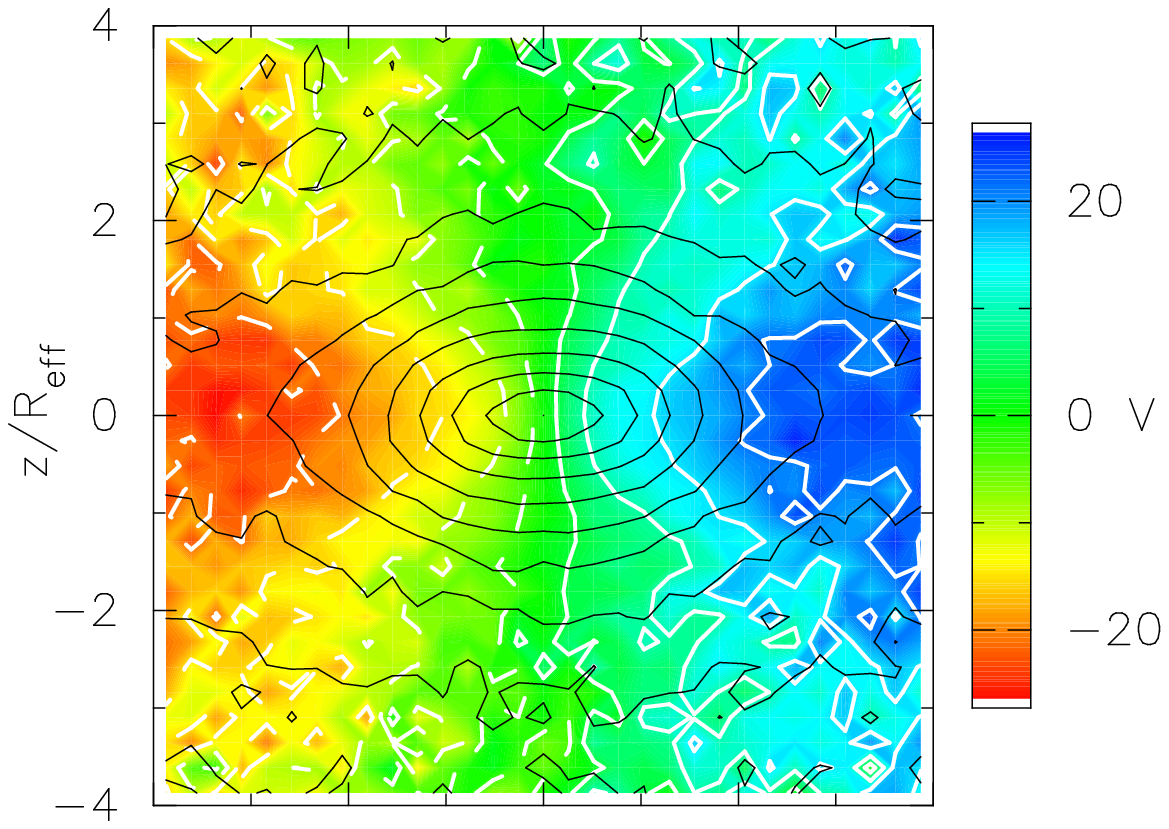} &
\includegraphics[width=0.325\hsize]{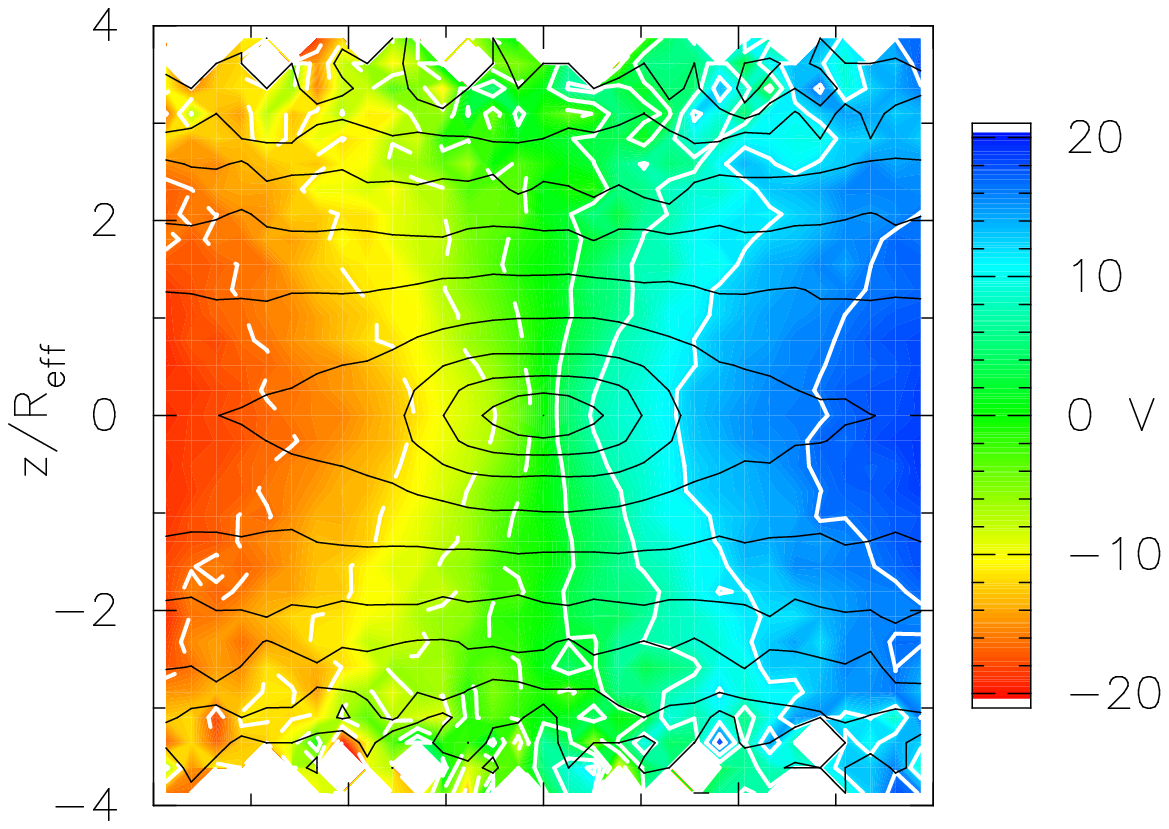} &
\includegraphics[width=0.325\hsize]{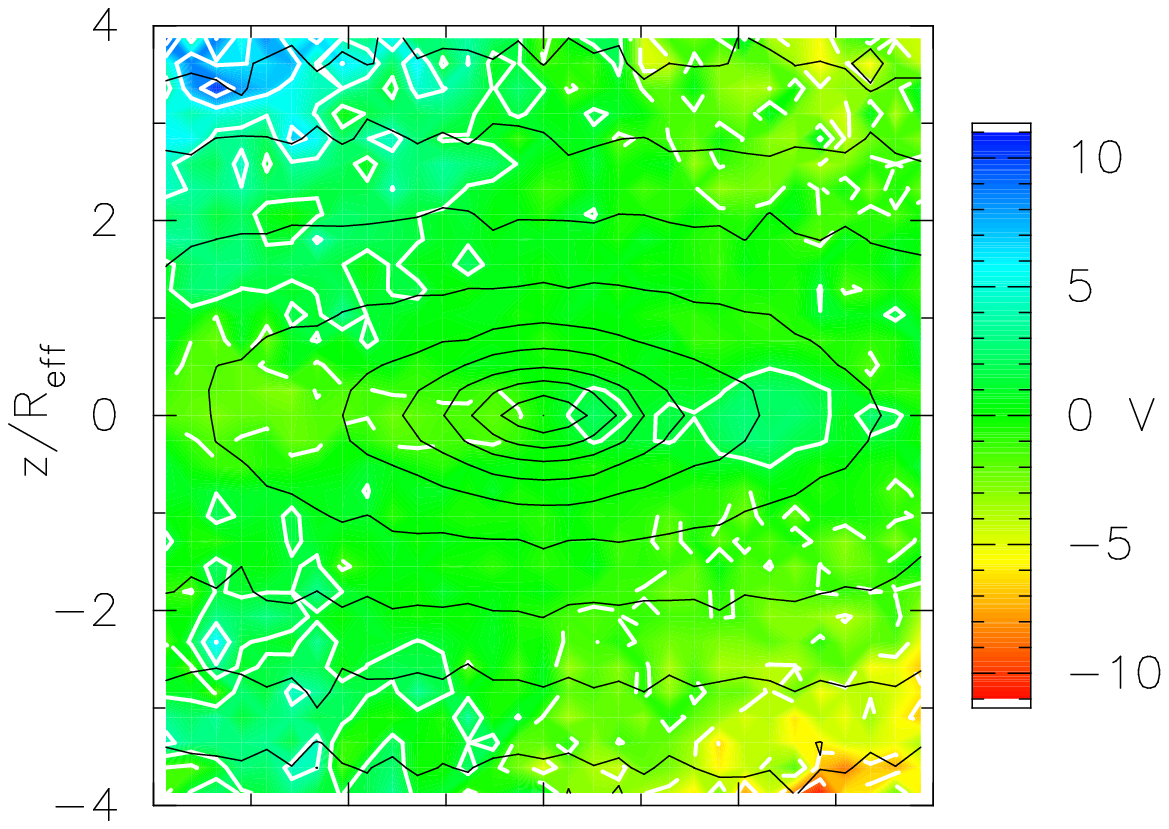}  \\
\includegraphics[width=0.34\hsize]{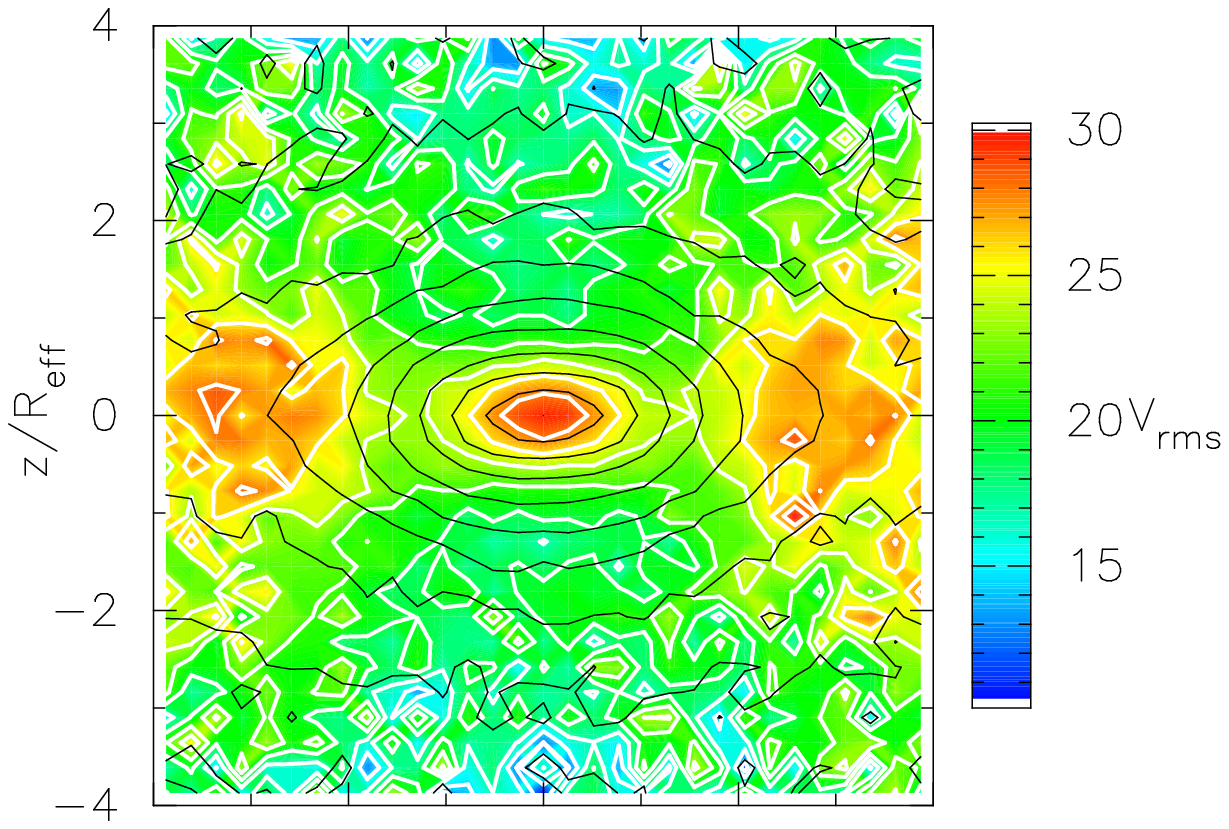} &
\includegraphics[width=0.34\hsize]{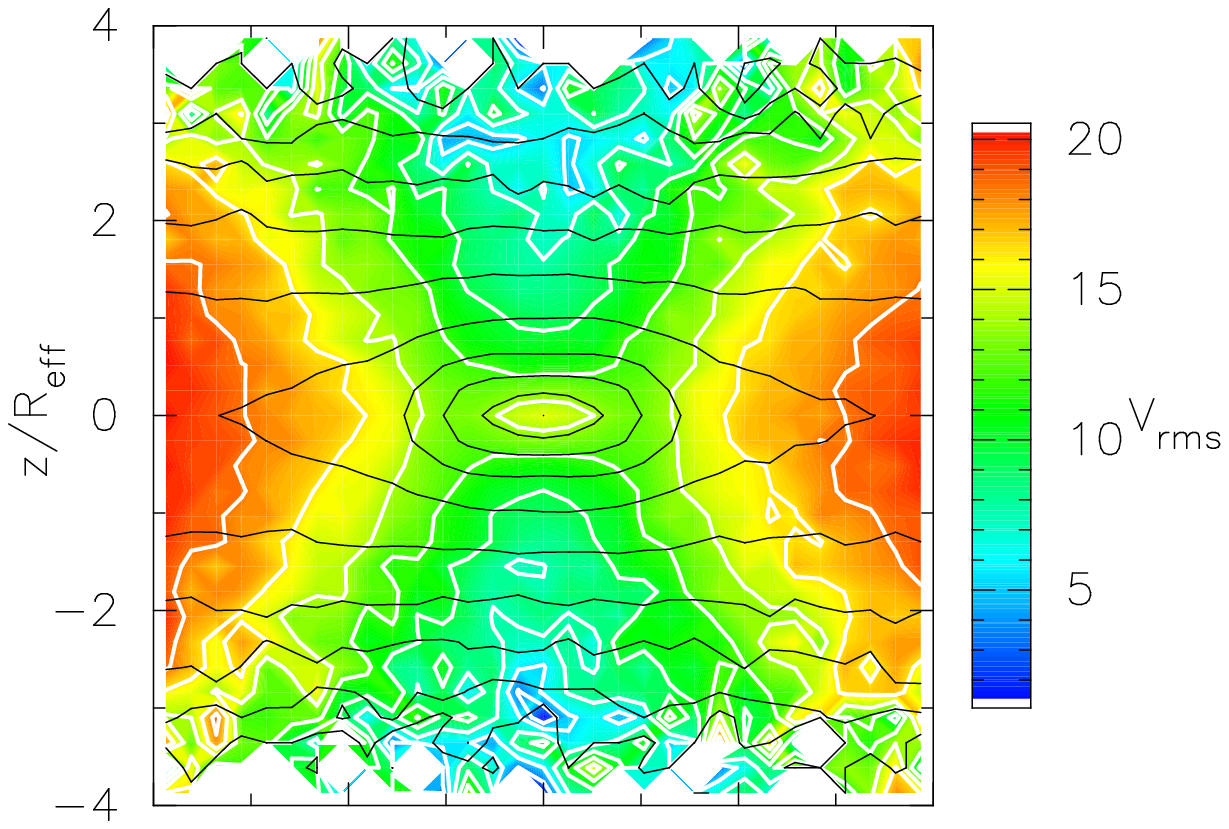} &
\includegraphics[width=0.34\hsize]{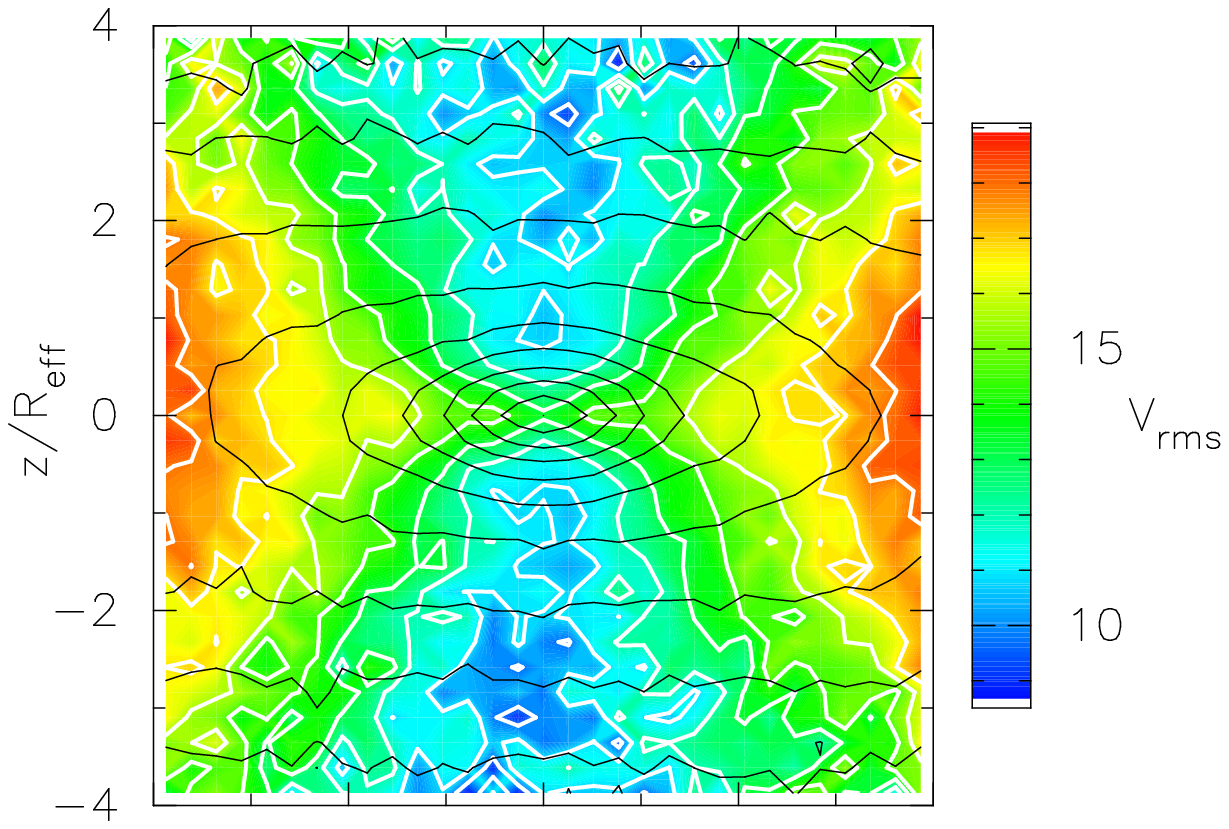} \\
\includegraphics[width=0.325\hsize]{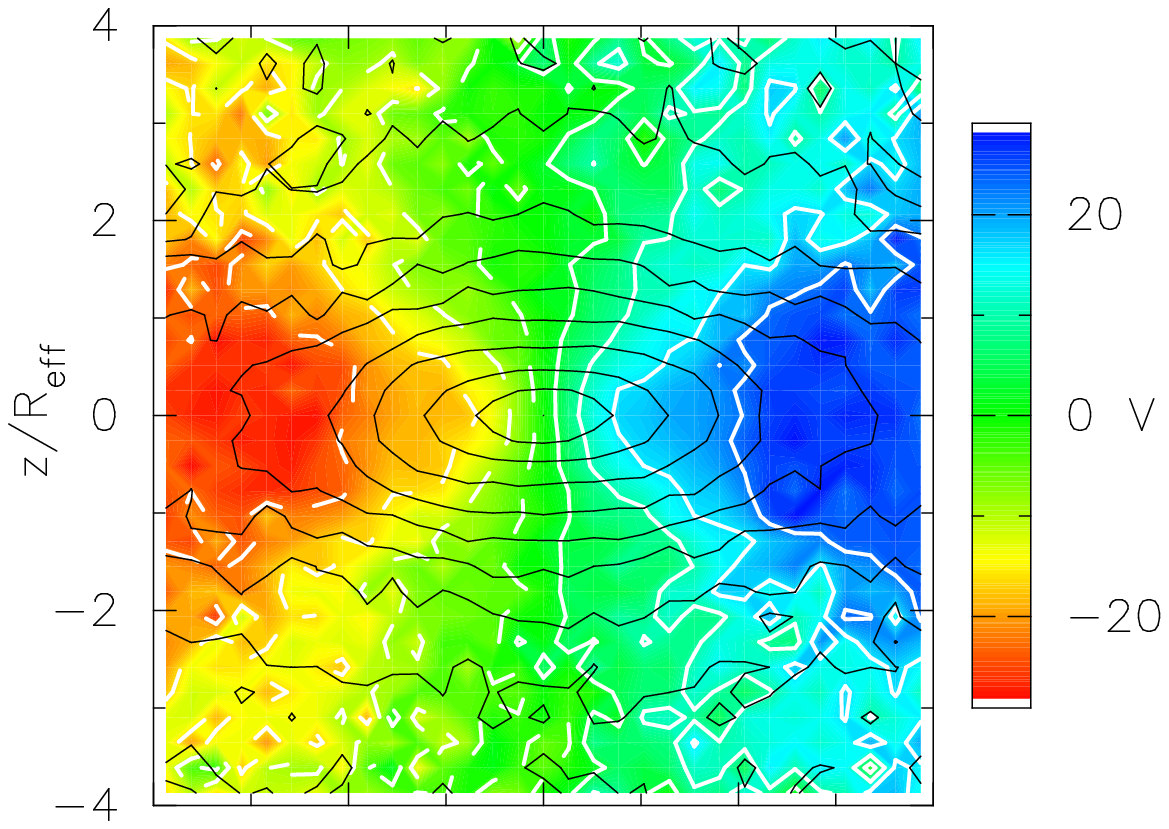} &
\includegraphics[width=0.325\hsize]{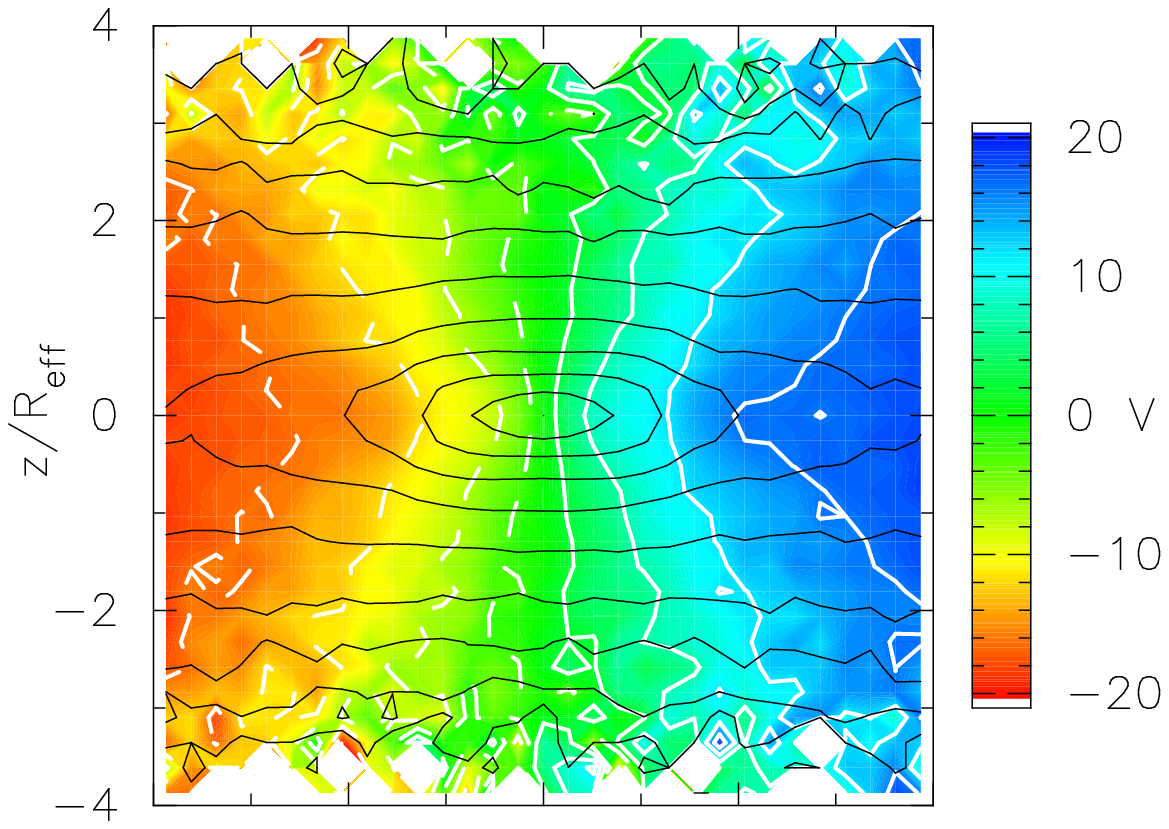} &
\includegraphics[width=0.325\hsize]{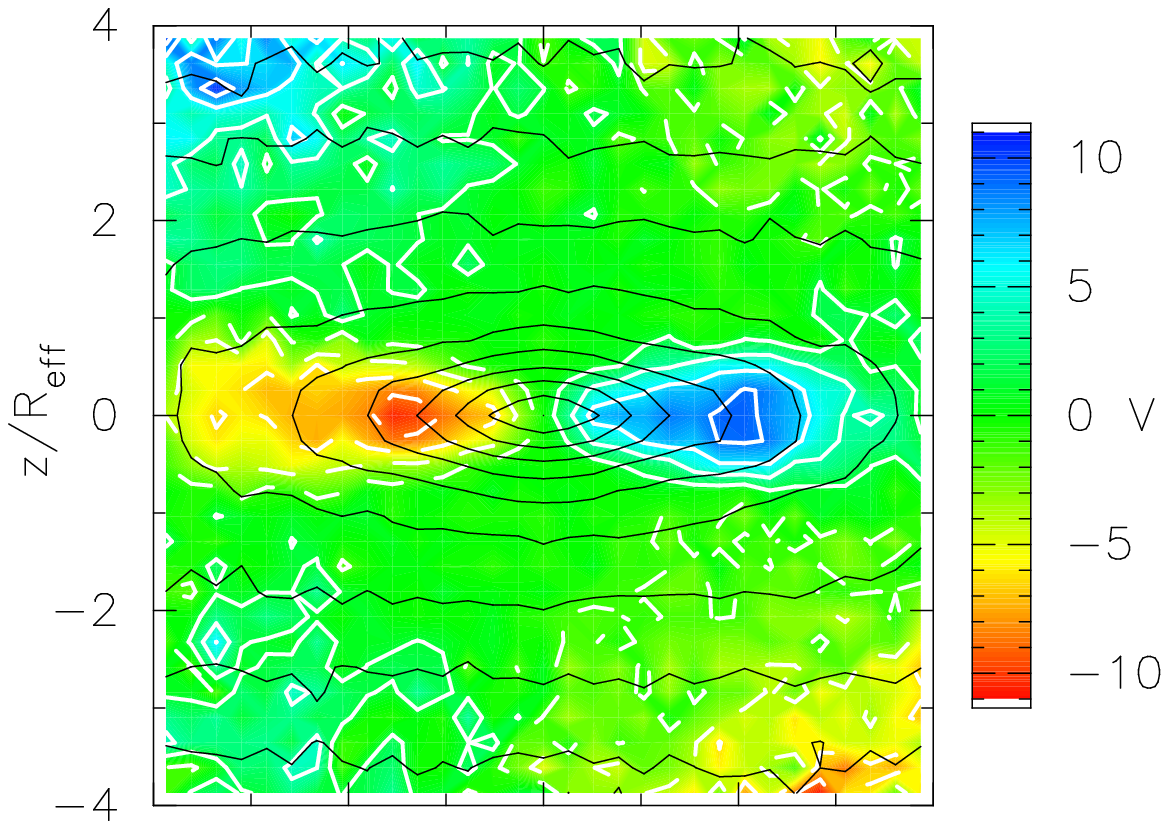}  \\
\includegraphics[width=0.34\hsize]{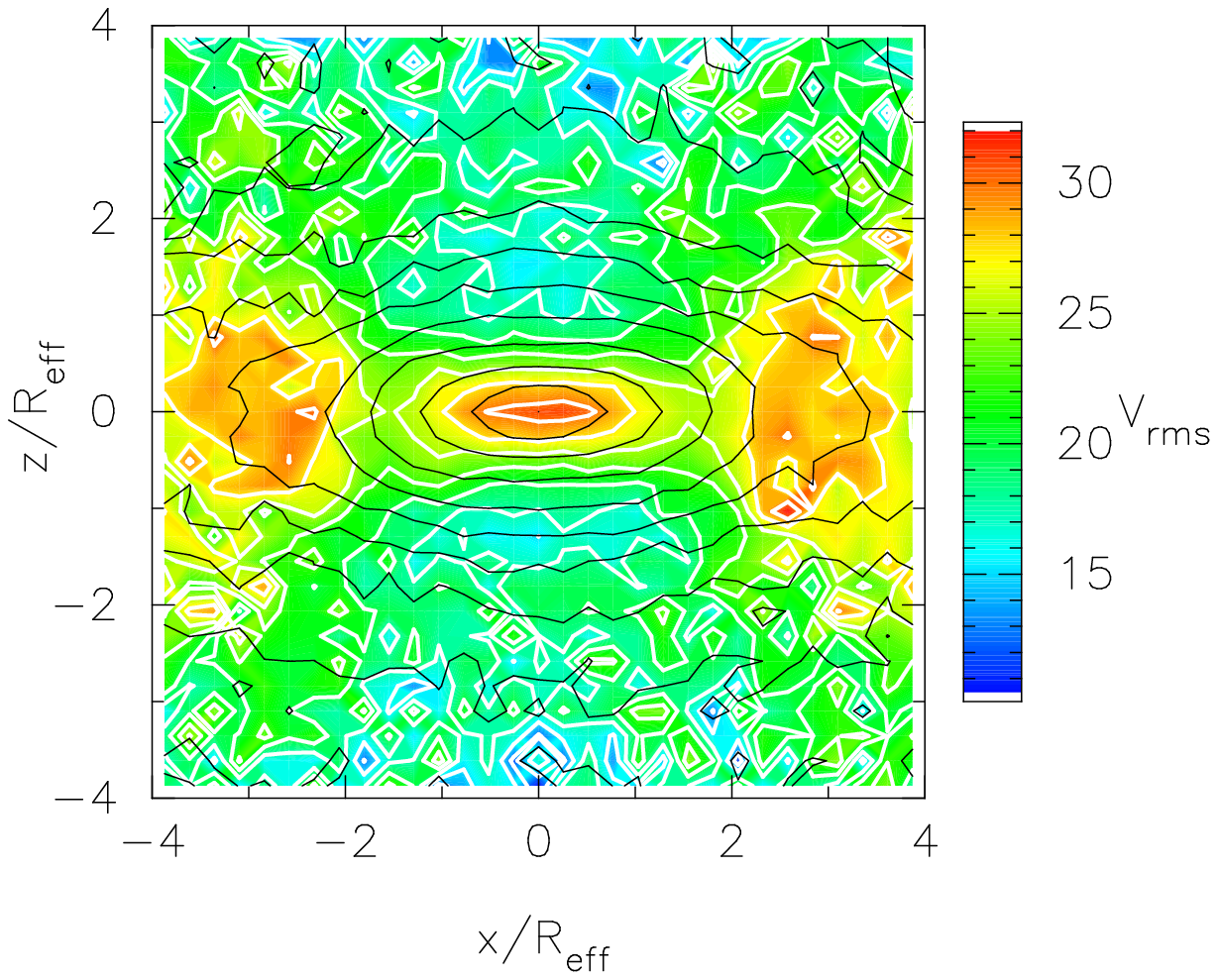} &
\includegraphics[width=0.34\hsize]{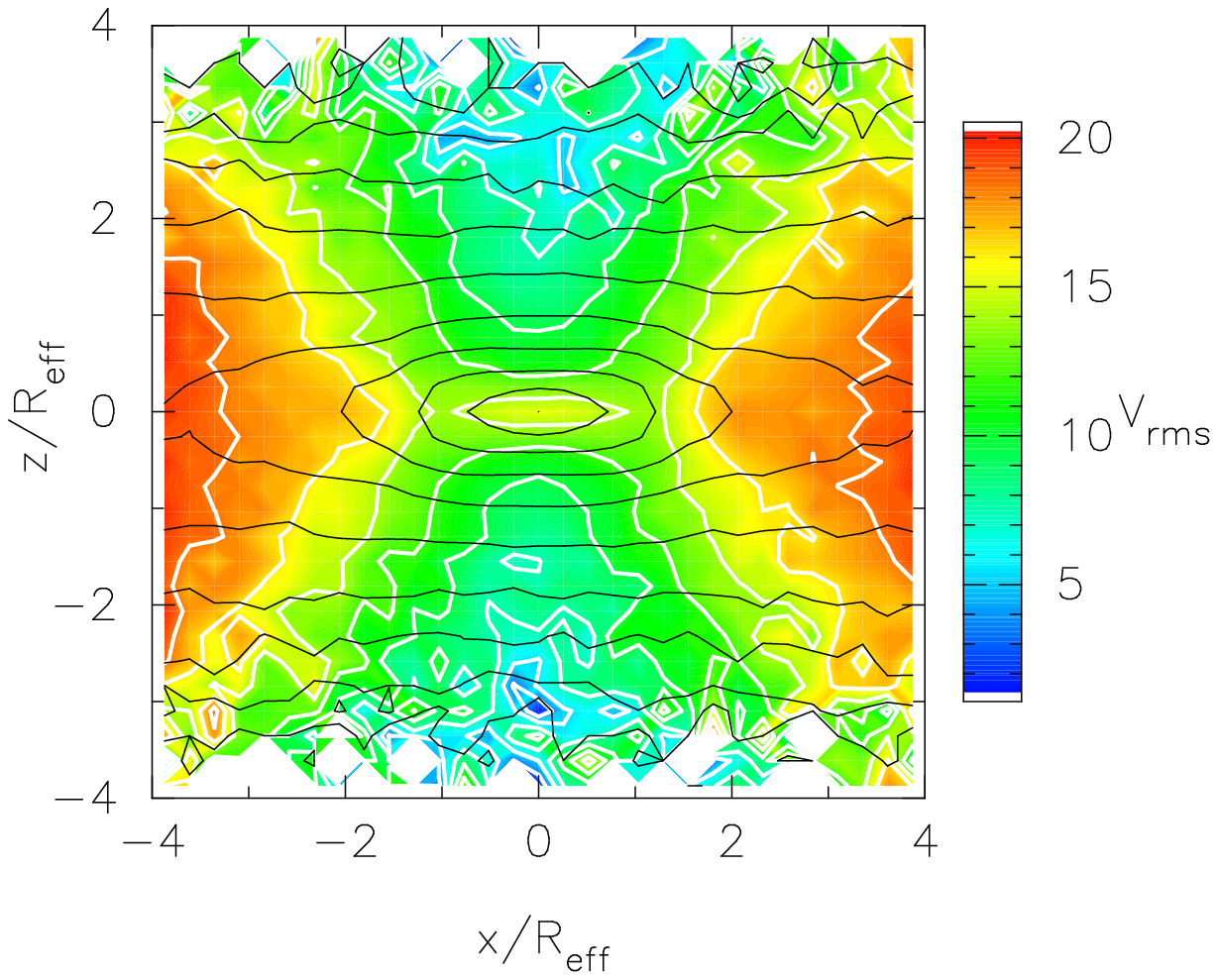} &
\includegraphics[width=0.34\hsize]{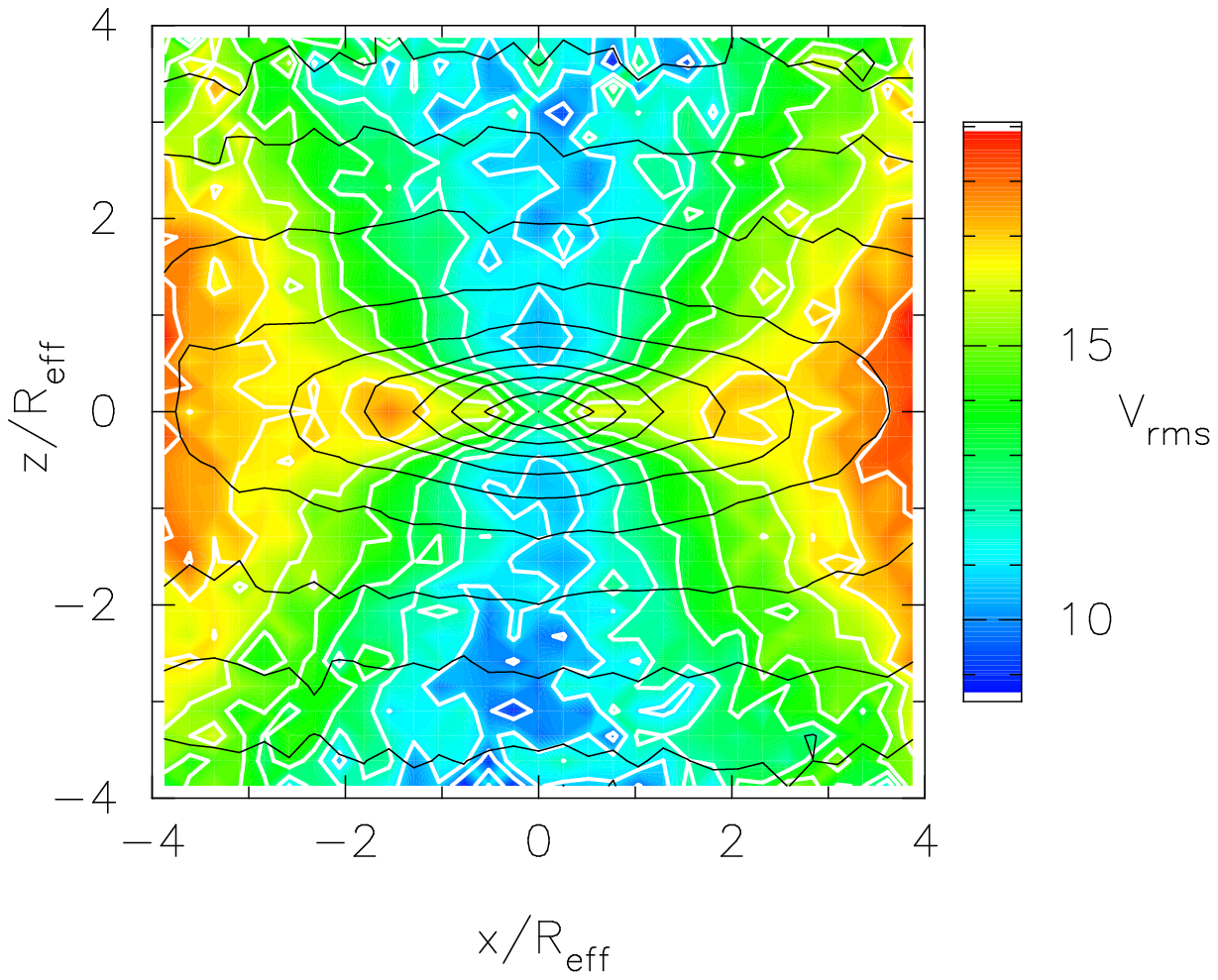} \\
\includegraphics[width=0.30\hsize,angle=270]{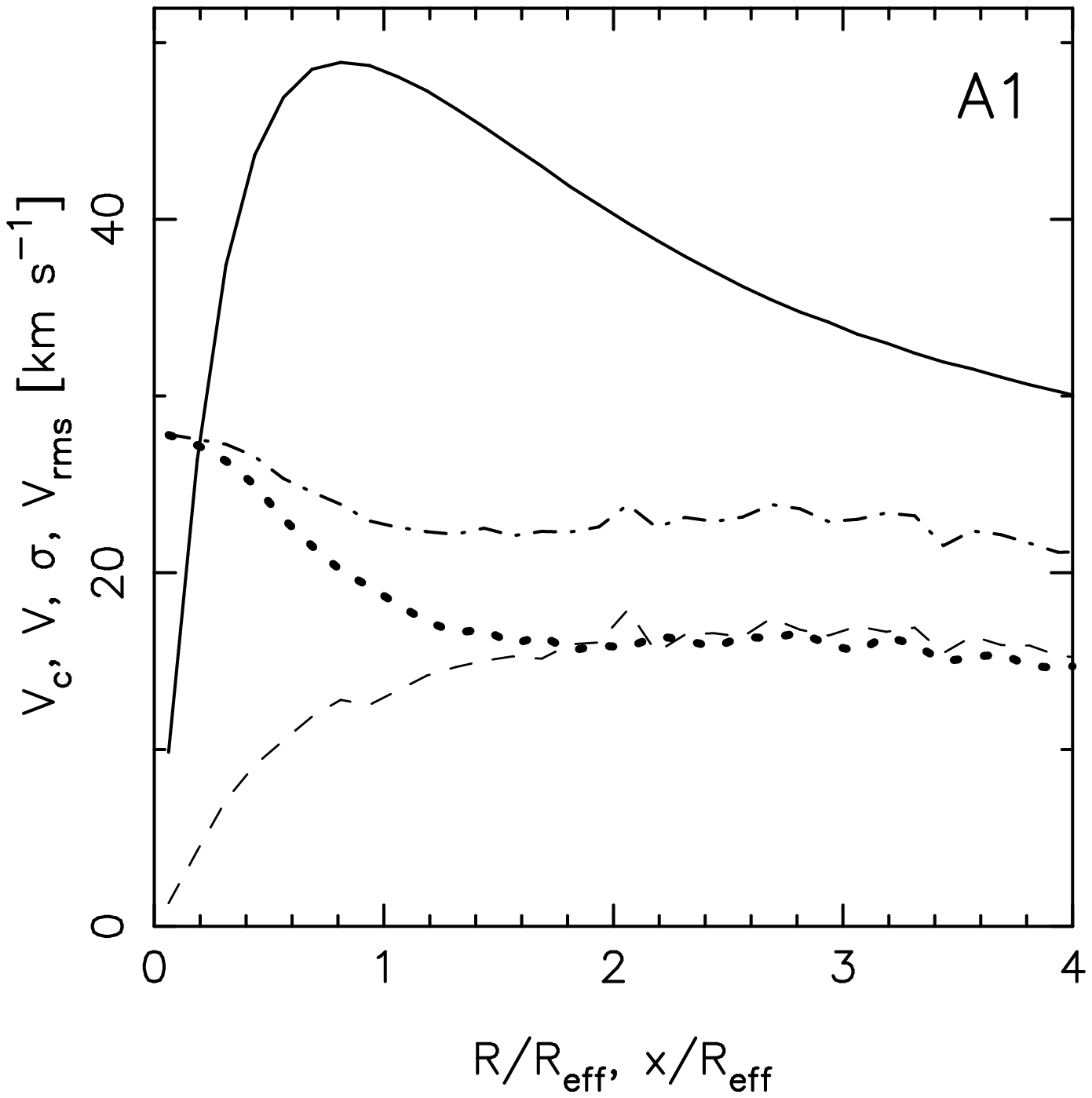} &
\includegraphics[width=0.30\hsize,angle=270]{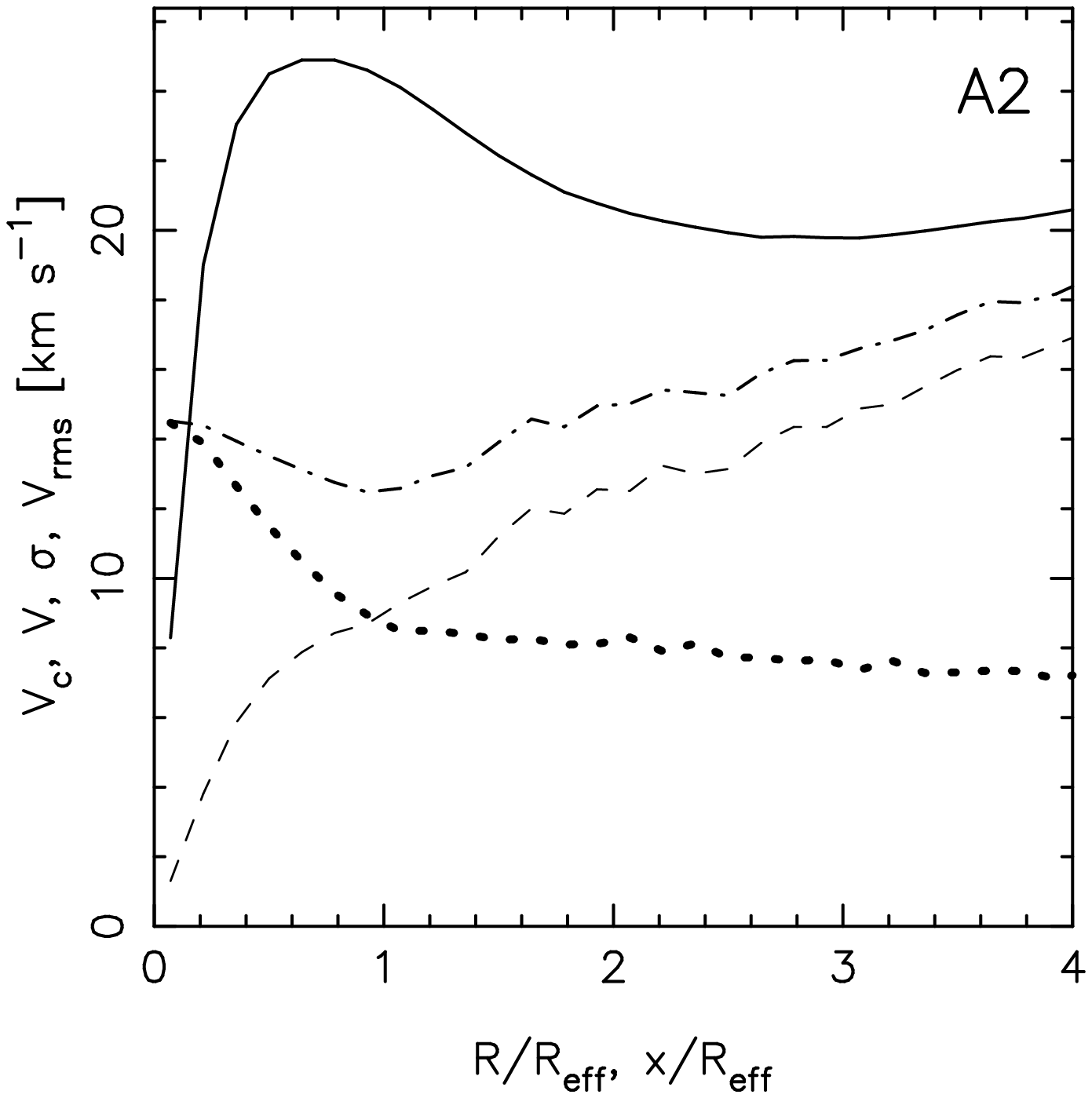} &
\includegraphics[width=0.30\hsize,angle=270]{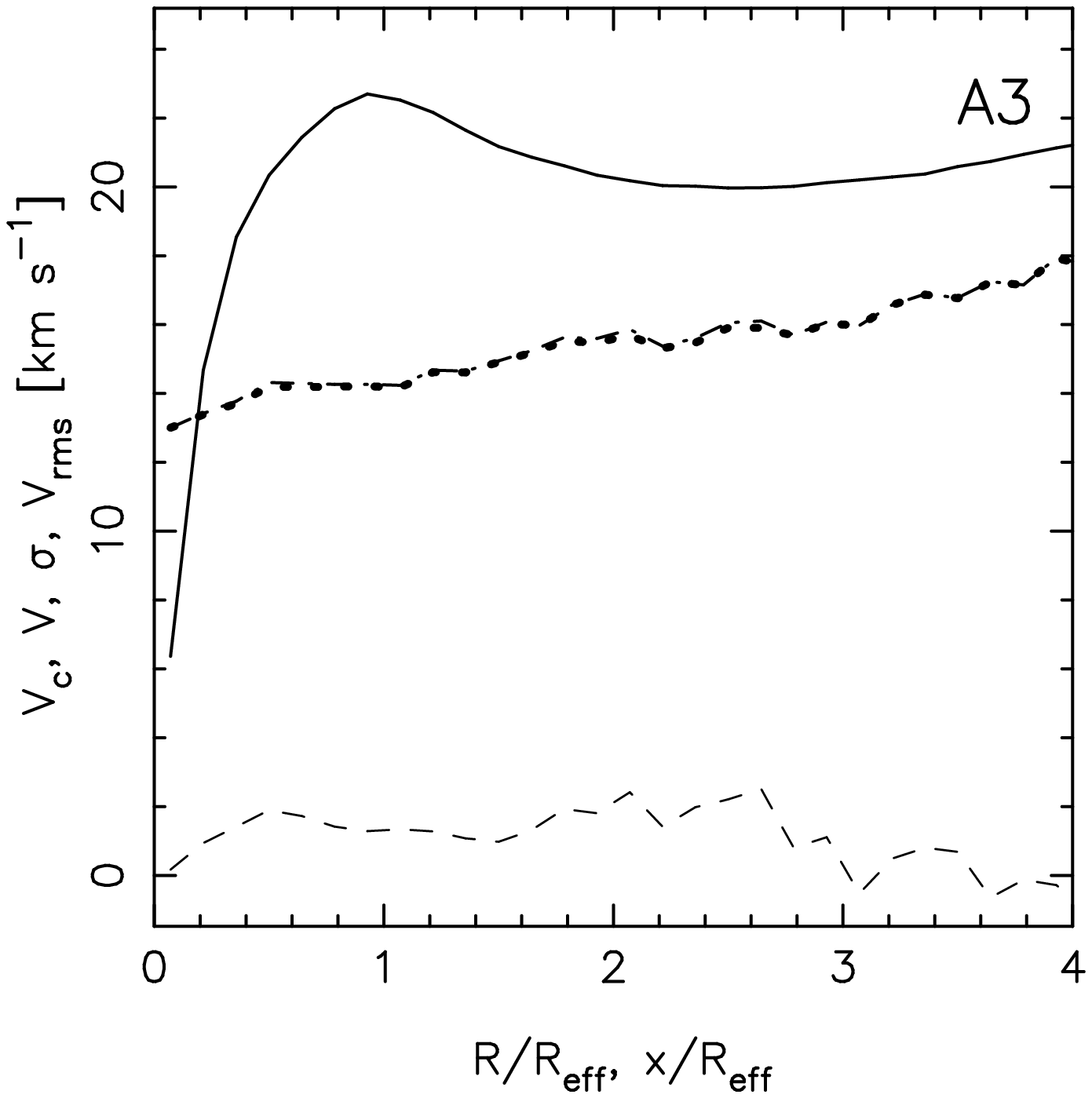}  \\
\end{tabular}
\caption{Kinematic fields for the accretion simulations A1 (left), A2
(middle) and A3 (right).  The top two rows adopt a mass-weighting
while the next two rows use luminosity-weighting, as described in the
text.  In the top four rows the black contours show log-spaced density while 
the white contours show the kinematic contours corresponding to each
panel.  Run A3 shows significant rotation only with
luminosity-weighting.  The bottom row shows the rotation curve $\rm
V_c\left(R\right)$ (solid lines), the line-of-sight velocities
$\rm V\left(x\right)$ (dashed lines), the line-of-sight velocity dispersions
$\rm \sigma\left(x\right)$ (dotted lines) and the root-mean-square 
velocities $\rm V_{rms}\left(x\right)$ (dashed-dotted lines) along the 
major axis.}
\label{fig:VFMA}
\end{figure*}

The initial SSC is isotropic and has no rotation and remains
unrotating after falling to the centre.  Figure~\ref{fig:CVSSH2} shows
the evolution of $\left(V/\sigma\right)_e$; after the first accretion
the merger remnant's rotation has already increased to
$\left(V/\sigma\right)_e \simeq 0.23$.  By the end of the
simulation the mass-weighted $\left(V/\sigma\right)_e \simeq 0.34$ 
($\simeq 0.41$ luminosity-weighted).  The velocity
field of the remnant; seen in Figure~\ref{fig:VFMA}, shows that the NC
is rotating.  However \vrms\ is still dominated by the velocity
dispersion at the centre even when luminosity weighting.  The
line-of-sight velocity $\rm V$ peaks at larger radii than
observed in NGC~4244 and M33.

\begin{figure*}
\centering
\begin{tabular}{ccc}
\includegraphics[width=0.33\hsize]{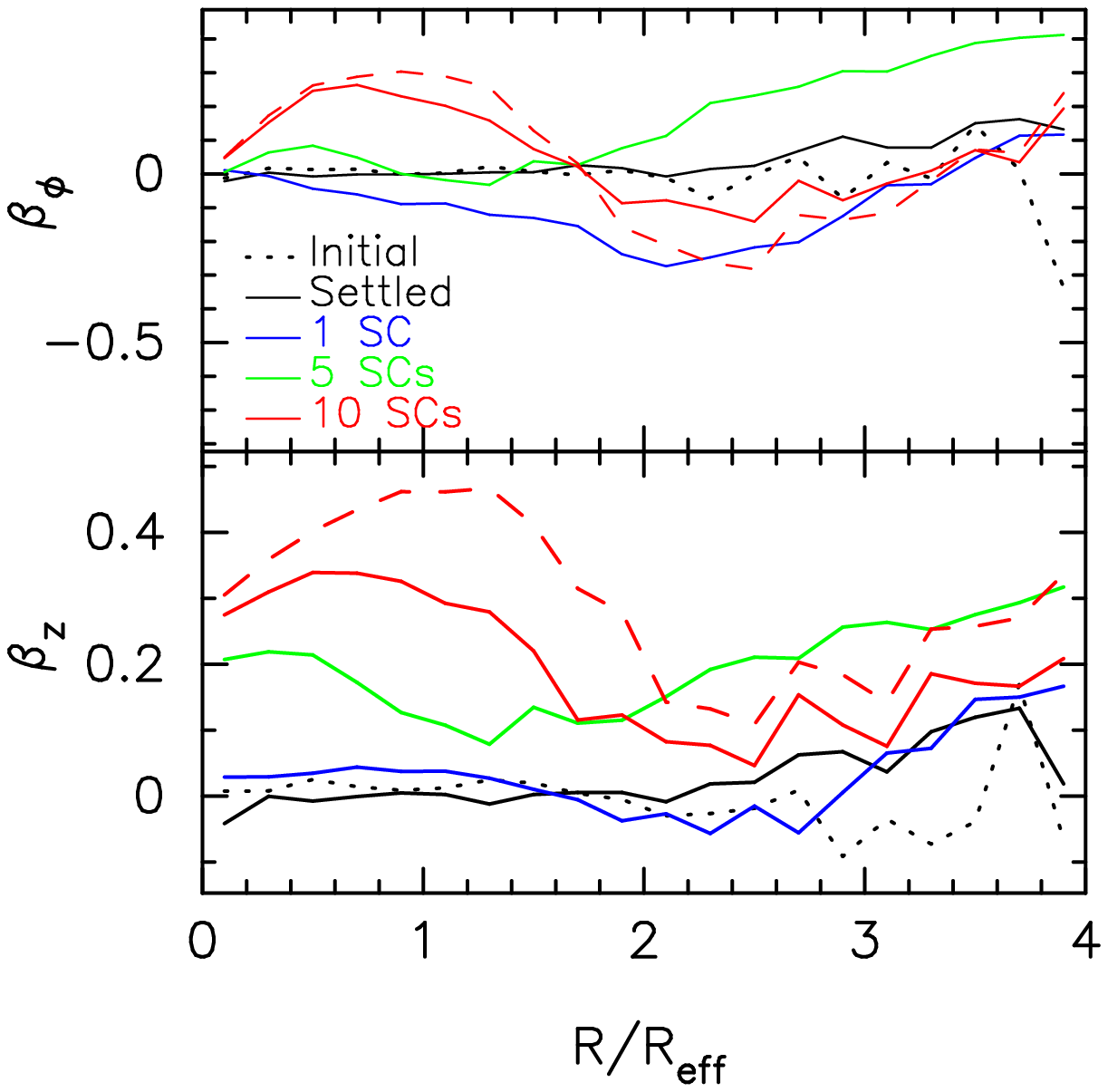} & 
\includegraphics[width=0.33\hsize]{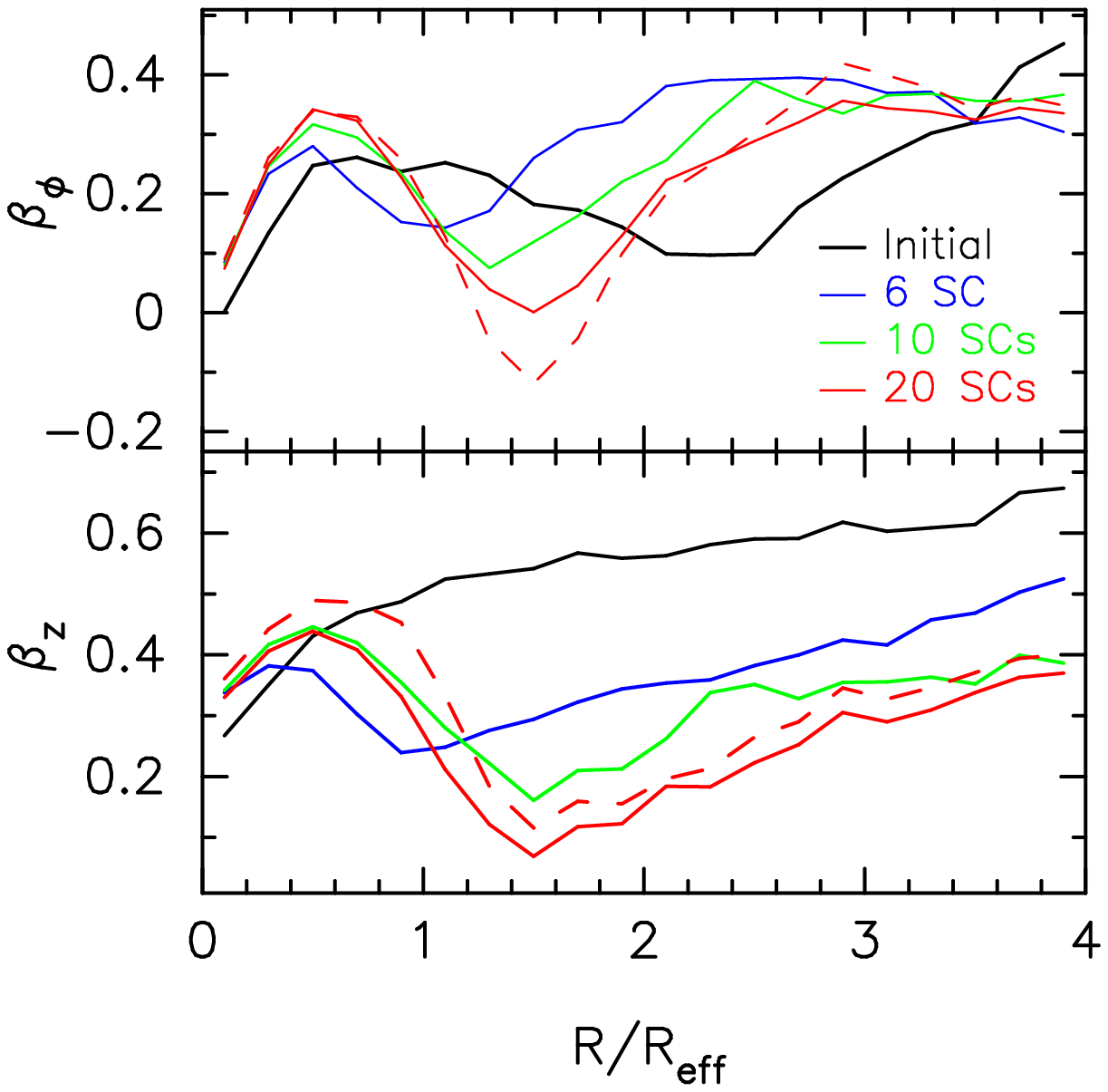} &
\includegraphics[width=0.33\hsize]{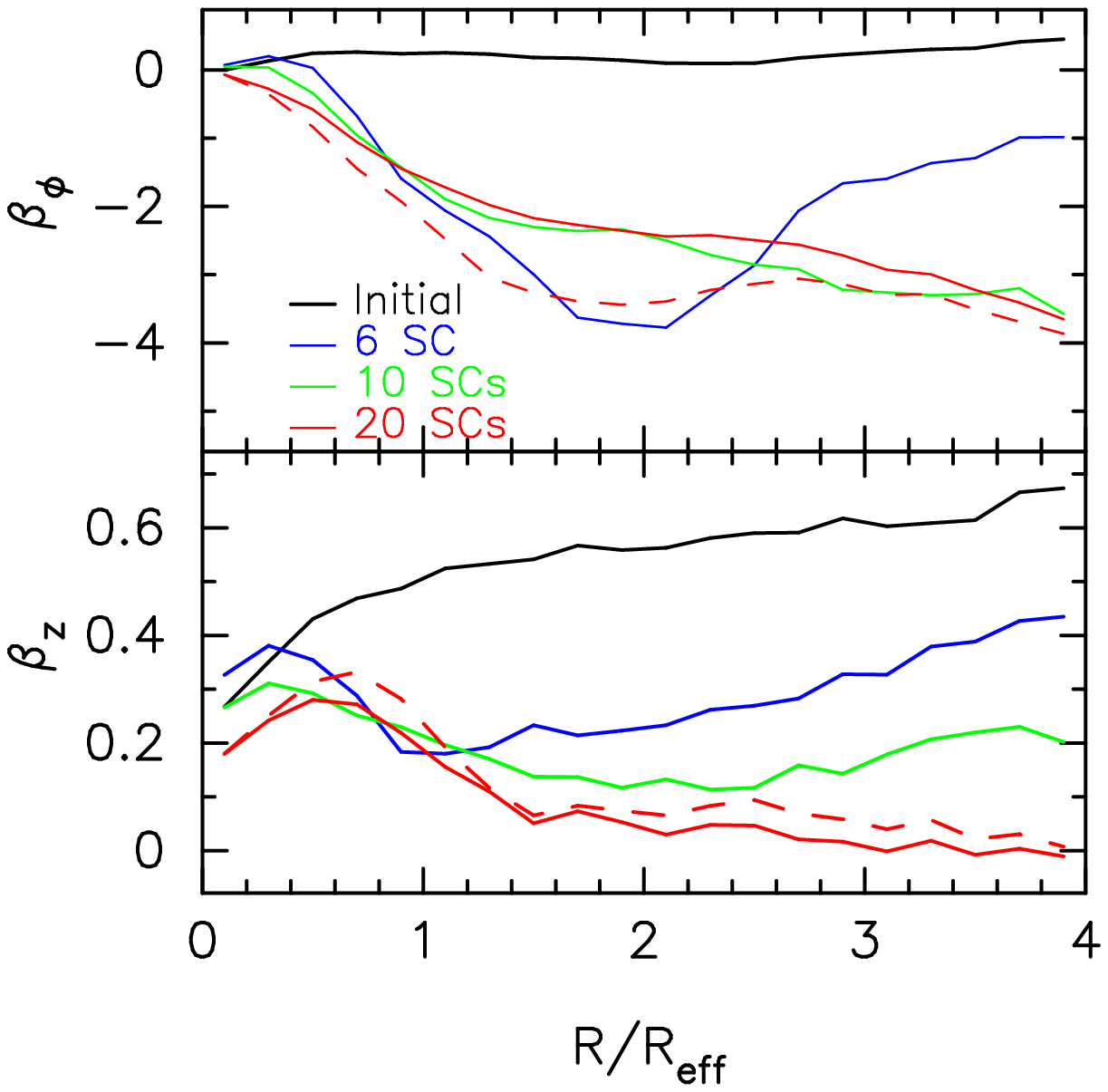} \\
\end{tabular}
\caption{The anisotropies $\beta_\phi$ (top) and $\beta_z$
(bottom) after different accretion events for runs A1 (left), A2
(middle) and A3 (right).  For A1 we show the initial super
star cluster by the black dotted line.  For each simulation the red 
dashed lines show the luminosity weighted anisotropies of the final NC.  
The lack of net angular momentum in run A3 drives $\beta_\phi$ to 
negative values, even when luminosity-weighting, whereas in run A2 
$\beta_\phi$ is positive.}
\label{fig:CylAniEvol}
\end{figure*}

The settling of the SC to the centre does not change its dispersion
within \re, since the bulge mass within \re\ changes only about
$0.6\%$.  The evolution of $\beta_\phi$ and $\beta_z$ is shown in
Figure~\ref{fig:CylAniEvol}.  The first accretion drives $\beta_\phi$
to negative values which slowly increases as more mass is
accreted. The final $\beta_z$ peaks within \re\ and declines further
out.

\section{Accretion onto bare NCDs}
\label{sec:accretions}

Here we consider the case of a bare NCD, without an initial NCS,
accreting multiple young SCs.  The initial NCD component has a mass of
$1\times10^6$~\Msun.  We use model C4 to represent the infalling SCs,
placing them at 63pc from the centre of the NC, since the NCD is
barely affected by the SC until it is well within this radius.  In run
A2 all SCs are on prograde circular orbits, whereas in A3 half of the
SCs have retrograde orbits.  In total the NCD accretes $20$~SCs,
corresponding to $4\times$ its own mass.  All SCs orbit in the plane
of the NCD without any vertical motions.  Each accretion event
requires $20 - 30$~Myrs to complete; we allow each accretion to finish
before introducing the next SC.  In total these simulations require
$1.1$ Gyrs.  Table \ref{tab:mergsims} gives further details of these
simulations.

\begin{figure*}
\centering
\begin{tabular}{ccc}
\includegraphics[width=0.32\textwidth]{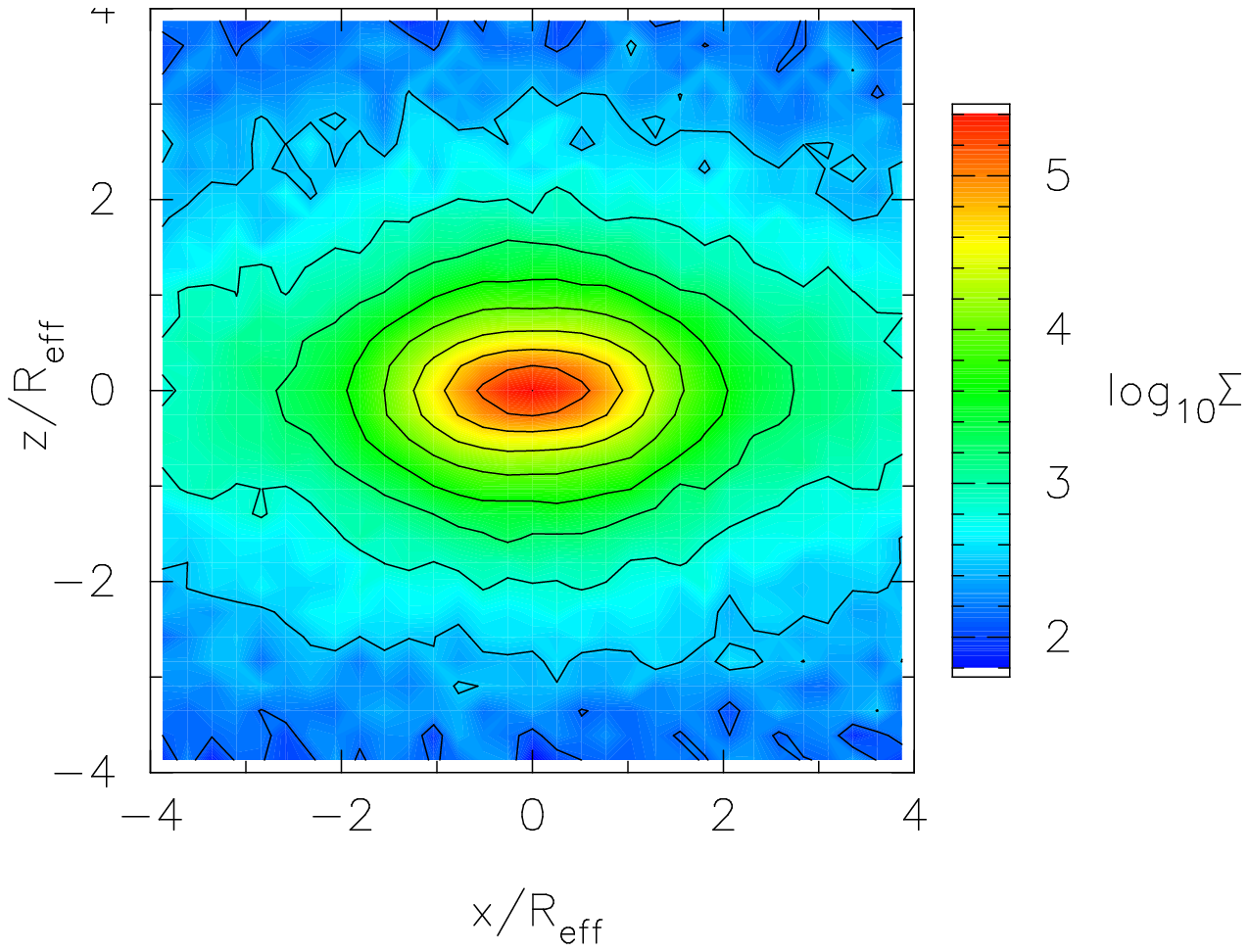} &
\includegraphics[width=0.32\textwidth]{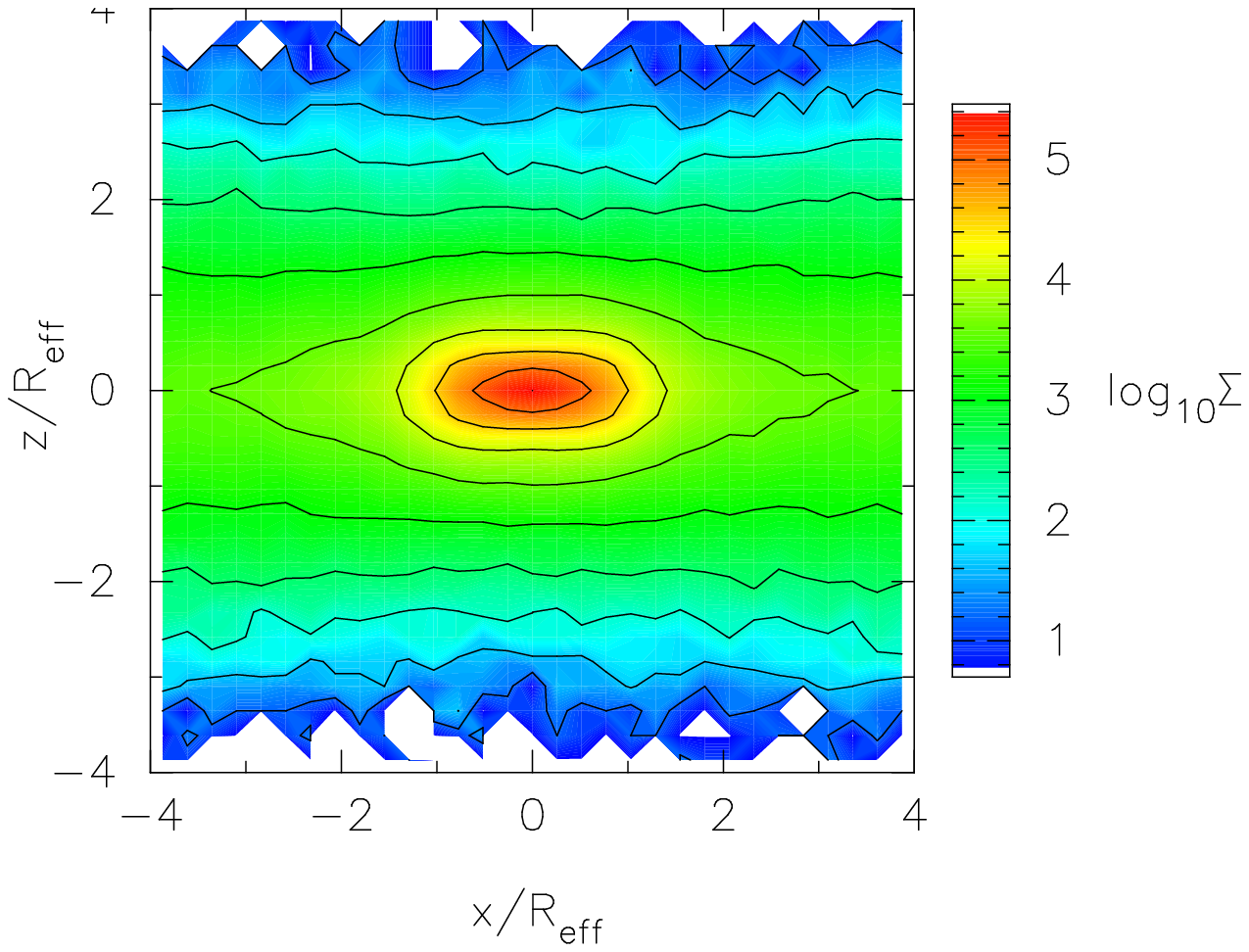} &
\includegraphics[width=0.32\textwidth]{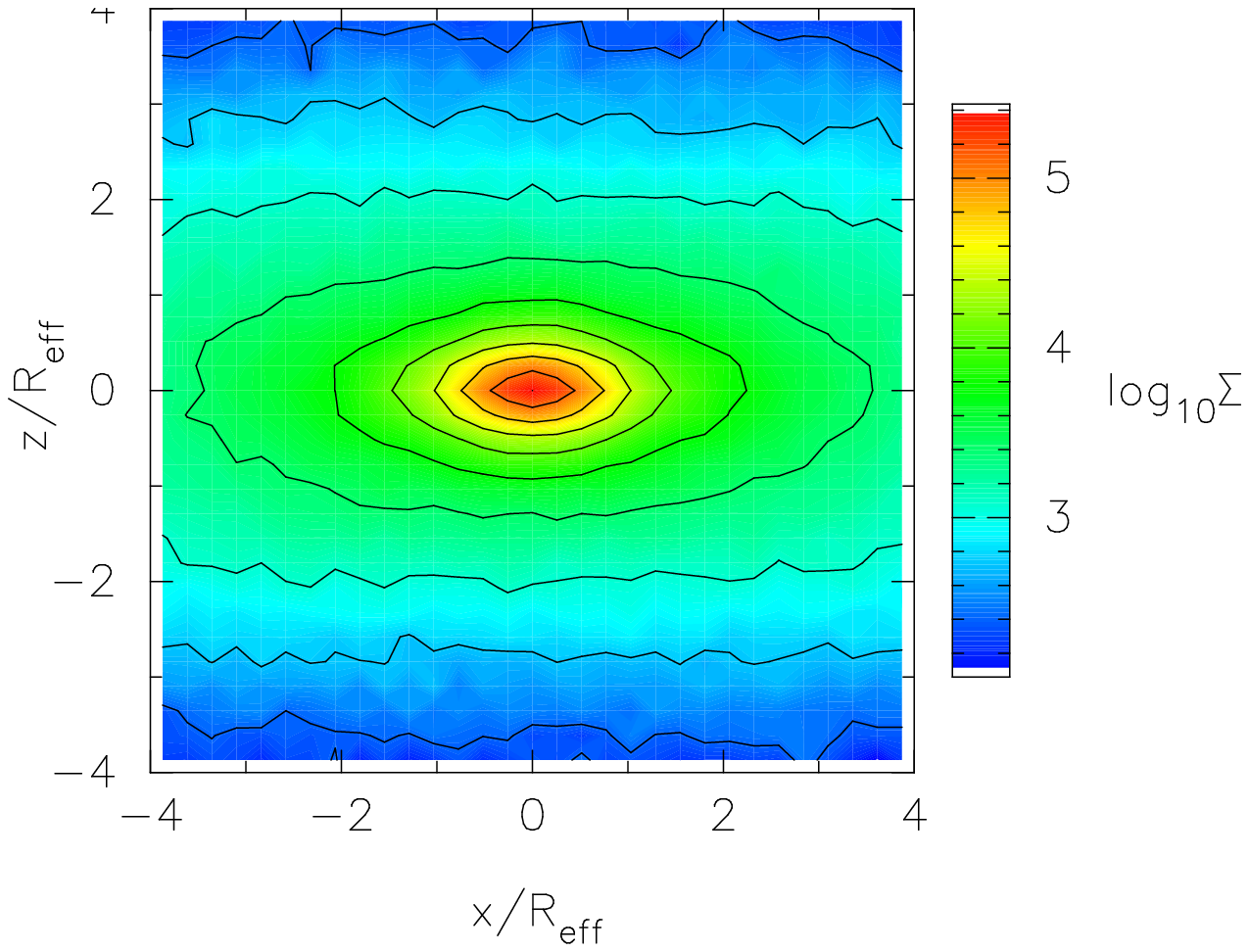} \\
\includegraphics[width=0.32\textwidth]{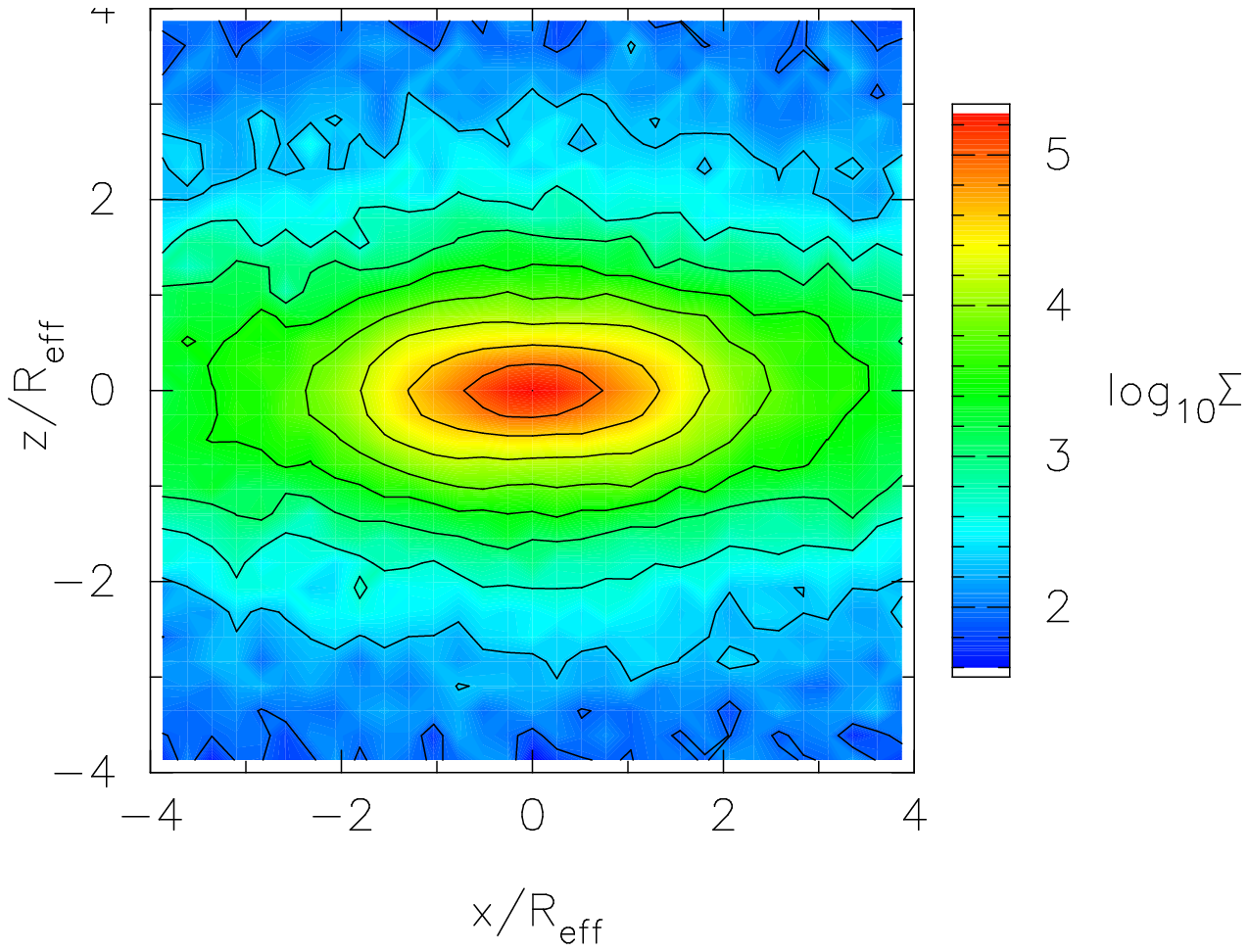} &
\includegraphics[width=0.32\textwidth]{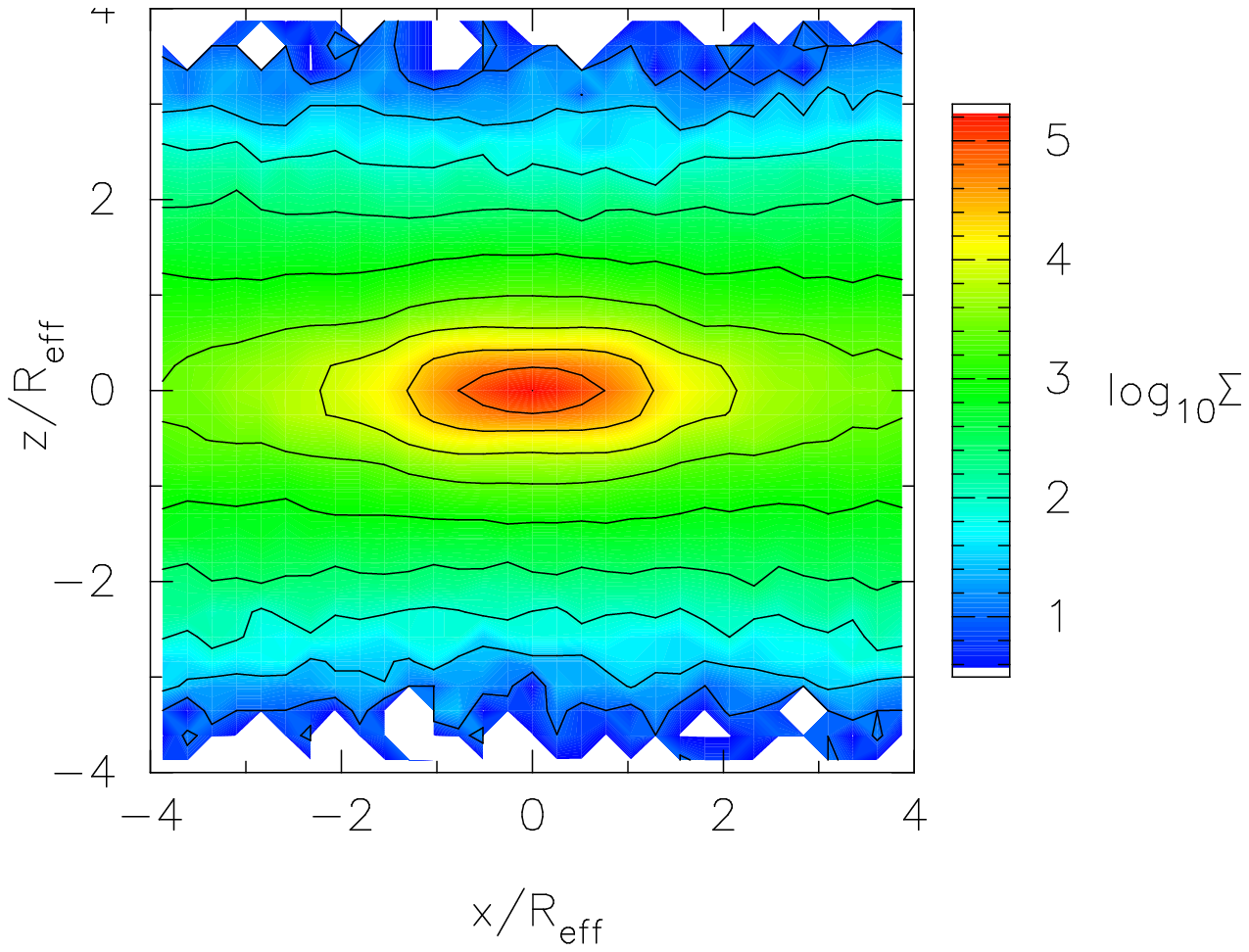} &
\includegraphics[width=0.32\textwidth]{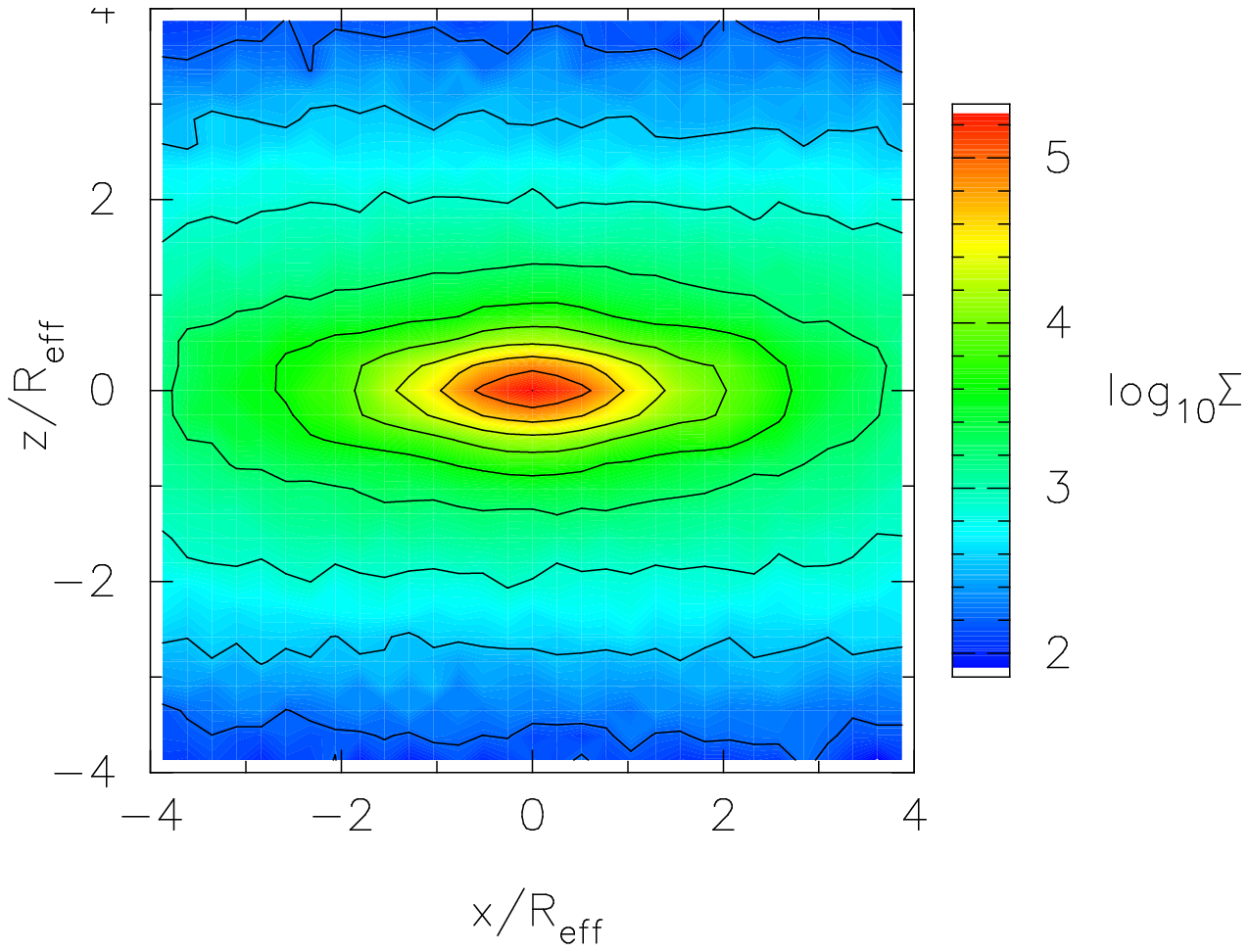} \\
\includegraphics[width=0.3\textwidth, angle=270]{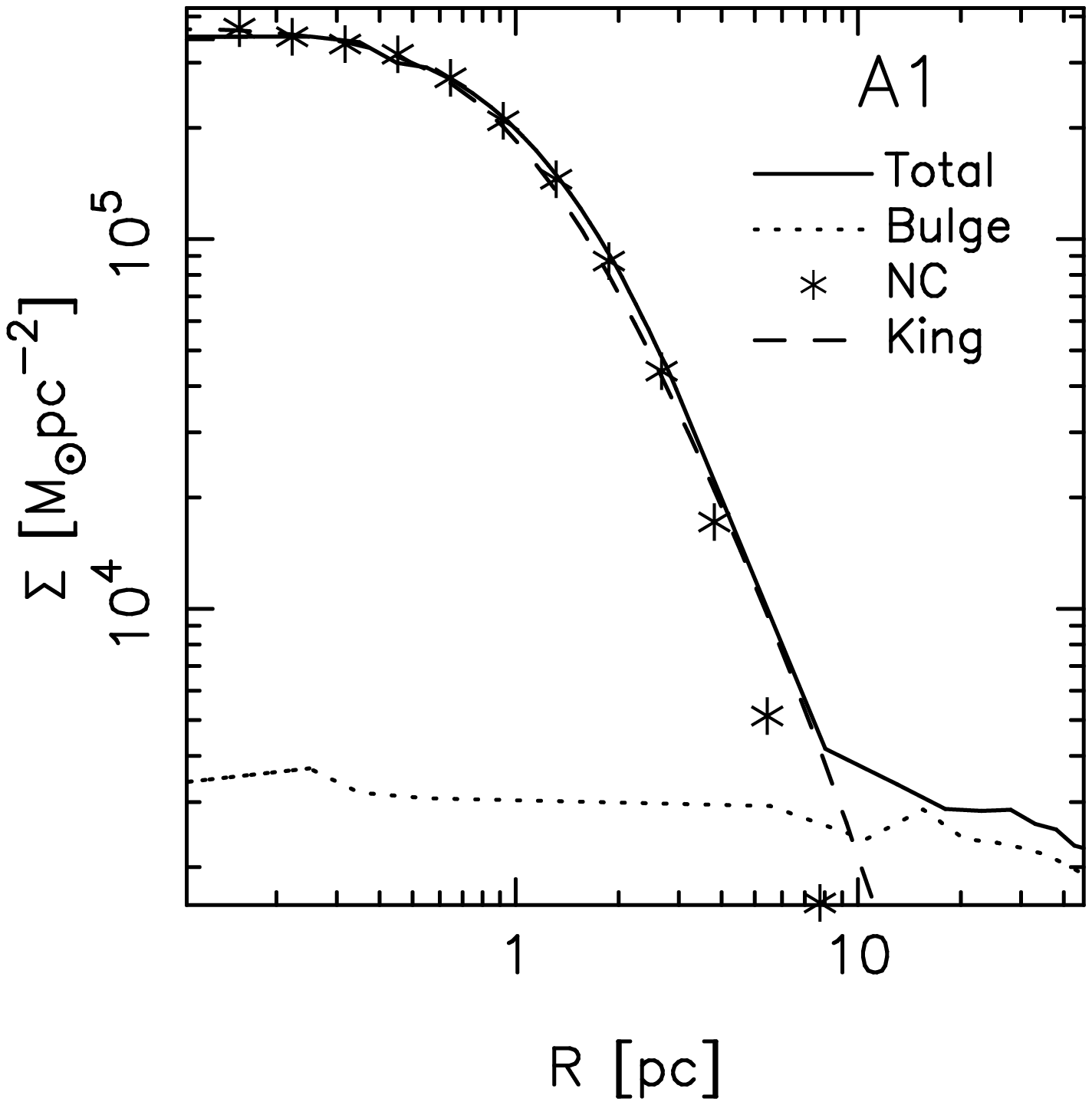} &
\includegraphics[width=0.3\textwidth, angle=270]{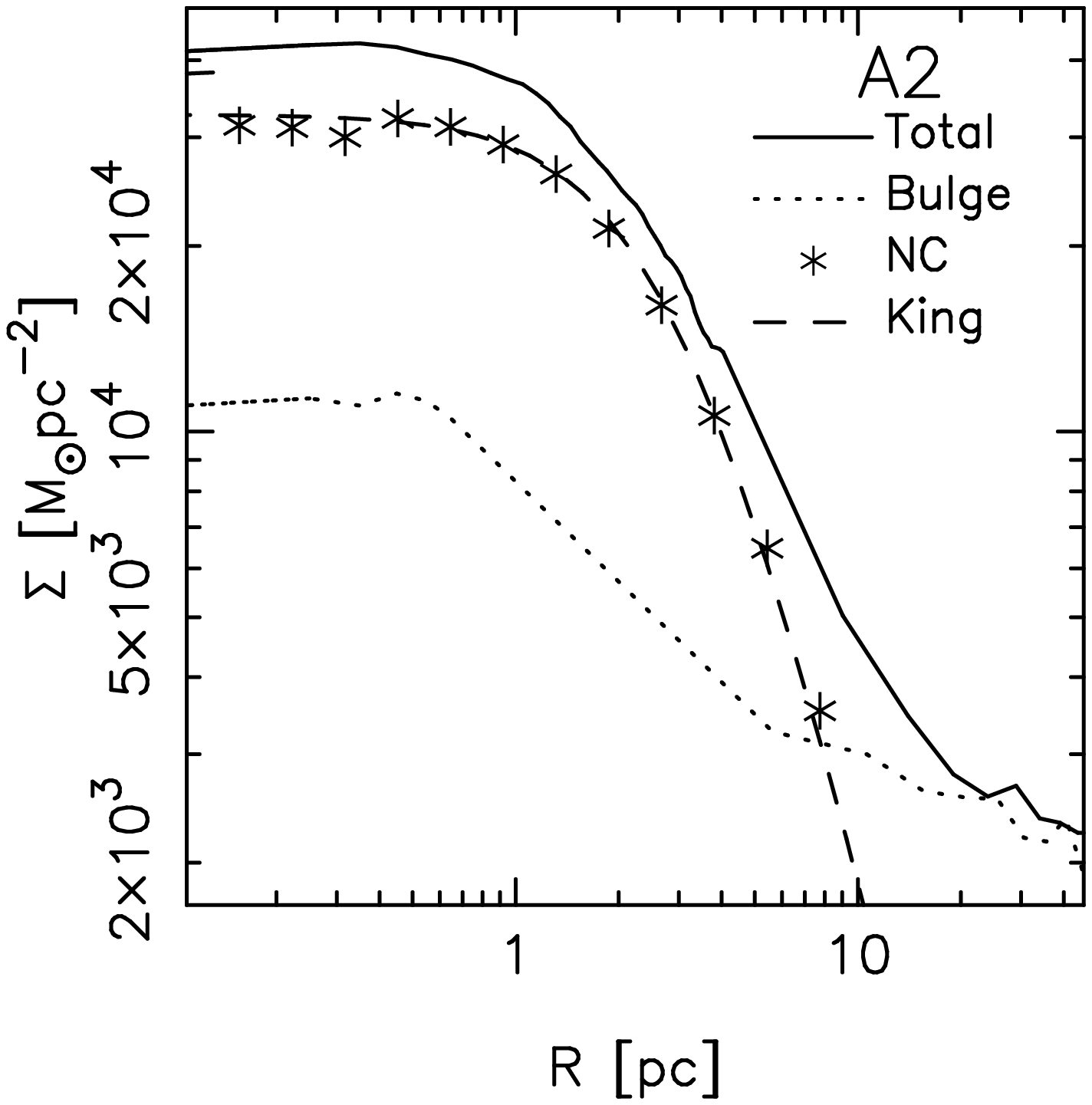} &
\includegraphics[width=0.3\textwidth, angle=270]{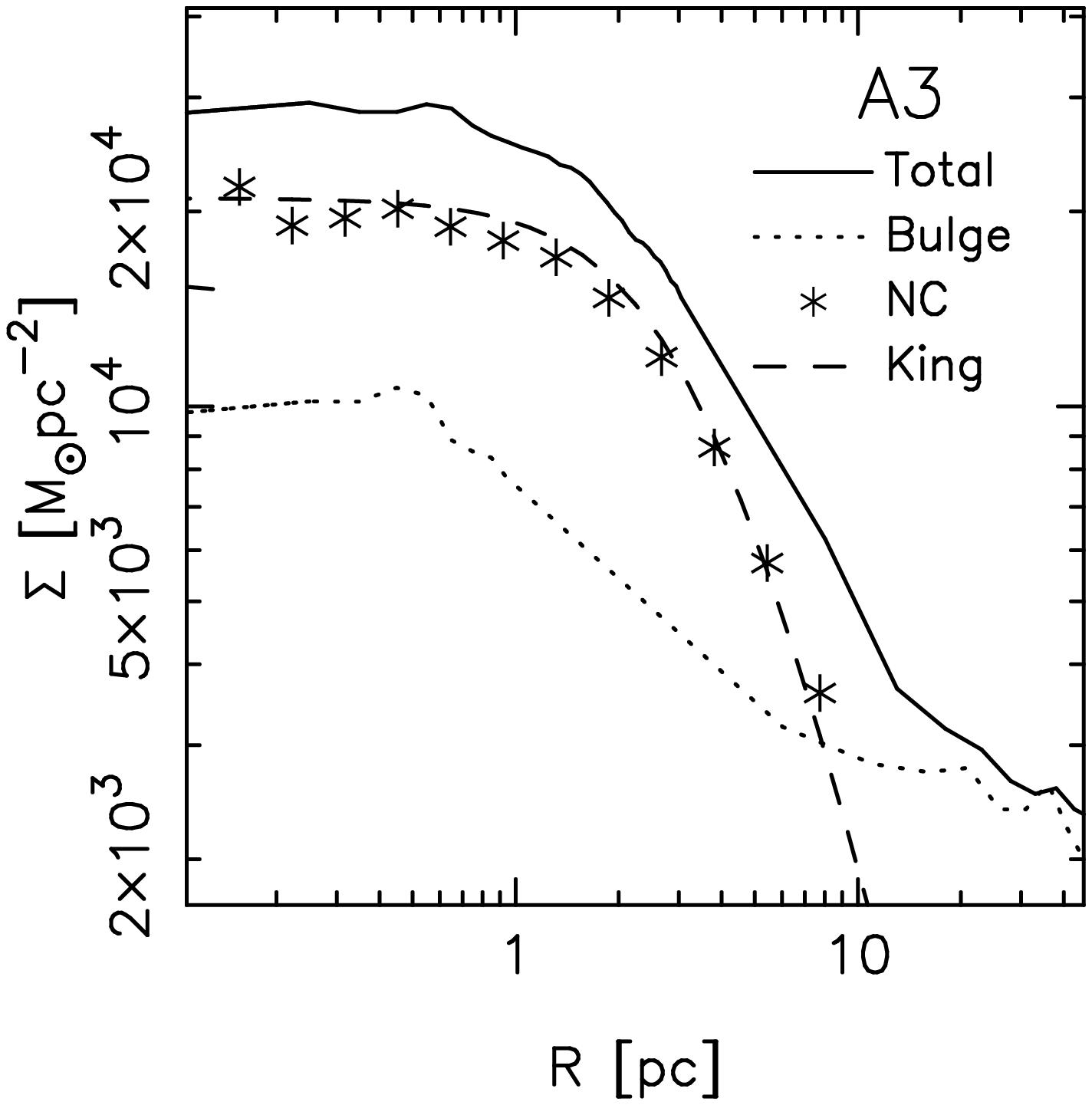} \\
\end{tabular}
\caption{Projected surface density maps seen edge-on for simulations
A1 (left), A2 (middle) and A3 (right).  The top row shows
mass-weighted maps and the middle row luminosity-weighted ones.  In
the bottom panel we show the face-on surface density profiles measured 
within circular annuli for the
NC as stars and the corresponding King profile by a dashed line.  The
bulge is shown by the dotted line and the combined surface density by the
solid line.}
\label{fig:surfacemultiaccretion}
\end{figure*}

The remnants both have $\re \simeq 11.1$~pc and structural properties
consistent with observed NCs as shown in Figure~\ref{fig:ScalRel},
which tracks the evolving NC.  The surface density increases after
each accretion event, evolving along the track of observed
NCs. When we continue to grow the NC in A2 by accreting a further 
10 SCs, the NC becomes denser than the infalling SC.   
Figure~\ref{fig:surfacemultiaccretion} shows the surface density profiles
seen edge-on.  The top panels show mass-weighted maps of the surface
density.  In the middle row of Figure~\ref{fig:surfacemultiaccretion}
we present luminosity-weighted surface density maps, which are more
discy than the mass-weighted ones.  In the bottom panel of Figure
\ref{fig:surfacemultiaccretion} we show the surface density profiles. 

Like the merger remnants, the accretion remnants can be triaxial.
The NC in run A2 is triaxial with a final $\varepsilon_{FO}\simeq0.22$
and a $T\simeq0.4$,  whereas run A3 is rounder with $\varepsilon_{FO}\simeq0.05$ 
and $T\simeq0.1$.  The final edge-on ellipticity at 3~\re\ is
$\varepsilon_{EO}\simeq0.60$ for run A2 and $\varepsilon_{EO}\simeq0.56$ 
for run A3 and a luminosity-weighted ellipticity $\varepsilon_{EO}\simeq0.73$ 
in A2 and $\varepsilon_{EO}\simeq 0.66$ in A3, which is in agreement with 
observed NCs.  Accretion leads to the formation of NCs with discy 
isophotes at large radii as can be seen in Figure~\ref{fig:harma}.  
As in NGC~4244, they show an increase in $B_4$ towards the centre 
and a decline further out.

\begin{figure}
\centering
\includegraphics[angle=270,width=\hsize]{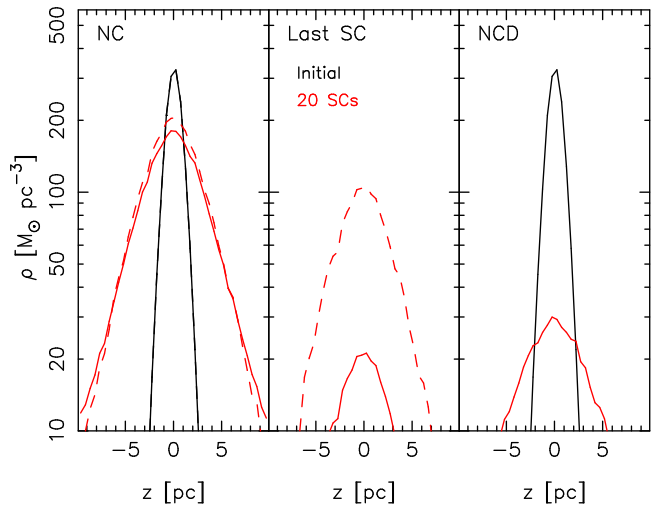}
\caption{The vertical density profile of simulation A2 for the initial NCD
(black) and after 20 accreted SCs (red).  The solid lines show mass-weighted 
and the dashed lines luminosity-weighted profiles.  We plot the density profile
for the total NC, the last accreted SC and the particles of the
initial NCD in the left, middle and right panel, respectively.}
\label{fig:VDP2}
\end{figure}

The vertical density profile of the NC in run A2 at \re\ is shown in 
Figure~\ref{fig:VDP2}.  The NCD is vertically heated by the accretion of 
SCs.  The last accreted SC is distributed in a thinner component than 
the initial NCD.  We again fit the luminosity-weighted density map with an 
elliptical King and an exponential disc profile as in 
\citet{Seth2006}. This gives a NCS with $\re\sim9.8$~pc and a flattening 
of $q\sim0.37$ and a NCD with $z_0\sim1.7$~pc and 
a scale-length $R_d\sim3.3$~pc.  The NCD accounts for $2\%$ of the NC 
mass,  which is $\sim3\times$ smaller than in NGC~4244.  The scale-height 
and scale-length of the NCD in run A2 is about the same of that observed 
in NGC~4244.  Thus, structurally, the NC is comparable to that in NGC~4244, 
provided that the accreted SC is young.  

\begin{figure}
\centering
\includegraphics[angle=270,width=0.9\hsize]{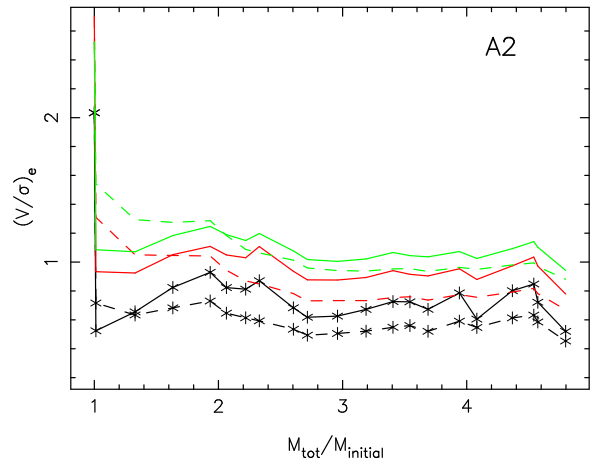} 
\caption{The evolution of the $\left(\rm V/\sigma \right)_{\rm e}$ within \re\
(black lines), 3~\re\ (red lines) and 5~\re\ (green lines) in run A2.
The dashed lines represent mass-weighted and the solid lines
luminosity-weighted measurements adopting the $M/L$ of the NCS and NCD
in NGC~4244.}
\label{fig:EVoS}
\end{figure}

The initial NCD is strongly rotating.  Accretion reduces the rotation
of the remaining NC.  Figure~\ref{fig:EVoS} shows the evolution of
$\left(V/\sigma\right)_e$.  The first accretion leads to a large
decrease in $\left(V/\sigma\right)_e$.  In run A2 the subsequent
accretions induce smaller changes, quickly asymptoting to
$\left(V/\sigma\right)_e \simeq 0.45$ ($\simeq 0.52$
luminosity-weighted).  $\left(V/\sigma\right)_e$ increases with radius
regardless of which weighting is used.  Run A3 instead drops to
$\left(V/\sigma\right)_e \simeq 0.10$ although a
luminosity-weighting gives the appearance of more rotation,
$\left(V/\sigma\right)_e \simeq 0.40$, comparable to that in
run A2.

\begin{figure}
\centering
\includegraphics[angle=270,width=0.9\hsize]{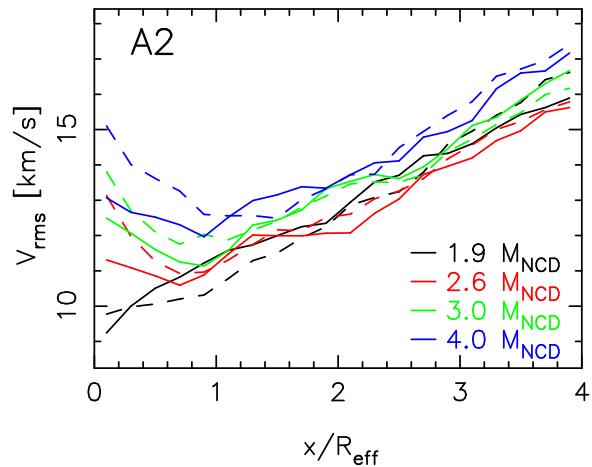} 
\caption{The \vrms\ profile after the NC has grown to
$1.9\times$ (black lines), $2.6\times$ (red lines), $3.0\times$ (green lines) 
and $4.8\times$ (blue lines) of the initial NCD's mass.  The solid lines show 
the $\vrms\left(x\right)$ of the NC seen side-on and the dashed 
lines seen end-on.}
\label{fig:VRMSA2}
\end{figure}

Figure~\ref{fig:VFMA} plots the first two moments of the line-of-sight
kinematics.  The remnant NC in A2 is significantly rotating while that
in A3 shows rotation only when luminosity-weighted.  The $\rm V$ 
shown in the bottom panel of Figure~\ref{fig:VFMA} peaks at larger 
radii compared to those in NGC~4244 and M33.  Unlike in runs
M1-M3 and A1, the NC remnant in run A2 and A3 has \vrms\ that
increases with radius, like the NC of NGC~4244. However the \vrms\ 
profiles in Fig.~\ref{fig:VFMA} shows that in run A2 
\vrms\ increases within \re.  In Fig.~\ref{fig:VRMSA2} 
we show that \vrms\ develops a central peak within \re\ after the 
initial NCD has doubled its mass.  Thus NGC~4244 cannot 
have accreted more than half of its mass as stars. 

\subsection{The vertical anisotropy}
As usual for rapidly rotating discs, the initial NCD is radially
biased, with $\beta_z$ initially large but this decreases with
accretion, although it remains positive.  Unlike with the kinematic
measurements, no important differences in $\beta_z$ and $\beta_\phi$
occur if we weight by luminosity as shown in Figure \ref{fig:EVoS}.  
Regardless of whether mass or luminosity weighting is applied, 
$\beta_z$ peaks within \re\ and declines beyond.

\begin{figure}
\centering
\includegraphics[width=0.9\hsize]{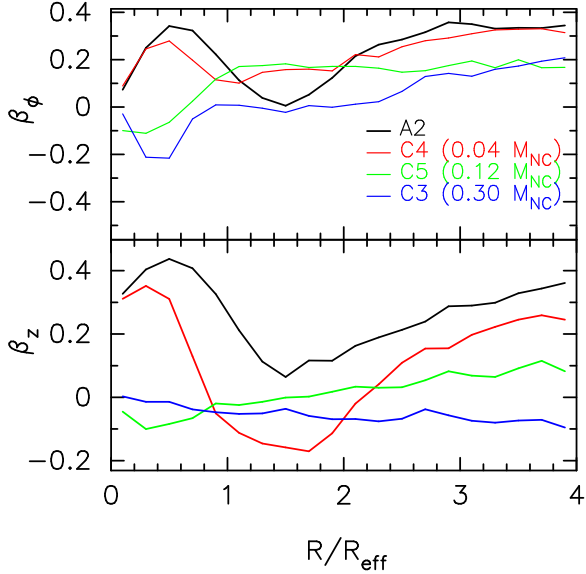} 
\caption{The anisotropies $\beta_\phi$ (top) and $\beta_z$
(bottom) for A2 after the accretion of 20 SCs and for the three test 
runs if we replace the final accreted SC with the indicated SC on a 
polar orbit.}
\label{fig:CylAniEvol058}
\end{figure}

The JAM model of the NC in NGC~4244 has a $\beta_z\simeq-0.2$.  We
have shown in model M2 that accretion of SCs with vertical motions
decreases $\beta_z$.  In run A2 the accreted SCs are in the plane of
the initial NCD.  In further tests we let the NC in A2 accrete the
20th SC on a polar orbit using SC models C4, C5 and C3 
($M=2\times10^5,6\times10^5,2\times10^6\Msun$).  In 
Figure~\ref{fig:CylAniEvol058} we show the anisotropy $\beta_\phi$ and 
$\beta_z$ of the remaining NC after these accretions.  Modest
accretion off the plane of the disc drives $\beta_z$ to negative
values within \re.  The accretion of SC C3 
drives $\beta_z<0$ within 4~\re\ and it also leads to $\beta_\phi<0$ 
within \re.  The accretion of SC C5 causes
$\beta_z<0$ within \re\ and $\beta_\phi<0$ only within $<0.5$~\re.  
Thus the observed negative $\beta_z$ requires that the NC accretes at 
least $\sim10\%$ of its mass directly as stars.

\section{JAM models of simulated NCs}
\label{sec:jams}

To compare the rotation of the real NCs and the simulated ones in a
fully consistent way, we applied the same JAM approach to measure the
rotation of the simulated NCs. The MGE models were fitted to
reconstructed images of all models listed in Table \ref{tab:mergsims}
at an edge-on orientation, for different viewing directions in the
disc plane. For each projection we fitted the anisotropy $\beta_z$ and
$M/L$ of the simulated NCs, and we then measured the rotation
parameter $\kappa$ at the best-fitting $(\beta_z,M/L)$. The inferred
parameters are given in Table~\ref{tab:anisotropy} and compared with the values measured
from the particles.  The comparison shows that the simple JAM models
capture the global anisotropy of the simulated NCs of the merger simulations and gives
confidence in the values we extracted from the real data.  In summary
for all simulations, as with the observations, we measure a degree of
rotation $\kappa\approx1$ within $5\%$ for models M1 and M2, and within $10\%$ 
for model M3.  Given the complex accretion process of the NC it is remarkable for 
the rotation to be so closely linked to the NC shape.  This result is similar to 
what was found for real galaxies in \citet{Cappellari2008} and suggests that a 
general process may be responsible for both observations.  For the models of 
the merger simulations we recover a weak anisotropy, with models for M1 and M3 
anisotropic with
$\beta_z\approx0.25$ and $\beta_z\approx0.17$ respectively, while model M2 is 
closer to fully isotropic with $\beta_z\approx0.04$ within 3~\re.  Contrary to models 
M1-M3, the JAM models do not represent acceptable models for the accretion 
simulations A1-A3. In fact the kinematics predicted from the simulated photometry 
under the assumption of constant $M/L$ and an oblate velocity ellipsoid, is 
qualitatively quite different from the simulated one. The main reason for 
this discrepancy is due to the strong variation in the $M/L$ in the simulated 
NC, which is not included in the JAM models.  Although it would be easy to 
construct JAM models using the gravitational potential directly measured from 
the simulations, this cannot be easily done from the observations.  
However in all accretion simulations the global rotation can robustly be 
determined by JAM.

We perform an independent measure of the degree of rotation of the
simulated NCs using Eqns. \ref{eq:vsigma} and \ref{eq:eps} to place
the simulations M1-M3 and A1-A3 on the
$\left(V/\sigma,\varepsilon\right)$ diagram. For the NCs M1-M3 
the diagram provides
results consistent with the JAM models, for both simulated and
observed NCs.  In particular both models M1 and M3 are in a location 
of the diagram which indicates significant anisotropy, while model M2 and 
the observed NC of M33 and NGC~4244 are close to the isotropic line 
(with a typical uncertainty of $\sim0.1$).  The measured values are shown in
Figure~\ref{fig:vs}. To first order the diagram shows that both the
simulated clusters and the real ones are rapidly rotating.  The simulated 
NCs in runs M1 and M3 are 
flatter than M2, but they have a similar $\left(\rm{V}/\sigma\right)_e$.  
The NCs in the accretion simulations A1-3 are similar flat than the merger 
simulations M1 and M3.  They have comparable 
$\left(\rm{V}/\sigma\right)_e$, although in A3 only the last accreted SC 
causes the rotation.  All these results are consistent with the finding 
of the JAM models.

\begin{table}
\centering
\begin{minipage}{0.5\textwidth}
\caption{Global cylindrical anisotropy and parameters from the JAM models 
for the simulations and observations.  We also present the values 
measured directly from the simulations.}
\centering
\begin{tabular}{@{}lccc|@{\vline}|cc@{}}
\hline
\hline
    &\multicolumn{3}{c@{\vline}}{JAM MGE}&\multicolumn{2}{c}{Simulation} \\
Run & $\beta_z$ & $\kappa$ & $\varepsilon_{EO}$ & $\left<\beta_\phi\right>$ & $\left<\beta_z\right>$  \\
    \hline
M1  &	 0.25	&  0.94  &  0.55 &	0.15	   &  0.38    \\
M2  &	 0.04	&  0.99  &  0.37 &	0.08	   &  0.10    \\
M3  &	 0.17	&  0.90  &  0.56 &	0.17	   &  0.26    \\
A1  &	 n.a.	&  n.a.  &  0.56 &	0.05	   &  0.35    \\
A2  &	 n.a.	&  n.a.	 &  0.73 &	0.10	   &  0.31    \\
A3  &	 n.a.	&  n.a.	 &  0.66 &     -1.92	   &  0.20    \\
\hline
NGC~4244 & -0.2 & 1.06 & 0.43 & &  \\
M33      &  0.0 & 0.84 & 0.28 & &   \\
\hline 
\label{tab:anisotropy}
\end{tabular}
\end{minipage}
\end{table}


\section{Discussion \& Conclusions}
\label{sec:conclusions}

\subsection{In situ formation versus accretion}

We have examined in detail the formation of nuclear clusters (NCs) via the 
mergers of star clusters (SCs).
This has been proposed as an important avenue for NC formation
\citep{Tremaine1975,Miocchi2006,Dolcetta2008,Agarwal2011}.  The main 
support for
this mechanism comes from the similarity of scaling relations between
SCs and NCs \citep{Lotz2001,Walcher2005}. In agreement with previous
studies \citep{Bekki2004,Dolcetta2008,Dolcetta2008b} we find that
such scaling relations are preserved after mergers.

As with previous studies \citep{Bekki2004}, the merger of SCs was found 
to produce triaxial NCs, but we showed that axisymmetry can
result from the presence of intermediate mass black holes (IMBH) or 
sufficient vertical motions.  In
the only observed galaxy where the face-on shape can be determined,
M33, we showed that the NC is most likely axisymmetric.  When the
simulated NCs are viewed edge-on, mergers produce boxy NCs,
unless the merger of SCs occurs onto a pre-existing super star 
cluster or a pre-existing nuclear cluster disc
at the centre.  The flattening is in the range of observed NCs in 
edge-on galaxies \citep{Seth2006}.

Our simulations indicate that a NC formed via merging of cored SCs 
leads to a cored NC \citep{Dehnen2005}.  During the merger, the NC density 
and mass increases and the NC evolves along the track defined by the 
observed density-mass relation for NCs.  However the increase in 
density saturates, when the 
${\rho}_{SC}\left(r_c\right)<3\rho_{NC}\left(r_c\right)$ \citep[Eqn. 8.92, ][]{BinneyTremaine}, 
where $r_c$ is the core radius, because the infalling 
SCs will be disrupted in the outer NC, leading to growth in mass and size but not in 
the mean density of the NC, similar to the evolution in mass and size of 
elliptical galaxies due to minor mergers \citep{Ferreras2009}. 
As seen in Figure~\ref{fig:ScalRel}, observed NCs and Milky Way globular 
clusters (GCs) overlap in the range $10^2<\Sigma_e<10^5\ \Msun\rm{pc}^{-2}$.  
Some observed NCs are denser than the present-day GC population found in the
Milky Way.  Thus if the main formation mechanism of NCs is the accretion of 
SCs, NCs have to form by the merger of denser and more massive SCs than those 
in the Galaxy.  Studies of young massive SCs in interacting 
galaxies \citep{Whitmore1995,McCrady2007} show that they have similar masses
($10^5-10^6$ \Msun) and similar sizes as the Milky Way GCs  
\citep[][and references therein]{Bastian2006}.  Due to the smaller infall times of 
massive SCs \citep{Milosavljevic2004,Neumayer2011}, 
the mergers of these massive SCs is more likely and would explain why NCs 
are denser than the present day GC population. 
 
Observed NCs contain thin, blue discs of young stars ($<$ 100 Myrs).
In NGC~4244, the mass of this disc is about $5\%$ of the total NC mass
- if this is typical for the lifetime of a galaxy then over a Hubble 
time dissipation and star formation is sufficient
to build the NC \citep{Seth2006}.  On the other hand we have 
shown that even if the accreted SCs confer no net angular momentum to 
the NC, because the last accreted young SC dominates the luminosity, the 
apparent rotation can be quite large.  We found that mergers can produce 
rapidly rotating NCs, having
$\left(\rm V/\sigma\right)_e$ values as large as those observed
\citep{Seth2008b}.  However, the second moment of the line-of-sight
velocity distribution \vrms, is centrally peaked, unlike in the
observations.  It is only if we introduce a rapidly rotating
($\left(\rm V/\sigma\right)_e\simeq 2.0$) nuclear disc at the centre of
the initial system that the subsequent evolution produced by infalling
SCs is able to qualitatively reproduce the observed \vrms\ field as
well as the observed isophotal shape, degree of rotation and
mass-density relation, provided no more than half of the NC's mass 
is accreted.  Thus a pure merger origin of NCs can be excluded.    

Our JAM model of the NC in NGC~4244 has a negative $\beta_z=-0.2$.
Because \citet{Cappellari2008} found that a decrease in inclination
leads to an increase in $\beta_z$, we tested the effect of inclination
on $\beta_z$ for the JAM model of NGC~4244.  For inclinations
$>75\degrees$, we found $\beta_z$ continues to be negative at a
$3\sigma$ level.  Thus the $\beta_z<0$ in NGC~4244 cannot be a
projection effect and must be intrinsic.  However, a NC disc formed
out of gas cooling would have $\beta_z>0$.  For example, the JAM model
of the NC in NGC~404 has $\beta_z\sim0.5$ \citep{Seth2010a}; in this
case the observed burst of star formation $\sim1$~Gyr ago could easily
have fed its NC and given rise to its positive $\beta_z$.  By
scattering box orbits, central black holes can lower $\beta_z$
\citep[e.g.][]{Merritt1998}.  Whereas \citet{Seth2008a} find indirect
evidence of supermassive black holes (SMBHs), with masses ranging from
$10-100\%$ in $\sim10\%$ of NCs, in NGC~4244 the radius of influence
for the largest possible intermediate mass black hole is $<1$~pc,
while the NC has $\re=5$~pc.  Therefore the presence of a intermediate 
mass black holes in the
centre of NGC~4244 cannot explain its $\beta_z<0$.  The only way we
were able to produce $\beta_z<0$ was through the accretion of SCs on
highly inclined orbits. Thus, at least $\sim10\%$ of the mass of the
NC of NGC~4244 needs to be accreted as SCs in order to obtain $\beta_z
< 0$, which constitutes a lower limit on the amount of mass accreted
in the form of star clusters.

\subsection{Nuclear Cluster Formation in Dwarf Elliptical Galaxies}

Our simulations can also make testable predictions relating to the 
formation of nuclei in dE galaxies.  \citet{Lotz2001} 
suggested that NCs in dE galaxies could have
formed by the merger of GCs.  They found a depletion of the most massive GCs
relative to the less massive ones in the inner
region of galaxies.  They interpreted this as being
due to shorter dynamical friction infall timescales for massive GCs
than for the lower mass ones.  In the ACS
Virgo Cluster Survey, \citet{Cote2006} found that, on average, NCs in
dE galaxies are $3.5$~mag brighter than the mean GC. So if
NCs form via the merger of GCs, on average about 25 GCs need to
merge.  Similar to what we found for NCs
in late-type spiral galaxies (Figure~\ref{fig:ScalRel}),
\citet[][see their Figure~18]{Cote2006} found that the NCs are
denser than the mean density of the GC population in dE galaxies.
They also found that the fraction of red GCs and the 
$\left(g-z\right)'_{AB}$ colour of NCs increases with the host 
galaxy luminosity
\citep{Peng2006,Cote2006}.  Using Monte Carlo simulations, and
assuming that these NCs formed via mergers, \citet{Cote2006} 
found that
the resulting scaling relation of the NC's colours is less steep than
those observed.  From spectroscopic data \citet{Paudel2010} found that
the metallicity of NCs in a sample of dE galaxies correlates with the
luminosity of the host galaxy.  They also found that the median
difference in age between the NC and the galactic main body is about
$\sim3.5$~Gyrs and that the difference is more prominent in dEs with
discy isophotal shapes. This implies that the formation of NCs in dEs
might be enhanced by the accretion of gas.  In contrast,
\citet{Paudel2010} found fairly old and metal-poor NCs in very faint
dEs, resembling the properties of the GC population.  This
suggests that NCs in faint dEs might have formed by different
processes than the NCs in brighter dEs.  

In summary both the accretion of gas with in situ star formation and the
merger of GCs could be at work to form NCs in dE galaxies.  If NCs in dEs 
form via merger of GCs, our simulations
indicate that they will have boxy shapes, have centrally peaked \vrms\ 
and be radially biased.  On the other hand, if the main formation mechanism is
dissipation, NCs will have discy isophotes and no centrally peaked \vrms.  
At present, no available observational data exists which is able to test these
predictions.

\subsection{The \Mcmo-\sig\ relation}
The star formation  histories in NCs of late-type
spiral galaxies are extended, with the youngest population of stars less than
100~Myrs old \citep{Walcher2006,Rossa2006,Seth2010a}.  NCs appear to be offset 
from the \Msig\ relation of  SMBHs
\citep{Ferrarese2006,Wehner2006}.  NCs have the same slope, but, for a given
velocity dispersion \sig, are $10\times$ more massive  than SMBHs. 
\citet{McLaughlin2006b} found that this offset can be explained by feedback from 
stars and supernovae.  Our simulations
indicate that at least  $10\%$ of the mass of NCs needs to be accreted by SCs 
and that at least $50\%$ of its mass needs to be accreted as gas.  Therefore 
a fraction of the star formation could occur outside the NC.  However, 
the accreted SCs have to be young and therefore still formed within the central region.  
SCs with masses in a range of $10^5-10^6\ \Msun$ have to be formed within $60-200$~pc,
otherwise their infall times are longer than 100~Myrs.  Therefore, even if
the NC has accreted half of its mass in young SCs, stellar
feedback occurs within the central region of the galaxy and therefore remains a
plausible explanation for the $\Mnc-\sig$.

\subsection{Summary}

We have studied the formation and evolution of stellar nuclear
clusters using $N$-body simulations.  Our main conclusions can be
summarised as follows:

\begin{itemize}

\item We find no evidence of non-axisymmetry in the nuclear cluster of
M33.  Its PA is consistent with that of its main disc and its apparent
ellipticity is consistent with a vertical flattening $q = 0.7$, which
is the average observed in the NCs of edge-on late-type galaxies
\citep{Seth2006}.  There is also only a small misalignment between the
photometric and kinematic PAs.  While this is the only galaxy where
this measurement can be done at present, it suggests that NCs are
generally axisymmetric.

\item The NC in NGC~4244 is nearly isotropic, with $\beta_z =
-0.2\pm0.1$, and is rapidly rotating.  It has a mass of $(1.1\pm0.2)
\times 10^7~\Msun$; if an IMBH is present its mass is less than $1\%$
that of the NC.  It is not possible from these models to distinguish
whether the rotation is present throughout the NC or is restricted to
the NCD.
 
\item The mergers of SCs produce NCs with density and sizes consistent
with observations, evolving along the track of observed NCs.  Remnant 
NCs can be axisymmetric.  Multiple accretions of
young SCs onto a pre-existing nuclear cluster spheroid can also
produce discy isophotal shapes.  Mergers can lead to rapidly rotating
NCs, as observed. They have $\beta_z \la 0.4$, where initial vertical
motions induce the smallest value.  However these NCs have centrally
peaked \vrms\, unlike observed NCs in late-type discs. 

\item Accretion of young SCs onto a bare NCD produces NCs with
densities, sizes and ellipticities comparable to those observed.  They
show discy isophotes and have rotations comparable to those in the NCs
of NGC~4244 and M33.  \vrms\ is dominated by dispersion if the accreted 
mass is greater than the initial mass of the NCD.  The formation of 
the NC in NGC~4244 therefore requires at least $50\%$ of the mass to be 
accreted as gas to match the observations. 

\item The negative $\beta_z$ in the NC of NGC~4244 requires at least 
$\sim10\%$ it's total mass to have been accreted from 
SCs.

\item We caution that even if the accreted SCs do not impart any net
angular momentum, the last accreted SC can dominate the apparent rotation when
luminosity-weighting and could be similar to the $\left(\rm{V}/\sigma\right)_e$ 
and \vrms\ fields of observed NCs.  

\item With the results presented in this paper, the 
simulations are now ahead of the observations with predictions of detailed 
observables that can be used to constrain the formation scenarios better. 
Integral-field observations of the kinematics of more NC are essential for 
further progress and we hope these will be obtained in the near future.
\end{itemize}

\section*{Acknowledgements} 

Part of the simulations in this paper were performed on the COSMOS 
Consortium supercomputer within the DIRAC Facility jointly funded by 
STFC, the Large Facilities Capital Fund of BIS.  Simulations have 
also been run at the Arctic Region Supercomputing
Center (ARSC) and the High Performance Computer Facility at the
University of Central Lancashire. We thank Jerry Sellwood for
providing us with the multi-mass bulge model and Jakob Walcher for
providing us with his observational data.  Markus Hartmann thanks 
Chris Brook and Rok Ro\v{s}kar for their support and help.  Michele Cappellari
acknowledges support from a Royal Society University Research Fellowship.


\bibliographystyle{mn2e}
\bibliography{nsc}

\label{lastpage}

\end{document}